\documentclass[a4paper,prd,noshowpacs,noshowkeys,notitlepage,nofootinbib,12]{revtex4}
\usepackage[T1]{fontenc}
\usepackage{ae,aecompl} 
\usepackage{amsmath,amssymb,mathrsfs}
\usepackage{graphicx}
\usepackage[]{units}
\usepackage[usenames,dvipsnames,svgnames]{xcolor} 
\usepackage{hyperref}
\hypersetup{
   colorlinks=true,       
    linkcolor=Blue,          
    citecolor=Plum,        
    filecolor=magenta,      
    urlcolor=YellowOrange           
}
\usepackage{ccnishi}
\usepackage[normalem]{ulem} 
\usepackage{setspace} 

\providecommand{\lag}{\mathscr{L}}
\providecommand{\tY}{\tilde{Y}}
\providecommand{\tV}{\tilde{V}}
\providecommand{\tK}{\tilde{K}}
\providecommand{\tX}{\tilde{X}}
\providecommand{\tI}{\tilde{I}}
\providecommand{\YY}{\mathbb{Y}}
\providecommand{\cM}{\mathcal{M}}
\providecommand{\XX}{\mathbb{X}}
\providecommand{\cV}{\mathcal{V}}
\providecommand{\MM}{\mathbb{M}}
\providecommand{\aver}[1]{\langle #1 \rangle}
\providecommand{\eq}[1]{\begin{equation} #1 \end{equation}}
\providecommand{\eqali}[1]{\begin{equation}\begin{aligned} #1
    \end{aligned}\end{equation}}
\providecommand{\xlink}[1]
  {\href{http://arxiv.org/abs/#1}{arXiv:#1}}
\DeclareMathOperator{\HS}{HS}
\DeclareMathOperator{\PL}{PL}
\DeclareMathOperator{\PE}{PE}

\begin{document}
\title{
Flavor invariants for the SM with one singlet vector-like quark
}
\author{E.~L.~F.~de~Lima}
\email{eduardo.lourenco@ufabc.edu.br}
\affiliation{Centro de Ci\^{e}ncias Naturais e Humanas\\
Universidade Federal do ABC -- UFABC, 09.210-170
Santo Andr\'{e}-SP, Brazil}
\author{C.~C.~Nishi}
\email{celso.nishi@ufabc.edu.br}
\affiliation{
Centro de Matemática, Computação e Cognição\\
Universidade Federal do ABC -- UFABC, 09.210-170
Santo André, SP, Brazil
}

\begin{abstract}
We study the flavor invariants of the SM augmented by one singlet vector-like quark.
Aided by the Hilbert series, we construct all the basic invariants with which any flavor invariant can be written as a polynomial.
In special, this theory contains one CP odd invariant of degree six which has degree much lower than the usual Jarlskog invariant of the SM.  
We find the nonlinear polynomial relations (syzygies) of lowest degrees involving these basic invariants, including the expression of the square of the CP odd invariant of lowest degree in terms of CP even invariants.
The $SU(3)$ identity underlying this syzygy is uncovered in terms of invariant tensors, which can be applied to rewrite any square of a CP odd invariant of the same form, involving three hermitean matrices of size three.
We demonstrate by an example that there is CP violation that is not detected by the CP odd invariants proposed in the literature so far but it can be detected with the full list of CP odd invariants found here.
\end{abstract}
\maketitle
\section{Introduction}

The nontrivial quark flavor structure of the SM has its origin in the two non-aligned and complex Yukawa couplings with the Higgs doublet.\,\footnote{
In the lepton sector, the basic origin is not yet clear.
}
Requiring at least 20 parameters once neutrino masses are included, the flavor sector is undoubtedly the sector that requires the most experimental input.
Explaining this puzzling structure with hierarchical masses and nontrivial mixing from an underlying theory constitutes the so-called \emph{flavor puzzle}\,\cite{flavor.puzzle}.
The established phenomenon of CP violation is also inextricably connected to the same source of flavor changing while flavor blind CP violation in the form of the $\bar\theta$ parameter is constrained to be unnaturally small.
Although showing up only in flavor changing processes, CP violation within the SM can be quantified by a single CP odd fully weak basis invariant known as Jarlskog invariant\,\cite{jarlskog,branco.gronau}:
\eq{\nonumber
\det([X_u,X_d])
=2i(y_u^2-y_c^2)(y_c^2-y_t^2)(y_t^2-y_u^2)(y_d^2-y_s^2)(y_s^2-y_b^2)(y_b^2-y_d^2)J\,,
}
where $J$ is the rephasing invariant of the CKM mixing matrix $V$, 
\eq{\nonumber
J=\im[V_{us}V_{cb}V^*_{ub}V^*_{cs}]=c_{12}c_{13}^2c_{23}s_{12}s_{13}s_{23}\sin\delta\,.
}
This CP odd invariant neatly embodies all the CP violation present in the SM but most of the information contained in it can be obtained from other CP even invariants\,\cite{manohar:hilbert,silva.trautner}.
The explanation is simpler for $J$, where the modulus $|J|$ measures half the area of the unitarity triangle which clearly indicates that it can be fully recovered from CP even quantities, i.e., the sides of the triangle\,\cite{branco:|V|}. The sign is thus the only new information provided by $J$ if we have access to a sufficient number of CP even quantities.

Once the SM is extended to solve some of its completeness issues, a generic flavored structure is highly constrained for new physics not far from the TeV scale, which leads to the \emph{flavor problem}\,\cite{isidori:lec12}.
That includes new sources of CP violation which, although desirable to help explain the matter-antimatter asymmetry of the universe, is highly constrained for quarks.
In this context, CP odd basis invariants are useful tools to detect and quantify CP violation in a basis (or rephasing) invariant manner.
In special, if one ensures that all CP odd invariants vanish, then the theory would conserve CP.
This idea was applied, for example, for the 2-Higgs-doublet model\,\cite{invs:2hdm}, which has a clear geometrical meaning when formulated in terms of field bilinears\,\cite{bilinear,nhdm:cp}.
Basis invariant necessary and sufficient conditions were also formulated for the 3-Higgs-doublet model (3HDM)
\cite{nhdm:cp}.
An alternative approach is to make maximal use of basis covariance which, e.g., leads to simpler relations to detect symmetries in the 3HDM\,\cite{beyond.invs}, allows writing the 3-loop renormalization group equations (RGE) for the 2HDM\,\cite{bednyakov:2hdm}, simpler RGEs for the SM Yukawas\,\cite{mannel} 
or the beta function form for the $\theta$ parameter in non-abelian gauge theories\,\cite{beta.theta}.
For other contexts of application of basis invariants, see Refs.\,\cite{lebedev,flavor.invs:other}.

As the theories become more complex, the procedure of finding and constructing invariants of ever higher degrees encounters the problem of when to stop.
This problem is solved by recognizing that the invariants form a ring and, when finitely generated,%
\footnote{
This is the case of reductive groups including the usual semi-simple groups.
}
any invariant can be written as a \emph{polynomial} of a finite set of \emph{basic} invariants whose number and degree can be found with the use of tools known as the Hilbert series (HS) and its plethystic logarithm (PL) \cite{manohar:hilbert,pouliot,lehman.1,gray.hanany,hanany:07,hanany.torri}.
As a counting tool of group invariants, one natural playground for its application is the Standard Model Effective Field Theory (SMEFT)\,\cite{smeft} where the Hilbert series was applied extensively to count and classify the independent higher dimensional operators\,\cite{lehman.1,hs:smeft,grojean:cp}.
The Hilbert series method can even be generalized to treat group covariants\,\cite{grinstein.merlo}.
For the applications of the Hilbert series in other contexts, see Refs.\,\cite{hs:multi-higgs,yu.zhou:inv.nu,hs:others}.

Here we tackle the problem of finding the basic flavor invariants for the SM extended by one singlet vector-like quark (VLQ). The addition of VLQs is a minimal extension of the SM that is well motivated by many aspects\,\cite{vlq:review}.
One of the motivations concerns the presence of additional sources of CP violation which can be much larger than in the SM.
Flavor invariants in this context were studied in Refs.\,\cite{vlq.inv,saavedra:vlq.inv} and more recently in Ref.\,\cite{albergaria} without the use of the Hilbert series.
In the context of Nelson-Barr models\,\cite{nelson.barr} to solve the strong CP problem, some CP odd invariants were studied in Refs.\,\cite{nb-vlq,vecchi.1}.

The outline of this article is the following: we set up the notation for the model composed of one singlet VLQ added to the SM in Sec.\,\ref{sec:spurions} and spell out the flavor transformations under which the couplings transform as spurions.
In Sec.\,\ref{sec:HS}, we show the Hilbert series counting the flavor invariants and also list the basic invariants with which all the invariants can be written as a polynomial.
Section \ref{sec:syzygy} discusses the construction of the basic invariants and the syzygies of the lowest degrees.
We find that some of these syzygies are directly generalizable for more than one VLQ.
We uncover in Sec.\,\ref{sec:su3.ff} the $SU(3)$ identity underlying the syzygy of lowest degree corresponding to writing the square of the lowest degree CP odd invariant in terms of CP even invariants.
This identity clarifies why the syzygy is not generalizable to more than one VLQ.
In Sec.\,\ref{sec:phys.param}, we make the connection between our invariants and physical parameters.
A one-to-one relation between physical parameters and primary invariants is discussed subjected to discrete ambiguities and, excluding some special points, a simpler one-to-one relation between physical parameters and a set of invariants can be written for which it is shown how to resolve the sign ambiguities.
Section \ref{sec:SU4} discusses the relation with the full weak basis transformations that mix the four righthanded quarks, including the VLQ. We find that the constructed basic invariants are essentially the same.
It is also shown that some additional CP odd invariants ---counted in our basic set--- may be necessary to detect CP violation for some special cases.
The summary is presented in Sec.\,\ref{sec:summary}.

\section{Spurions for one singlet VLQ of down-type}
\label{sec:spurions}

We consider the SM adjoined with one VLQ $B_L,B_R$ with the same gauge quantum numbers as the singlet down quarks $d_{iR}$.
The relevant Lagrangian including the SM Yukawa interactions is
\begin{equation}
\label{lag:vlq}
      -\lag = \overline{q}_{iL}\Tilde{H} Y^u_{ij}  u_{jR} + \overline{q}_{iL} H Y^d_{ij} d_{jR} +
      \overline{q}_{iL} H Y^B_i B_{R} + \overline{B}_{L} M B_{R}+ h.c.,
\end{equation}
where the sum over repeated flavor indices is implicit for $i,j=1,2,3$.
The case of one singlet up-type VLQ is analogous.
We have already chosen the basis in the space $(d_{iR},B_R)$ appropriately to eliminate some terms and we call this basis the VLQ mass basis.
Other possible weak basis choices can be found in Ref.\,\cite{vlq:review}.
The matrices $Y^u,Y^d$ are complex $3\times 3$ matrices in flavor space whereas $Y^B$ is a $3\times 1$ complex matrix.
This basis is well suited for matching to the SMEFT after integrating out the VLQ\,\cite{delAguila:2000rc}.

In the absence of $Y^u,Y^d,Y^B$, the theory is characterized by the flavor symmetry group 
\begin{equation}
\label{GF:vlq:MB}
    U(3)_q\otimes U(3)_u\otimes U(3)_d \otimes U(1)_{\text{VLQ}},
\end{equation}
where the group $U(3)_q\otimes U(3)_u\otimes U(3)_d$ is the same as in the SM:
\eq{
q_L\to U^qq_L\,,\quad
u_R\to U^u u_R\,,\quad
d_R\to U^d d_R\,,\quad
}
with $U^q\in U(3)_q$, $U^u\in U(3)_u$, $U^d\in U(3)_d$.
In the same limit, the theory including the bare mass term $M$ in \eqref{lag:vlq}, is also invariant by $U(1)_{\text{VLQ}}$ corresponding to the simultaneous (vector) phase rotation of the VLQ:
\eq{
\label{vlq:number}
  B_L \to e^{i\theta} B_L,\quad     B_R \to e^{i\theta} B_R\,.
}

When the Yukawa couplings $Y^u,Y^d,Y^B$ are turned on, the group \eqref{GF:vlq:MB} is explicitly broken.
Invariance is restored if the Yukawa couplings $Y^u,Y^d,Y^B$ are spurions transforming as 
\eq{
\label{spurions:Y}
Y^u\to U^qY^u (U^u)^\dag\,,\quad
Y^d\to U^qY^u (U^d)^\dag\,,\quad
Y^B\to U^q Y^B e^{-i\theta}\,.
}
We can disregard $U(3)_u$ and $U(3)_d$ by considering
\eq{
X_u\equiv Y^u(Y^u)^\dag\,,\quad
X_d\equiv Y^d(Y^d)^\dag\,.
}
Instead of \eqref{GF:vlq:MB}, the relevant flavor group is just
\eq{
\label{GF}
G_F=SU(3)_q\otimes U(1)_{\rm VLQ}\,,
}
where we drop the unnecessary $U(1)_q$.
We then take our basic spurions as $X_u,X_d$ and $Y$, where we use $Y$ in place of $Y^B$ to simplify the notation.
Under $G_F$, these spurions transform as
\eq{
\label{spurions}
X_u\sim (\bs{3}\otimes\bs{\bar{3}},0)\,,\quad
X_d\sim (\bs{3}\otimes\bs{\bar{3}},0)\,,\quad
Y\sim (\bs{3},-1)\,,\quad
Y^\dag\sim (\bs{\bar{3}},+1)\,.
}
It is clear that $YY^\dag$ transforms in the same way as $X_u$ or $X_d$ but we need to consider $Y$ and $Y^\dag$ as separate spurions to correctly track their degrees of freedom as $3\times 1$ (or $1\times 3$) complex matrices instead of $YY^\dag$ which is a $3\times 3$ hermitean matrix constrained to be rank one.
Because of that, within traces, we only need to consider the first power of $YY^\dag$ since its second power is proportional to the original matrix as
\eq{
\label{YY2=YY}
(YY^\dag)^2=(Y^\dag Y)YY^\dag\,.
}

\section{Hilbert series in VLQ mass basis} 
\label{sec:HS}

Considering the basic spurions $X_u,X_d,Y,Y^\dag$ in \eqref{spurions}, we can calculate the \emph{unrefined} Hilbert series 
$H(q)$ through the Molien-Weyl formula\,\cite{gray.hanany,hanany.torri,lehman.1} with the variable $q$ representing our four spurions $X_u,X_d,Y,{Y}^\dag$. Note the difference with respect to the \emph{physical degree} for which $X_u$ would have the same degree as the product $YY^\dag$.
After performing the residue integrals, we obtain
\eqali{
\label{HS:vlq.mass}
H(q)&=
\frac{{1 + 2q^4 + 4q^5 + 5q^6 + 2q^7 + 2q^8 + 2q^9 + 5q^{10} + 4q^{11} + 2q^{12} + q^{16}}}
{(1-q)^2\left(1-q^2\right)^4\left(1-q^3\right)^6\left(1-q^4\right)^3}
\cr
&=1+2q+7q^2+18q^3+44q^4+98q^5\cdots\,.
}
See appendix \ref{app:HS} for the details. 
There we also show the Hilbert series in terms of the physical degree\footnote{%
We thank the anonymous referee for suggesting the replacement $X_u\to q^2,X_d\to q^2$.
Compared to the series in \eqref{HS:vlq.mass:q^2}, the series \eqref{HS:vlq.mass} has the advantage 
that its PL shows no cancellations between positive and negative terms.
}
in \eqref{HS:vlq.mass:q^2} and the refined (graded) Hilbert series as well.
The denominator of the $H(q)$ carries the information about the primary invariants, i.e., those 
invariants that are algebraically independent. There are 15 factors in the denominator of the $H(q)$ 
in \eqref{HS:vlq.mass}, which means there are a total of 15 primary flavor invariants, three 
invariants of degree four, six invariants of degree three, four invariants of degree two and two 
invariants of degree one. 
The case of one up-type singlet VLQ would lead to an identical Hilbert series.

The number of primary invariants should also coincide with the number of physical parameters contained in $X_u,X_d,Y$.
As in the SM, $X_u$ and $X_d$ contains 10 physical parameters in the basis where, e.g., $X_u$ is diagonal.
They correspond to three up-type masses, three down-type masses, one CP odd and three CP even mixing parameters. 
The VLQ Yukawa $Y$ contains five physical parameters because one phase can be removed by rephasing using \eqref{vlq:number}. The total is correctly 15.
See Sec.\,\ref{sec:phys.param} for more details.

We can also gain information about the basic invariants, i.e., the generating set of the ring of invariants with which we can write any invariant as a polynomial.
The plethystic logarithm (PL) of the Hilbert series contains such a information and is given by
\begin{equation}
    \PL[H(q)]= \sum_{r=1}^\infty \frac{\mu(r) \log H(q^r)}{r}\,,
\end{equation}
where $\mu(r)$ is the Möbius function.
Explicit calculation leads to the series
\eqali{
\label{PL:vlq.mass}
\PL[H(q)]&=2 q+4 q^2+6 q^3+5 q^4+4 q^5+5 q^6+2 q^7-q^8-6 q^9-15 q^{10}
+\cdots\,.
}
Since the series is nonterminating, this ring of invariants is a non-complete 
intersection\,\cite{yu.zhou:inv.nu,derksen}.
The first positive terms of the PL usually indicate the number of basic 
invariants, with the caveat that cancellations may take place with the following negative terms related to the syzygies\,\cite{yu.zhou:inv.nu}.\footnote{%
We are not aware of a proof about the relation between the first positive coefficients of the PL with the number of basic invariants.
The PL in \eqref{PL:vlq.mass:q^2} written in terms of the physical degree also exhibits a cancellation between one positive term and one negative term.
}
So the number of basic invariants should be always validated by explicit construction and check of 
reducibility of further invariants.
Barring the possibility of cancellations,
the first negative coefficients indicate the presence of one syzygy of degree 8 and six syzygies of 
degree 9.
From the positive terms of the PL in \eqref{PL:vlq.mass}, 
we infer the presence of
28 basic invariants in the generating set, which we constructed and list in Table 
\ref{tab:flavor_invariants}, with 9 CP-odd and 19 CP-even invariants. 
We employ the notation $\aver{A}=\tr[A]$ for traces.
The 9 CP-odd invariants are not primary because their square can be written in terms of other CP-even invariants.%
\footnote{This is the case of one singlet VLQ. See comments after \eqref{syzygy:ff} and in Sec.\,\ref{sec:su3.ff} for more than one VLQ.}
CP violation is manifested in the CP-odd invariants which are expressed as a subtraction between two terms (or a commutator).
They are purely imaginary because their second terms are the complex conjugate of the first terms.
For the last CP odd invariants without CP even counterparts, changing the subtraction to summation render them reducible, i.e., expressible in terms of previously defined invariants. 
Further discussion on how we obtained the list of basic invariants in Table \ref{tab:flavor_invariants} is given in the next section.

Table~\ref{tab:flavor_invariants} contains only traces with no $YY^\dag$ or with only one $YY^\dag$ as higher powers are reducible due to \eqref{YY2=YY} while multiple insertions split the trace into more than one trace.
The invariants with no $YY^\dag$ are denoted by $I_{(2n)(2m)}$, representing an invariant of degree $n$ in $X_u$ and degree $m$ in $X_d$, so the labels effectively track the degree in $Y^u,Y^d$ respectively.
In form, these are the same invariants as in the SM\,\cite{manohar:hilbert}.\footnote{%
Note, however, that $X_d$ can deviate from the SM values when considering the VLQ. 
}
The invariants with one $YY^\dag$ are denoted by $K_{(2n)(2m)}$, with the same degree in $X_u,X_d$ as $I_{(2n)(2m)}$.
When there are more than one invariant with the same degree, we also include the label $\pm$ depending on if it is a sum or subtraction of two terms.
This sign also coincides with the CP parity. 
In the table we also list two types of degrees for each invariant.
The physical degree corresponds to the degree where each of $Y^u,Y^d,Y$ carries one unit.
This contrasts with the degree in the Hilbert series \eqref{HS:vlq.mass} and in the PL \eqref{PL:vlq.mass} where each of $X_u,X_d,Y$ (or $Y^\dag$) count as degree one.
The CP odd invariant of lowest degree $K_{22}^-$ is of physical degree 6 which has degree much lower than the Jarlskog invariant $I^-_{66}=\det[X_u,X_d]=\aver{[X_u,X_d]^3}/3$ of the SM which is degree 12.
The invariant $K_{22}^-$ was already considered in the literature to estimate 3-loop contributions to $\bar{\theta}$ in Nelson-Barr models\,\cite{vecchi.1}.
\begin{table}[h]
\centering
\begin{tabular}{|c|c|c|c|}
\hline
Flavor Invariant & Phys. degree & degree in \eqref{PL:vlq.mass} & CP \\
\hline
$ I_{20} = \aver{X_u} $ & 2 & 1 &+ \\
$ I_{02} = \aver{X_d} $ & 2 & 1 &+ \\
$ K_{00} = \aver{YY^\dagger} $ & 2 & 2 & + \\
$ I_{40} = \aver{X^2_u} $ & 4 & 2 & + \\
$ I_{04} = \aver{X^2_d} $ & 4 & 2 & + \\
$ I_{22} = \aver{X_u X_d} $ & 4 & 2 & + \\
$ K_{20} = \aver{X_u YY^\dagger} $ & 4 & 3 & + \\
$ K_{02} = \aver{X_d YY^\dagger} $ & 4 & 3 & + \\
$ I_{60} = \aver{X^3_u} $ & 6 & 3 & + \\
$ I_{06} = \aver{X^3_d} $ & 6 & 3 & + \\
$ I_{42} = \aver{X^2_u X_d} $ & 6 & 3 & + \\
$ I_{24} = \aver{X_u X^2_d} $ & 6 & 3 & + \\
$ K_{40} = \aver{X_u^2YY^\dagger} $ & 6 & 4 & + \\
$ K_{04} = \aver{X_d^2YY^\dagger} $ & 6 & 4 & + \\
$ K^+_{22} = \aver{\{X_u,X_d\} YY^\dagger }$ & 6 & 4 & + \\
$ K^-_{22} = \aver{[X_u,X_d]YY^\dag} $ & 6 & 4 & - \\
$ I_{44} = \aver{X_u^2X_d^2} $ & 8 & 4 & + \\
$ K^-_{42} = \aver{[X^2_u, X_d]YY^\dag} $ & 8 & 5 & - \\
$ K^+_{42} = \aver{\{X_u^2,X_d\}YY^\dag} $ & 8 & 5 & + \\
$ K^-_{24} = \aver{[X_d^2,X_u]YY^\dag} $ & 8 & 5 & - \\
$ K^+_{24} = \aver{\{X^2_d,X_u\}YY^\dag} $ & 8 & 5 & + \\
$ K^-_{44} = \aver{[X^2_u,X^2_d] YY^\dag} $ & 10 & 6 &- \\
$ K^+_{44} = \aver{\{X^2_u,X^2_d\} YY^\dag} $ & 10 & 6 &+ \\
$ K_{62}^- = \aver{X_u^2X_dX_uYY^\dag} - \aver{X_uX_dX_u^2YY^\dag} $ & 10 & 6 &- \\
$ K^-_{26} = \aver{X_d^2X_uX_dYY^\dag}-\aver{X_dX_uX_d^2YY^\dag} $ & 10 & 6 &- \\
$ I_{66}^- = \aver{X_u^2X_d^2X_uX_d}-\aver{X_d^2X_u^2X_dX_u} $ & 12 & 6 & - \\
$ K^-_{64} = \aver{X_u^2X_d^2X_uYY^\dag}-\aver{X_uX_d^2X_u^2YY^\dag} $ & 12 &7 &- \\
$ K^-_{46} = \aver{X_d^2X_u^2X_dYY^\dag}-\aver{X_dX_u^2X_d^2YY^\dag} $ & 12 & 7&- \\
\hline
\end{tabular}
\caption{\label{tab:flavor_invariants}%
Summary of the 28 basic flavor invariants where $\aver{A}\equiv\tr[A]$. Along with their physical degree, the degree in the PL \eqref{PL:vlq.mass}, and CP parity. 
We use the commutator $[X_u, X_d] \equiv X_u X_d - X_d X_u$ and the anticommutator $\{X_u, X_d\} \equiv X_u X_d + X_d X_u$.}
\end{table}

\section{Reduction of invariants and syzygies}
\label{sec:syzygy}

Here we complement the discussion on how to construct the basic invariants in table~\ref{tab:flavor_invariants}.
The invariants without $YY^\dag$ follow the same structure of the flavor invariants within the SM, discussed previously in the literature\,\cite{manohar:hilbert}.
So we only need to discuss invariants with one $YY^\dag$ in the trace.
The basic structure is the trace of a chain of matrices,
\eq{
\label{K:chain}
\aver{A_1A_2\cdots A_rYY^\dag}\,,
}
where each $A_i$ can be one of 
\eq{
\label{Ai:K}
\{X_u,X_d,(X_u)^2,(X_d)^2\}\,,
}
with neighboring matrices being distinct.
Let us denote the number $r$ of matrices $A_i$ in \eqref{K:chain} as the \emph{$K$-length} of the invariant of $K$-type.
The cubic powers $(X_u)^3$ or $(X_d)^3$, or higher powers, are excluded because these powers can be reduced in terms of other traces due to the Cayley-Hamilton theorem (CHT) for a $3\times 3$ matrix.
Another consequence of the CHT is that\,\cite{manohar:hilbert}
\eq{
\label{manohar.id}
\aver{ABAC}+\aver{A^2BC}+\aver{BA^2C}=\text{reducible}\,,
}
where $A,B,C$ are any $3\times 3$ matrix and \emph{reducible} here means that it is a combination of terms with the product of at least two traces (invariants).
This identity allows us to exclude invariants with repeating $(X_u)^2$ or $(X_d)^2$ in any part of the chain since the fourth power is reducible.
If $X_u$ or $X_d$ is the repeating matrix, then an invariant with a chain of $k$ matrices among \eqref{Ai:K} can be written in terms of invariants with $K$-length at most $k-1$.

In table~\ref{tab:flavor_invariants}, the $K$ invariants of $K$-length up to two are exhausted from the top up to $K^+_{44}$. The rest of the invariants in the table has $K$-length three and only the CP odd ones are kept because the CP even ones are reducible again by \eqref{manohar.id}.
They exhaust the CP odd invariants of $K$-length three.
For $K$-length four, without repeating the matrices in \eqref{Ai:K}, we can construct the following CP odd invariants:
\eqali{
\label{inv.u3d3y2}
K_{66}^{(1)}&=\aver{X_u^2X_d^2X_uX_dYY^\dag}-c.c.,
\cr
K_{66}^{(2)}&=\aver{X_d^2X_u^2X_dX_uYY^\dag}-c.c.,
\cr
K_{66}^{(3)}&=\aver{X_u^2X_dX_uX_d^2YY^\dag}-c.c.,
\cr
K_{66}^{(4)}&=\aver{X_uX_d^2X_u^2X_dYY^\dag}-c.c.,
}
where $c.c.$ denotes the complex conjugate.
Note that $\aver{AB}^*=\aver{A^\dag B^\dag}$.
One can check that all the invariants in \eqref{inv.u3d3y2} are related by the identity \eqref{manohar.id}, modulo reducible terms, considering one of $X_u$ in $X_u^2$ as a repeating matrix or, similarly, considering $X_d$.
They can be more easily related from the fact that $\aver{A^2BAC}+\aver{ABA^2C}$ is reducible due to the reducibility of $\aver{(A+B+C)^5}$.
However, the reduction of any of \eqref{inv.u3d3y2} cannot be achieved directly by the identity \eqref{manohar.id}.
Their reduction involves the reduction of $\aver{[X_u,X_d]^3YY^\dag}$ through the CHT.
It is interesting to note that $\aver{[X_u,X_d]^3}=3I_{66}^-$ is the SM Jarlskog invariant and it is not reducible.
It becomes reducible only accompanied by $YY^\dag$ within the trace.
Further details on how to reduce the invariants \eqref{inv.u3d3y2} can be found in appendix \ref{app:reduction} where we also show the explicit reduction of $K_{66}^{(3)}$.
For larger $K$-length, it is clear that some matrix in \eqref{Ai:K} will be repeated and then the invariant will be reducible.

\subsection{Syzygy of lowest degree}
\label{sec:lowest}

The PL in \eqref{PL:vlq.mass} tells us that the syzygy of least order is of degree $q^8$ in the HS.
As expected, it corresponds to the reduction of $\big(K^{-}_{22}\big)^2$ in terms of CP even invariants.
So the syzygy is of degree $X_u^2X_d^2(YY^\dag)^2$, i.e., of physical degree 12.
The shorter expression is
\eqali{
\label{syzygy:ff}
\big(K^{-}_{22}\big)^2
&=\big(\tK^+_{22}\big)^2
   -4 \tI_{22}
   \tK_{02} \tK_{20}+4 \tK_{20} \tK^+_{24}+4 \tK_{02}\tK^+_{42}-4 \tK_{04} \tK_{40}
   -2 \tI_{40}\tK_{02}^2-2 \tI_{04} \tK_{20}^2
  \cr&\quad
  +2K_{00}^2(\tI_{04} \tI_{40}-2\tI_{44})-4 K_{00}(\tI_{42} \tK_{02}+\tI_{24} \tK_{20}- \tK^+_{44})
   \cr&\quad
   -2K_{00}(\tI_{40}\tK_{04}+\tI_{04}\tK_{40})\,,
}
where all the invariants with tilde on the righthand side have the same form as the respective 
invariants but they are calculated with $Y,Y^\dag$ and the traceless parts 
$\tX_u\equiv X_u-\ums{3}\aver{X_u}\id$ and $\tX_d\equiv X_d-\ums{3}\aver{X_d}\id$ 
(corresponding to their \emph{octet} components).
To obtain the full syzygy with the trace part, one should replace 
$\tX_u\to X_u-\ums{3}\aver{X_u}\id$ and $\tX_d\to X_d-\ums{3}\aver{X_d}\id$ within the traces.
The full expression is shown in \eqref{syzygy:ff:full}.
Note that $K_{22}^{-}=\tK_{22}^-$ in the lefthand side is the same irrespective of using $X_u,X_d$ 
or their traceless parts.
So the righthand side of the full expression also must lead to the same result whether we calculate it with the traceless parts or not.
The expression, however, is not valid for the traceless part of $X=YY^\dag$; see dicussion below 
\eqref{syzygy:ff:full}.

Another important remark is that the syzygy \eqref{syzygy:ff} is only valid when $X=YY^\dag$ is rank one.
When we test the identity for $X$ being a generic positive definite hermitean matrix of rank two or three, we discover that it is not valid.
So the syzygy in this form is not generalizable for two or more VLQs.
The validity for only rank one $X$ can be understood in terms of the underlying $SU(3)$ relation discussed in Sec.\,\ref{sec:su3.ff}.
In fact, using the found $SU(3)$ relation, we conclude that the generalization of \eqref{syzygy:ff} will \emph{not} correspond to a syzygy but to a \emph{reduction} of a higher degree CP even invariant.
Therefore, although it is always true that the square of a CP odd invariant can be written as a polynomial of CP even invariants, such a relation will not always correspond to a syzygy as a higher degree CP even invariant may appear \emph{linearly}.

\subsection{Some syzygies of physical degree 14 and 16}
\label{sec:syz:14-16}

Here we list the six syzygies of degree $q^9$ in the PL \eqref{PL:vlq.mass}.
Explicit calculation shows that there are two syzygies of degree $X_u^3X_d^2Y^4$, i.e., of physical degree 14, one in the CP odd sector and another in the CP even sector.
Another syzygy resides in the CP odd sector of degree $X_u^4X_d^3Y^2$, i.e., of physical degree 16.
The rest of the syzygies are obtained by simply exchanging $X_u\leftrightarrow X_d$.

Syzygy of degree $X_u^3X_d^2Y^4$ (physical degree 14), CP odd sector:
\begin{equation}
\begin{aligned}
0&=(\ums{2} K^+_{22} - I_{02} K_{20}) K^-_{42}
  + (I_{02} K_{40} -\ums{2} K^+_{42}) K^-_{22}
\cr&\quad
  + (K_{00} I_{02} - K_{02}) K^-_{62}
  + K_{20} K^-_{44} - K_{00} K^-_{64} + K_{40} K^-_{24}\,.
\end{aligned}
\end{equation}
One obtains another syzygy by replacing $X_u\leftrightarrow X_d$.
Note that all the terms have the form (CP even)$\times$(CP odd).
This syzygy is only valid for $X=YY^\dag$ of rank one and it is not generalizable for more than one VLQ.

Syzygy of degree $X_u^3X_d^2Y^4$ (physical degree 14), CP even sector:
\begin{equation}
\begin{aligned}
0&=3 K_{00}^2 \Big[ I_{20}^3 (I_{02}^2 -  I_{04}) + 2 I_{20}^2 (I_{24} - I_{02} I_{22}) 
  -I_{02}^2 I_{20} I_{40} + I_{20} I_{04} I_{40} + 2 I_{02} I_{20} I_{42} - 2 I_{20} I_{44}\Big]
\cr
&\quad
+2 K_{00} \Big[I_{20}^3 (K_{04} - 3 I_{02} K_{02}) - 3 I_{20}^2 K_{24}^+ 
  -3 K_{20} I_{44} - 3 I_{20} K_{02} I_{42} + 3 I_{02} K_{20} I_{42}
\cr
&\qquad
   + 3 K_{40} (I_{02} I_{22} - I_{24}) + 3 I_{02} I_{20}^2 K_{22}^+ - 
   K_{04} I_{60} + 3 I_{02} I_{20} K_{02} I_{40} + 
   3 I_{20}^2 K_{02} I_{22} 
\cr
&\qquad
   -3 I_{02} I_{20} K_{42}^+ + 3 I_{20} K_{44}^+ 
    +\ums[3]{2} I_{20}^2  K_{20} (I_{04} - I_{02}^2) 
    +\ums[3]{2} K_{20} I_{40} (I_{04} - I_{02}^2 )\Big]
\cr
&\quad
  +2 K_{02}^2 (2 I_{20}^3 - 3 I_{20} I_{40} + I_{60}) 
  +6  K_{02} K_{20} (I_{02} I_{40} - I_{42})
\cr
&\quad
  +6 K_{40} (K_{02} I_{02} I_{20} + I_{02}^2 K_{20} 
  -I_{04} K_{20} - K_{02} I_{22} -I_{02} K_{22}^+ + K_{24}^+ -2 I_{20} K_{04})
\cr
&\quad
  +6 K_{04} (I_{20}^2 K_{20} - K_{20} I_{40}) + 6 K_{02} I_{20} (K_{42}^+ - I_{20} K_{22}^+) 
  +6 K_{20} (K_{44}^+ - I_{20} K_{42}^+)
\cr
&\quad
+3 K_{22}^+ K_{42}^+ - 3 K_{22}^- K_{42}^-\,.
\end{aligned}
\end{equation}
One obtains another syzygy by replacing $X_u\leftrightarrow X_d$.
Here almost all the terms have the form (CP even)$\times$(CP even), except for the last term that has the form (CP odd)$\times$(CP odd).
This syzygy is only valid for $X=YY^\dag$ of rank one.

Syzygy of degree $X_u^4X_d^3Y^2$ (physical degree 16),  CP odd sector:
\begin{equation}
\begin{aligned}
0&= (I_{20} K_{00} - 3 K_{20}) I_{66}^-
    + (I_{04} I_{20}^3 - 3 I_{20}^2 I_{24} - I_{04} I_{20} I_{40} + 4 I_{20} I_{44} - I_{04} I_{60}) K_{22}^- 
\cr&\quad    
    + (I_{02} I_{20}^3 - 3 I_{20}^2 I_{22} - I_{02} I_{20} I_{40} + 4 I_{20} I_{42} - I_{02} I_{60}) K_{24}^- 
\cr&\quad    
    + (-2 I_{04} I_{20}^2 + 5 I_{20} I_{24} + 3 I_{04} I_{40} - 6 I_{44}) K_{42}^-
\cr&\quad  
    + (2 I_{02} I_{20}^2 - 5 I_{20} I_{22} - 3 I_{02} I_{40} + 6 I_{42}) K_{44}^- 
\cr&\quad  
    + (-I_{20} I_{40} + 3 I_{60}) K_{26}^- 
    + (I_{20}^2 - 3 I_{40}) K_{46}^- + (I_{04} I_{20} - 3 I_{24}) K_{62}^- + (-I_{02} I_{20} + 3 I_{22}) K_{64}^-
\,.
\end{aligned}
\end{equation}
By replacing $X_u\leftrightarrow X_d$, we obtain another syzygy.
Note that all the terms have the form (CP even)$\times$(CP odd).
In contrast to the syzygies found so far, this syzygy remains valid even if we replace $X=YY^\dag$ by a generic positive definite hermitean matrix of rank two or three.
So it is generalizable for more than one VLQ.
In the first term, it is interesting to note the presence of the Jarlskog invariant of the SM, $I_{66}^-=\det([X_u,X_d]^3)$, which did not appear in the syzygies so far.

\section{$SU(3)$ relation for syzygy}
\label{sec:su3.ff}

The syzygy \eqref{syzygy:ff} of degree $(X_uX_dYY^\dag)^2$ corresponds to writing $(K^{-}_{22})^2$ in terms of CP even invariants.
In terms of $SU(3)$ tensors in the adjoint, the commutator in $K^{-}_{22}$ involves the Lie algebra structure constant $f_{ijk}$ contracted to the adjoint vectors:
\eq{
\label{8:udB}
(\bs{8}_u)_i\equiv \tr[t_iX_u]\,,\quad
(\bs{8}_d)_i\equiv \tr[t_iX_d]\,,\quad
(\bs{8}_B)_i\equiv \tr[t_iYY^\dag]\,.
}
The syzygy we seek involves two tensors $f_{ijk}f_{i'j'k'}$ contracted to 
$(\bs{8}_u)_i(\bs{8}_u)_{i'}(\bs{8}_d)_j(\bs{8}_d)_{j'}(\bs{8}_B)_k(\bs{8}_B)_{k'}$, thus the relevant tensor is
symmetrized in the pairs $(ii'),(jj'),(kk')$. In birdtrack notation\,\cite{Cvitanovic:1976am,cvitanovic}, it can be written as
\eqali{
\label{Tff}
(T_{f\!f})_{ii'jj'k'k}=
\raisebox{-3.2em}{\includegraphics[scale=.2]{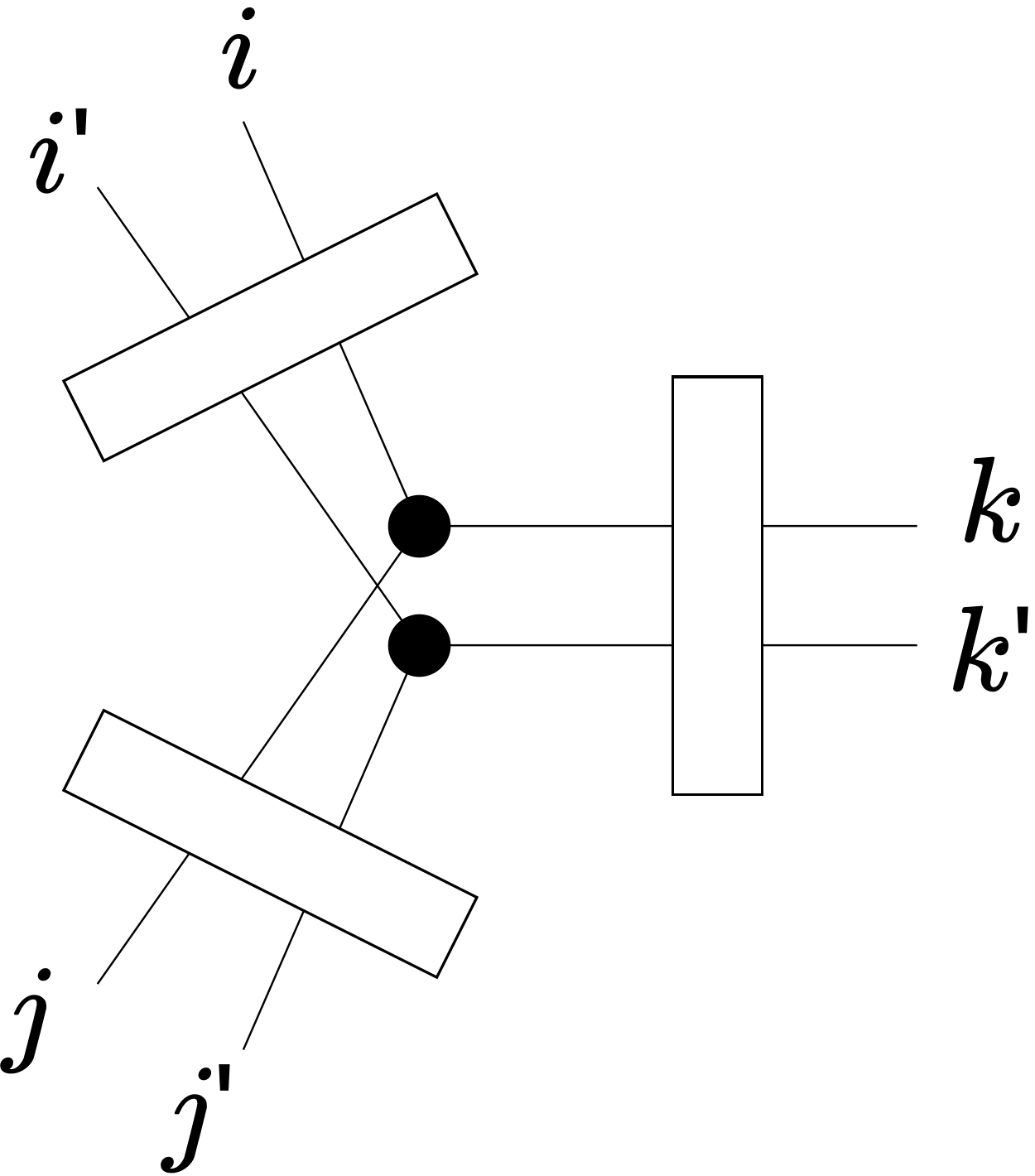}}~
=S_{ii',a'a}S_{jj',b'b}S_{kk',c'c}[iC_{abc}][iC_{a'b'c'}]\,.
}
The indices should be read in the counter-clockwise direction.\footnote{%
Our convention of reading indices in counter-clockwise order follows \cite{cvitanovic} and differs from the usual convention of birdtracks for QCD applications\,\cite{keppeler}. As in \cite{cvitanovic}, we also use thin lines to denote the adjoint $\bs{8}$ instead of gluon lines.}
The tensor $C_{abc}$ is the tensor $f_{ijk}$ normalized differently as in
\eqali{
\raisebox{-1.5em}{\includegraphics[scale=.2]{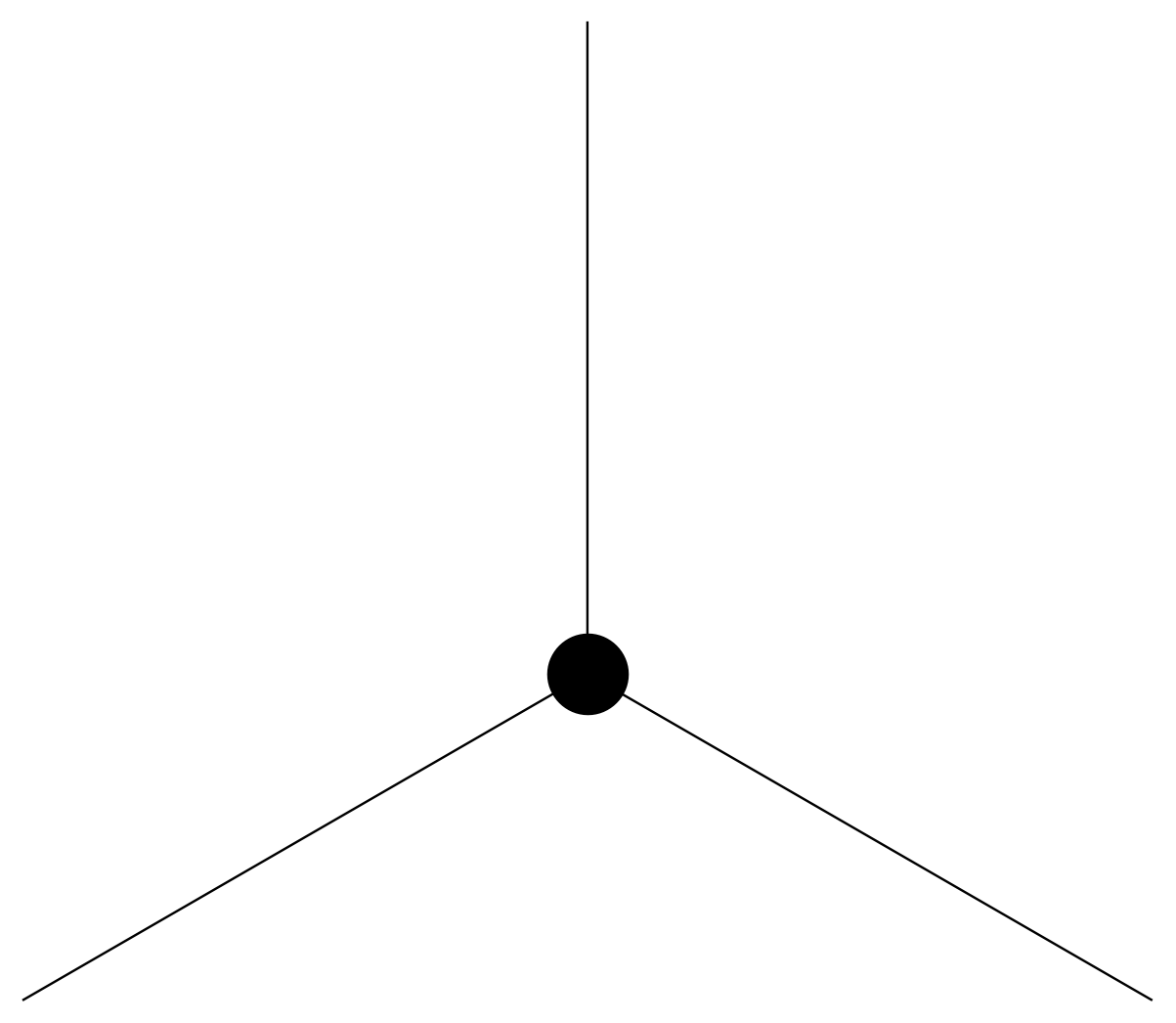}}~
=iC_{ijk}=\tr\big[[t_i,t_j]t_k\big]\,,
\qquad
\raisebox{-1.5em}{\includegraphics[scale=.2]{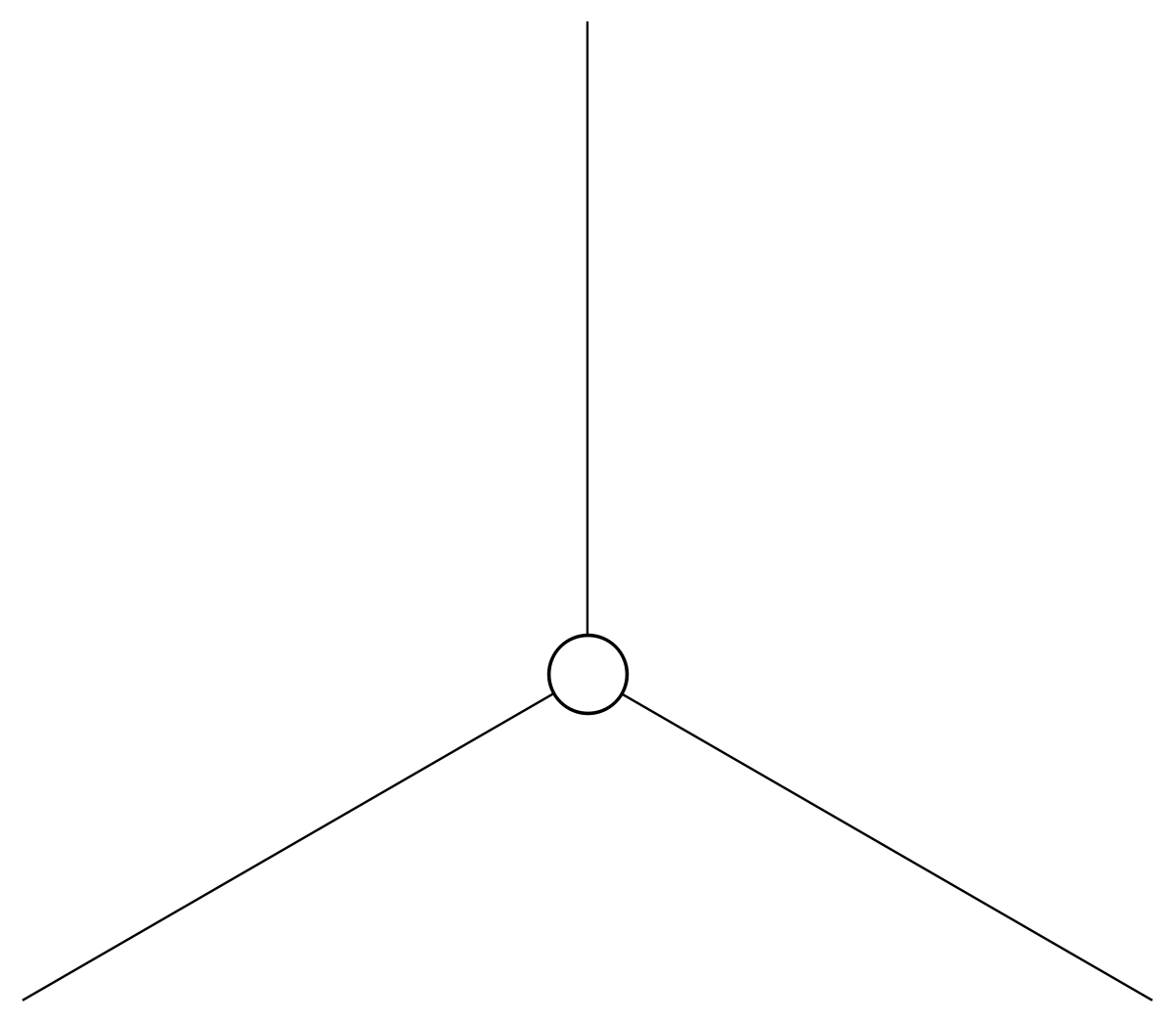}}~
=d'_{ijk}=\tr\big[\{t_i,t_j\}t_k\big]\,,
}
where the normalization $\tr[t_it_j]=\delta_{ij}$ is assumed.\footnote{%
Our definition of $d'_{ijk}$ differs from \cite{cvitanovic} by a factor 1/2.
}
We have also defined the totally symmetric cubic tensor $d'_{ijk}$.
For generators $T_i$ with a different normalization, 
\eq{
\tr[T_iT_j]=a\delta_{ij}\,,
}
one should replace
\eq{
t_i\to \frac{T_i}{\sqrt{a}}\,.
}
The relation to the usual tensors $f_{ijk}$ and $d_{ijk}$ of $SU(3)$ calculated using Gell-mann matrices is $C_{ijk}=\sqrt{2}f_{ijk}$ and $d'_{ijk}=\sqrt{2}d_{ijk}$.
For example, $iC_{123}=i\sqrt{2}$ and $d'_{118}=\sqrt{2/3}$.
The tensor $S_{ii',a'a}$ is the symmetrizer operator denoted by a hollow stripe:
\eqali{
\label{P:symm}
S_{ii',a'a}=
\raisebox{-1.1em}{\includegraphics[scale=.19]{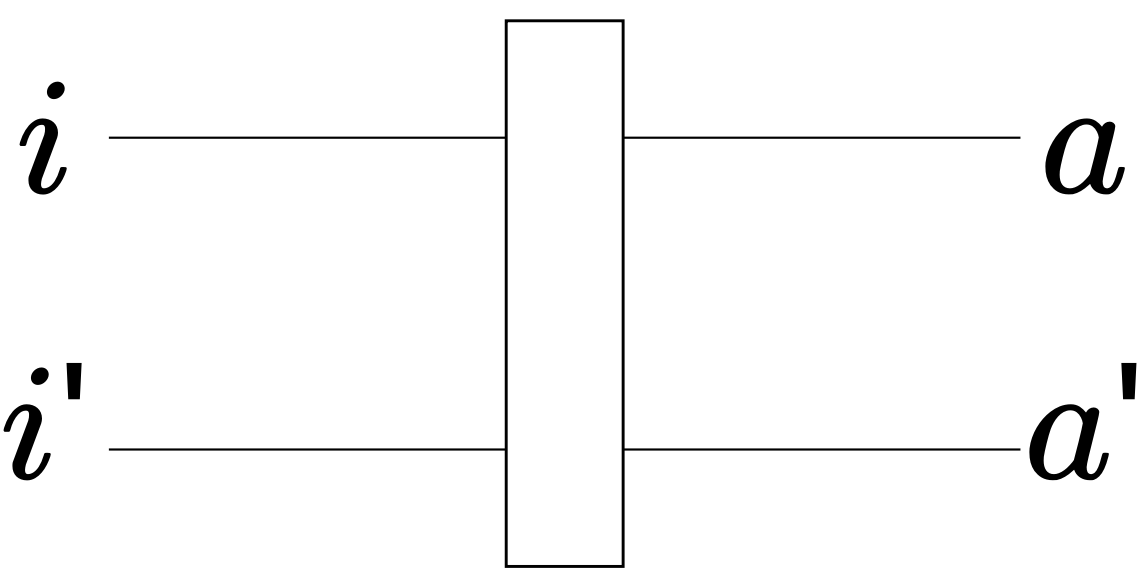}}~
=
\ums{2}\,\raisebox{-.7em}{\includegraphics[scale=.21]{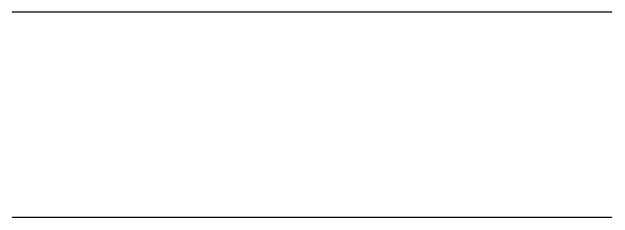}}
+\ums{2}\,\raisebox{-.7em}{\includegraphics[scale=.21]{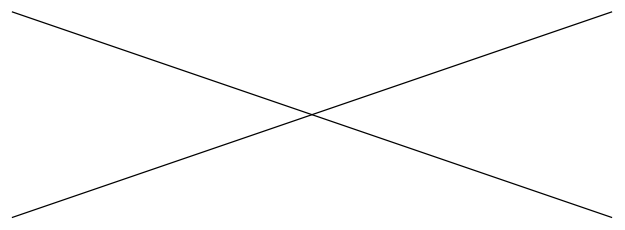}}
=\ums{2}\delta_{ia}\delta_{i'a'}+\ums{2}\delta_{ia'}\delta_{i'a}\,.
}

To find a relation for \eqref{Tff}, we need a basis of invariant tensors in the space $\bs{8}^{\otimes 6}$ of 6-legged tensors.
As shown in Ref.\,\cite{dittner}, the total number of $k$-legged independent invariant tensors in $\bs{8}^{\otimes k}$ coincides with the number of invariants (singlets) in the branching of $\bs{8}^{\otimes k}$.
For $k=6$, this number is 145\,\cite{dittner} and using such a basis would be impractical.

Fortunately, pairs of legs of \eqref{Tff} are symmetrized and we only need to consider $[(\bs{8}\otimes\bs{8})_s]^3$ for which
$(\bs{8}\otimes\bs{8})_s$ has much lower dimension.
The number of singlets in $[(\bs{8}\otimes\bs{8})_s]^3$ is just 21 and a possible basis for this space is discussed in appendix \ref{app:21.tensors}.
Here we follow an even simpler route observing that the tensor \eqref{Tff} has a permutation symmetry where the symmetrized pairs of indices can be interchanged without changing the tensor.
Therefore, we only need a smaller set of 8 tensors which is shown in Table~\ref{tab:8.tensors}.
Note that they are all symmetric by permutation of pairs of legs and they involve only the symmetric cubic $d'_{ijk}$.
\begin{table}[h]
\[
\begin{array}{|c|c|}
\hline
\text{Tensor(s)} &  
\\
\hline
\rule{0em}{2.5em}
T_{1} & \raisebox{-1.9em}{\includegraphics[scale=.12]{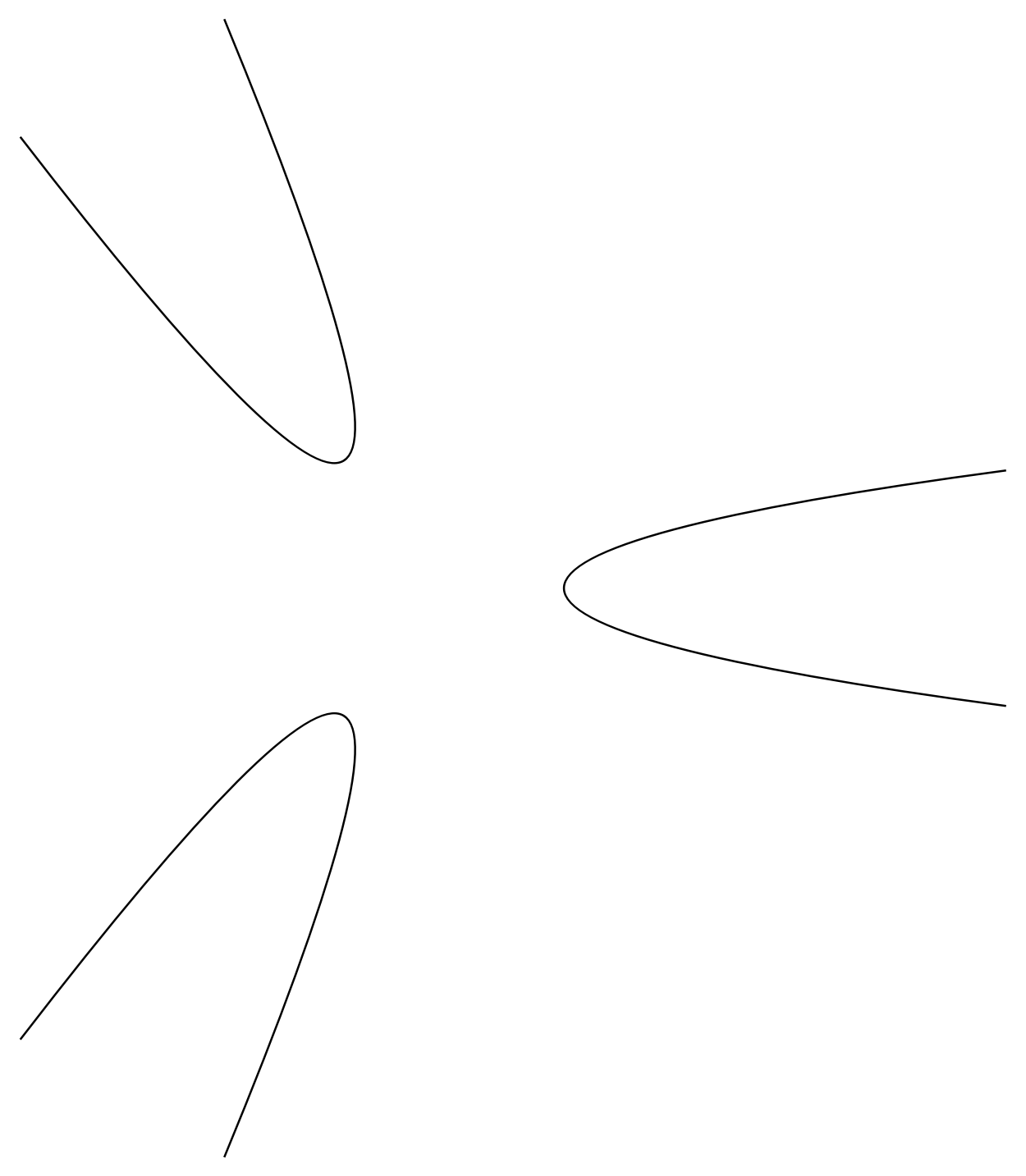}}
\cr
\hline
\rule{0em}{2.5em}
T_{2} & \raisebox{-1.9em}{\includegraphics[scale=.12]{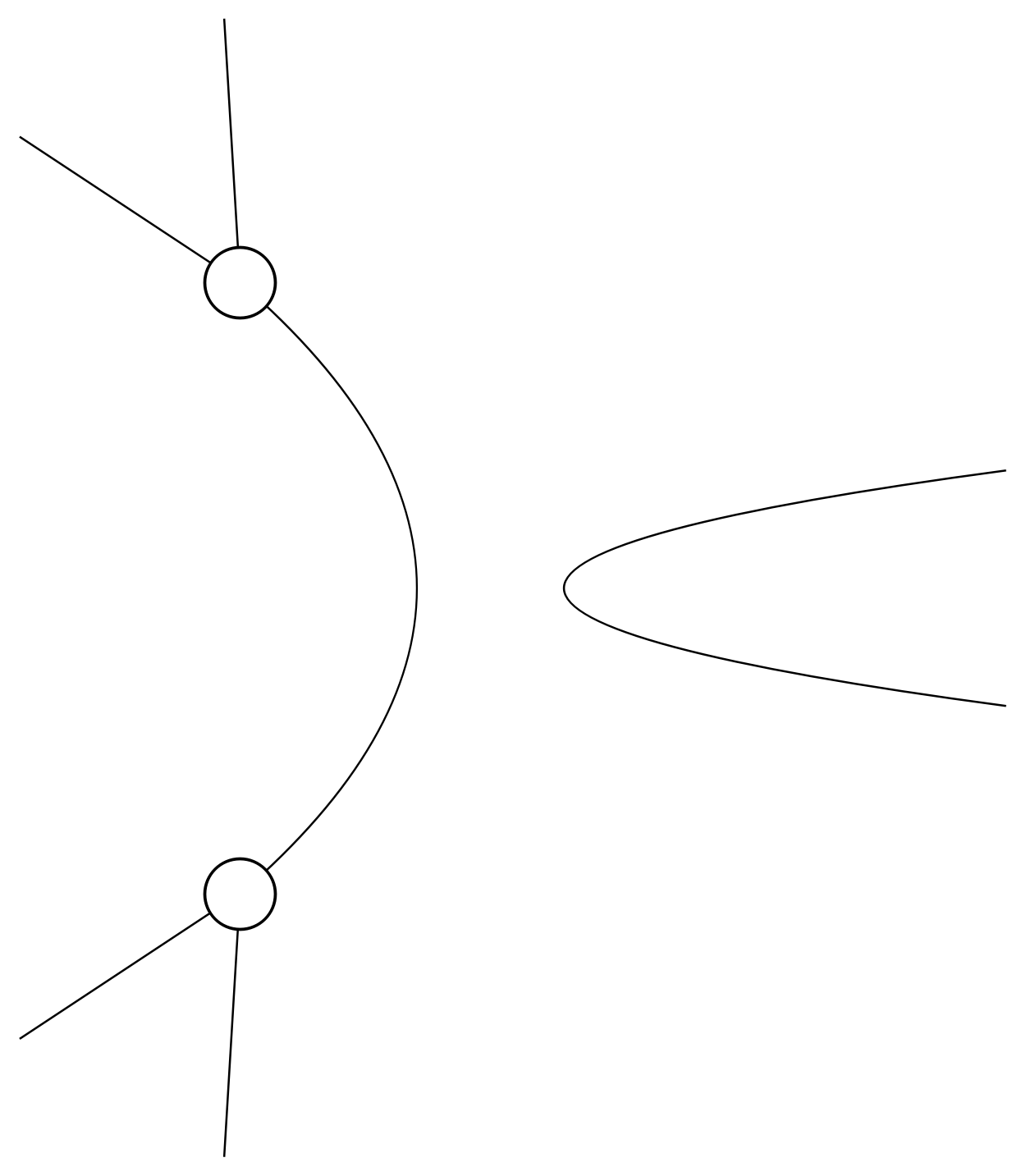}} +
  \raisebox{-1.9em}{\includegraphics[scale=.12]{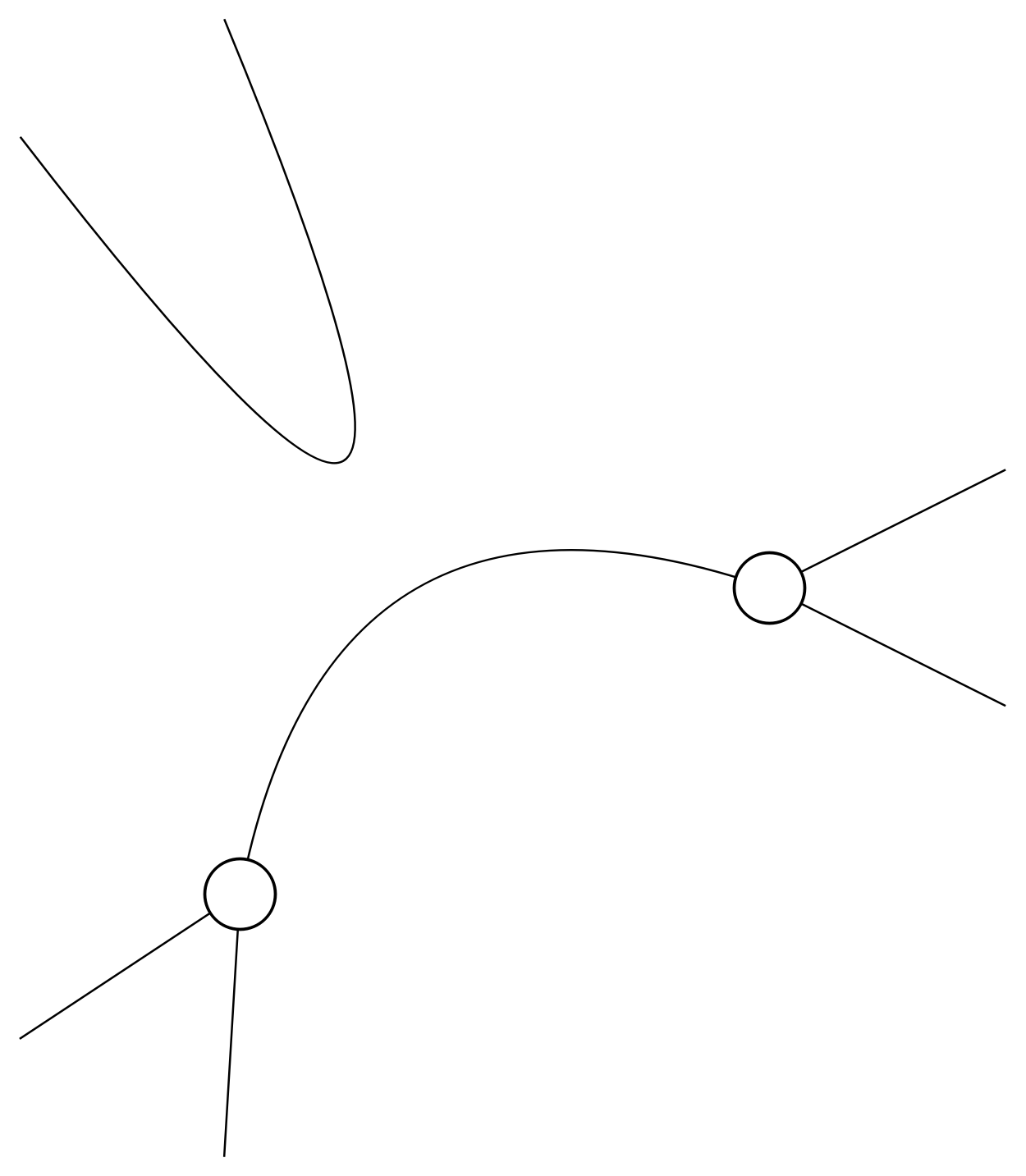}} +
  \raisebox{-2em}{\includegraphics[scale=.12]{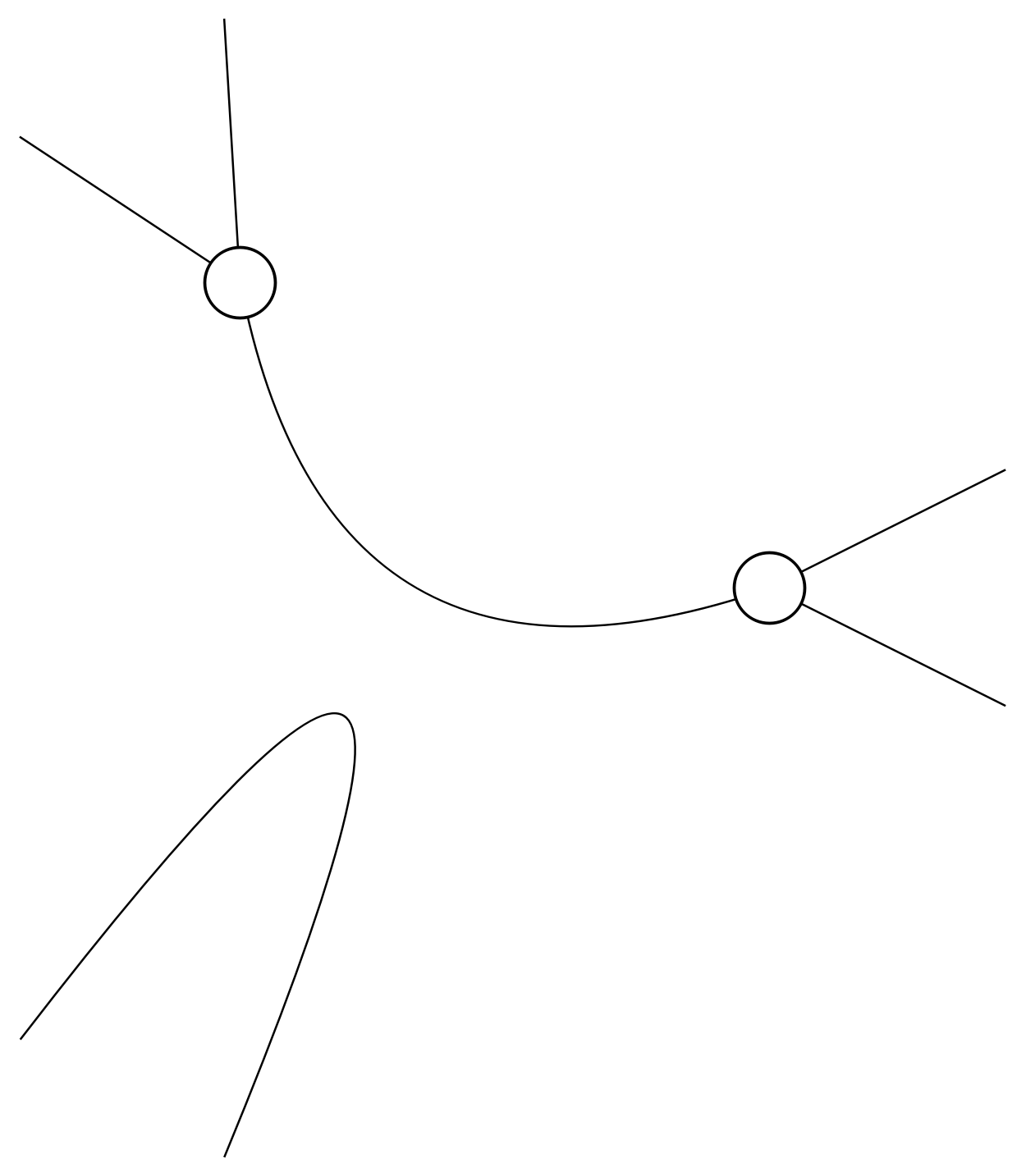}} 
\cr
\hline
\rule{0em}{2.5em}
T_{3} &
\raisebox{-2.2em}{\includegraphics[scale=.12]{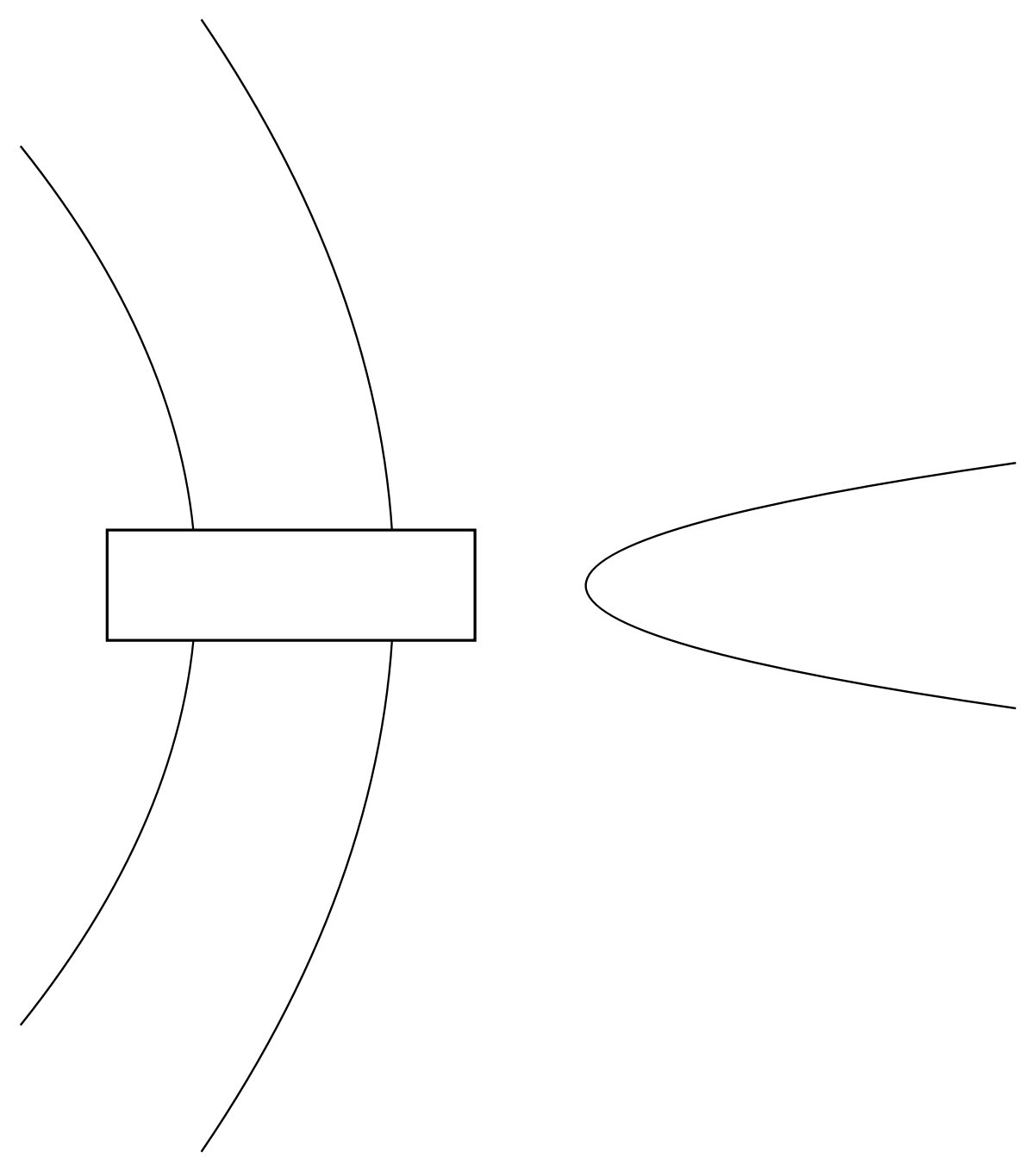}} +
  \raisebox{-1.9em}{\includegraphics[scale=.12]{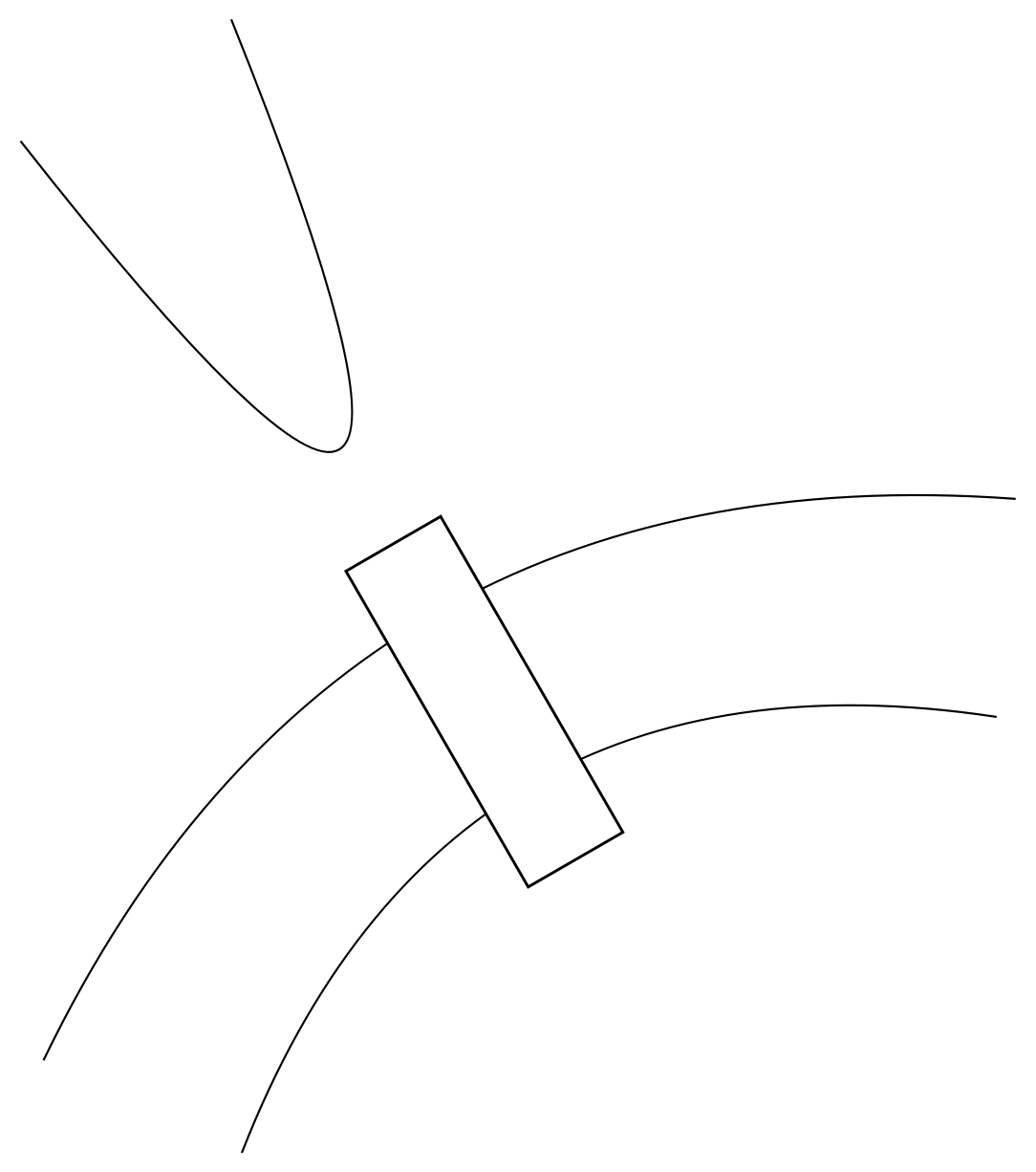}} +
  \raisebox{-1.9em}{\includegraphics[scale=.12]{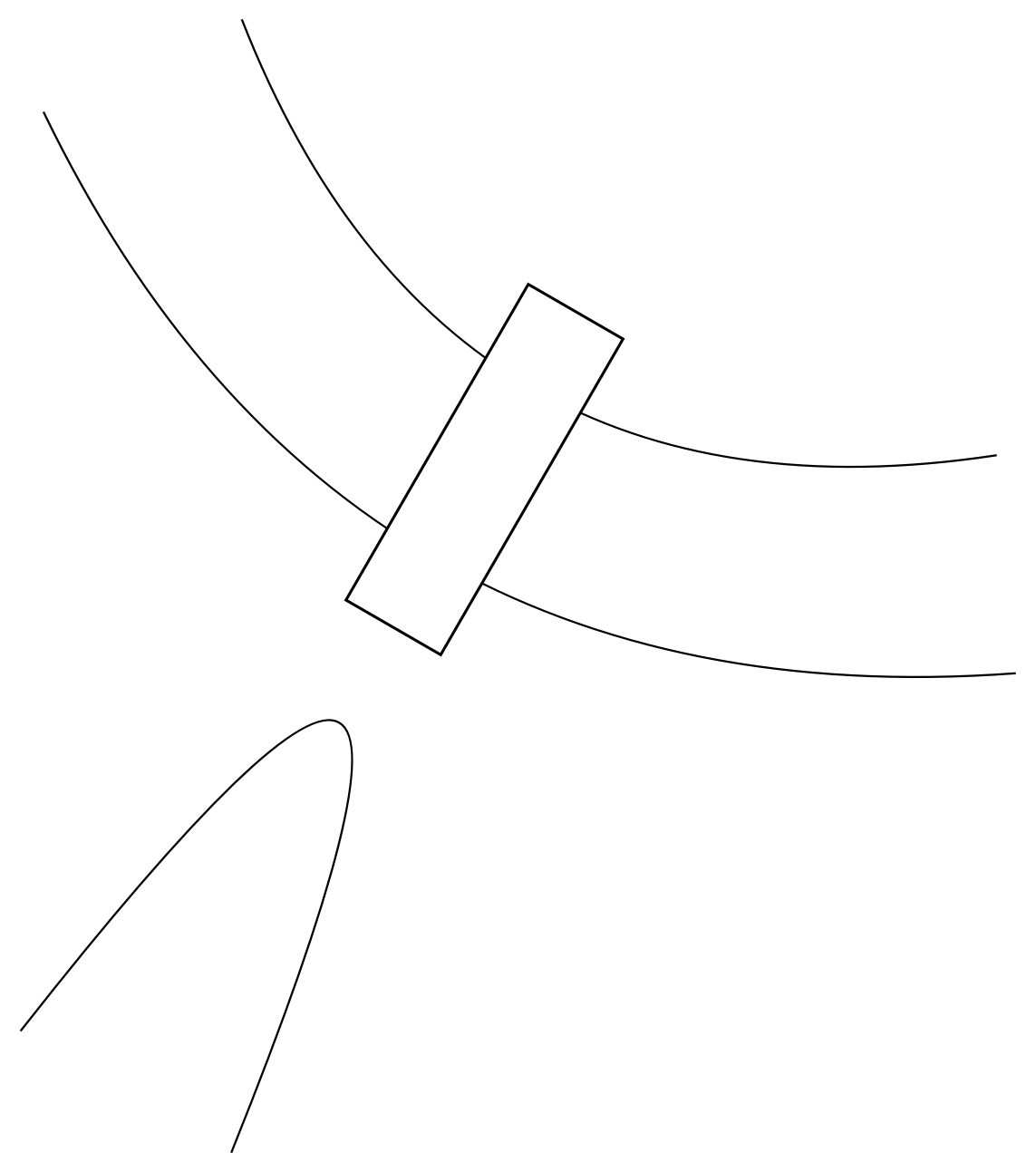}} 
\cr
\hline
\rule{0em}{2.5em}
T_{4a} &
\raisebox{-1.9em}{\includegraphics[scale=.12]{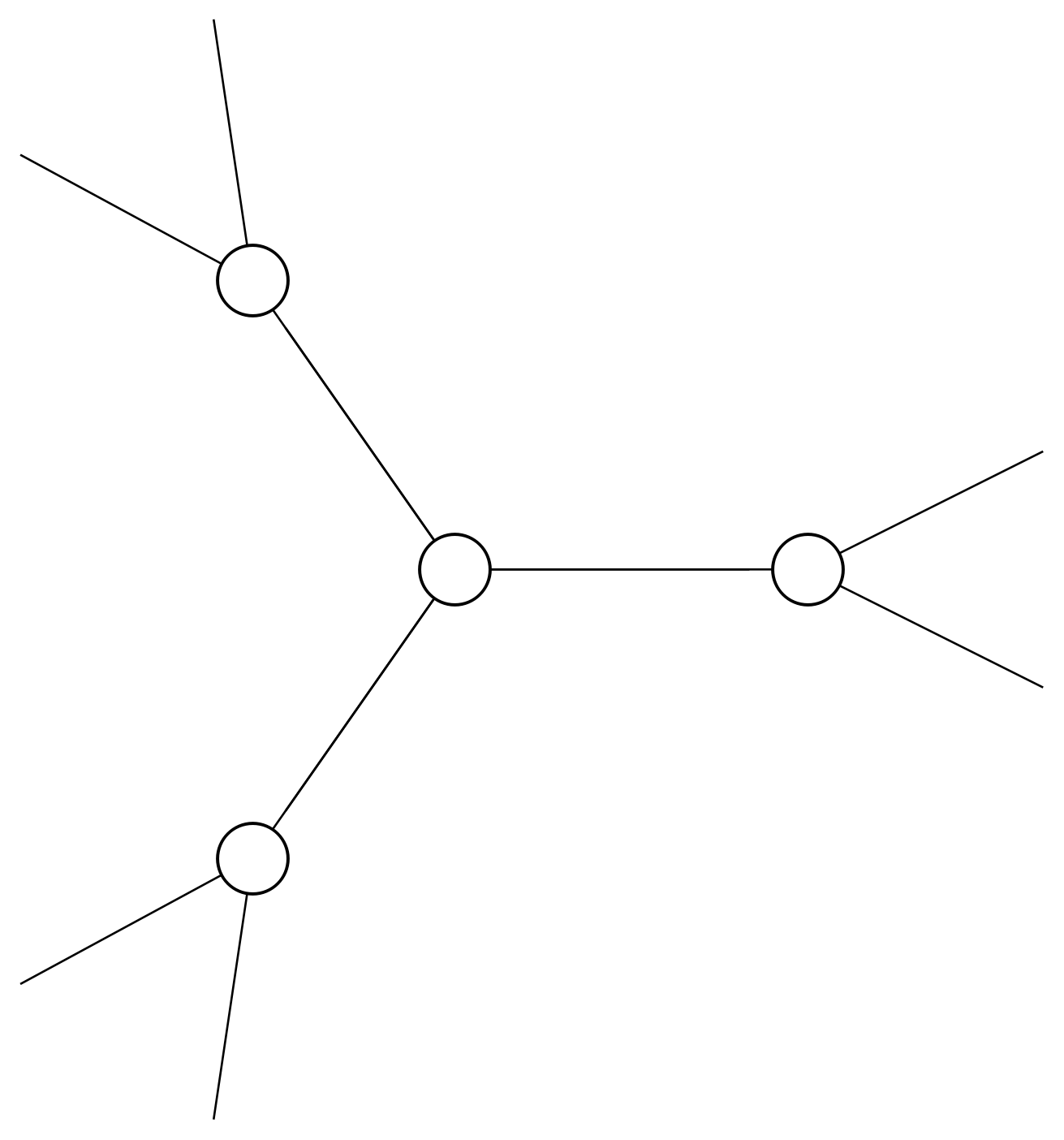}} 
\cr
\hline
\rule{0em}{2.6em}
T_{5} &
  \raisebox{-2em}{\includegraphics[scale=.12]{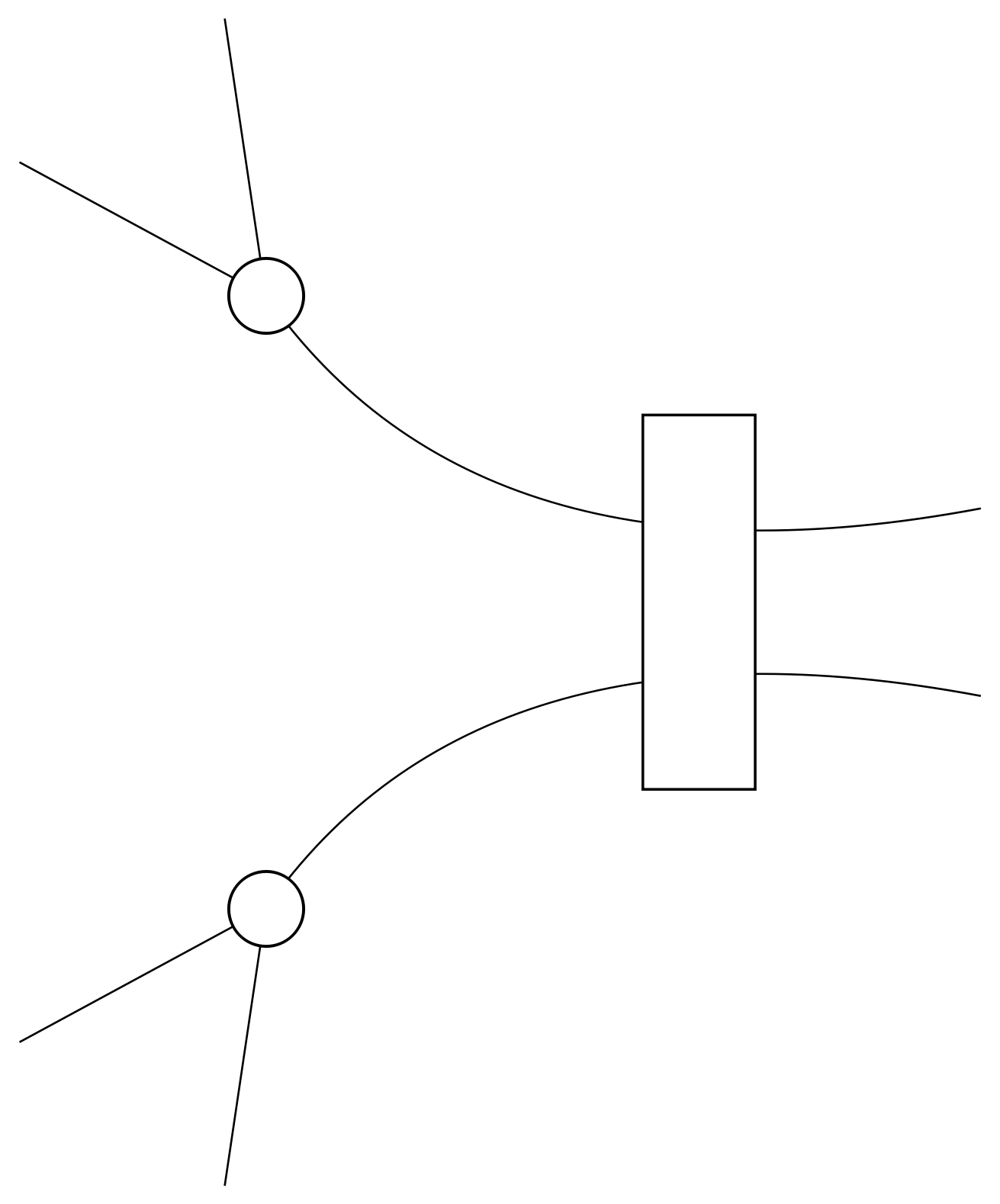}} +
  \raisebox{-1.8em}{\includegraphics[scale=.12]{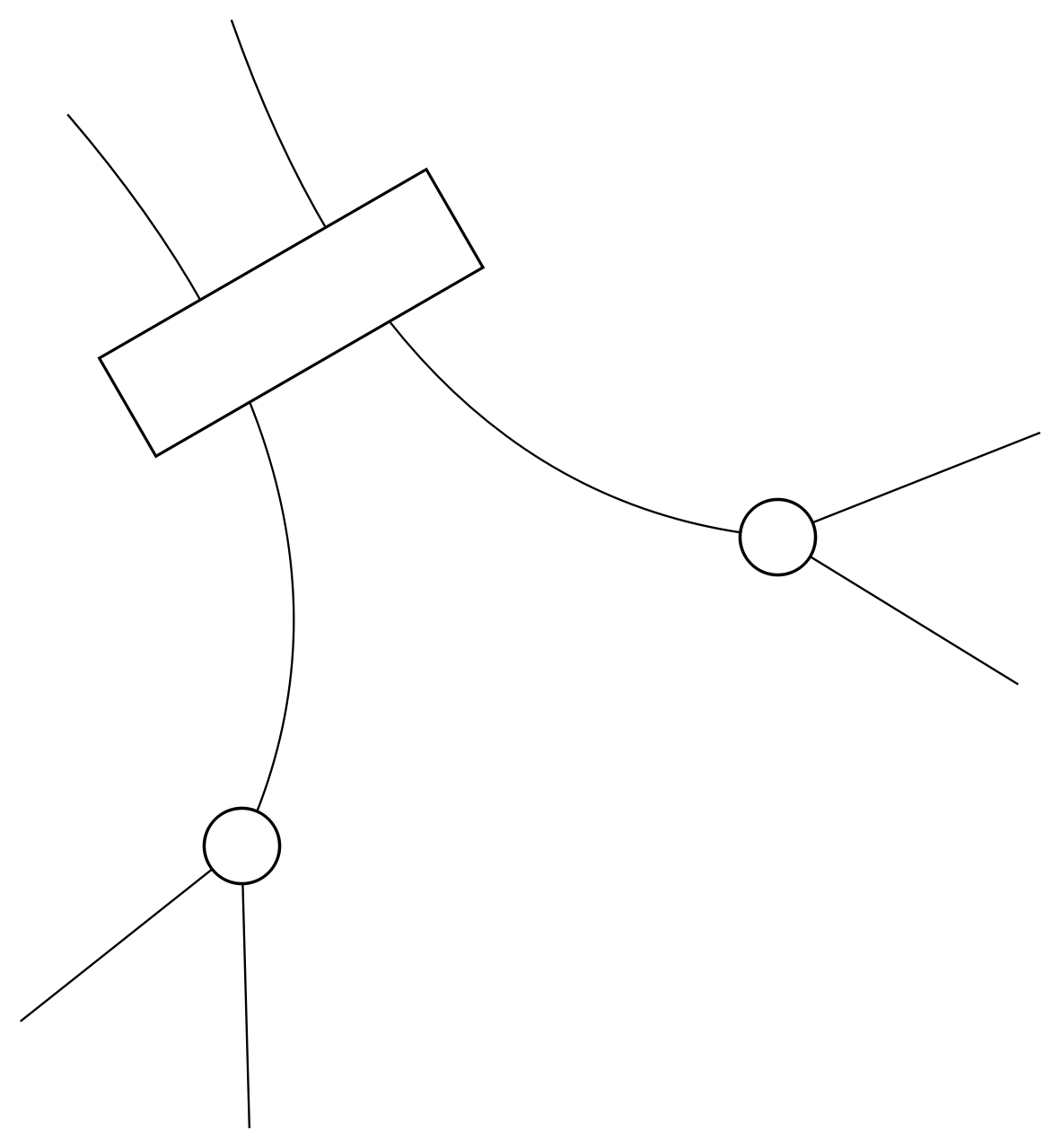}} +
  \raisebox{-1.8em}{\includegraphics[scale=.12]{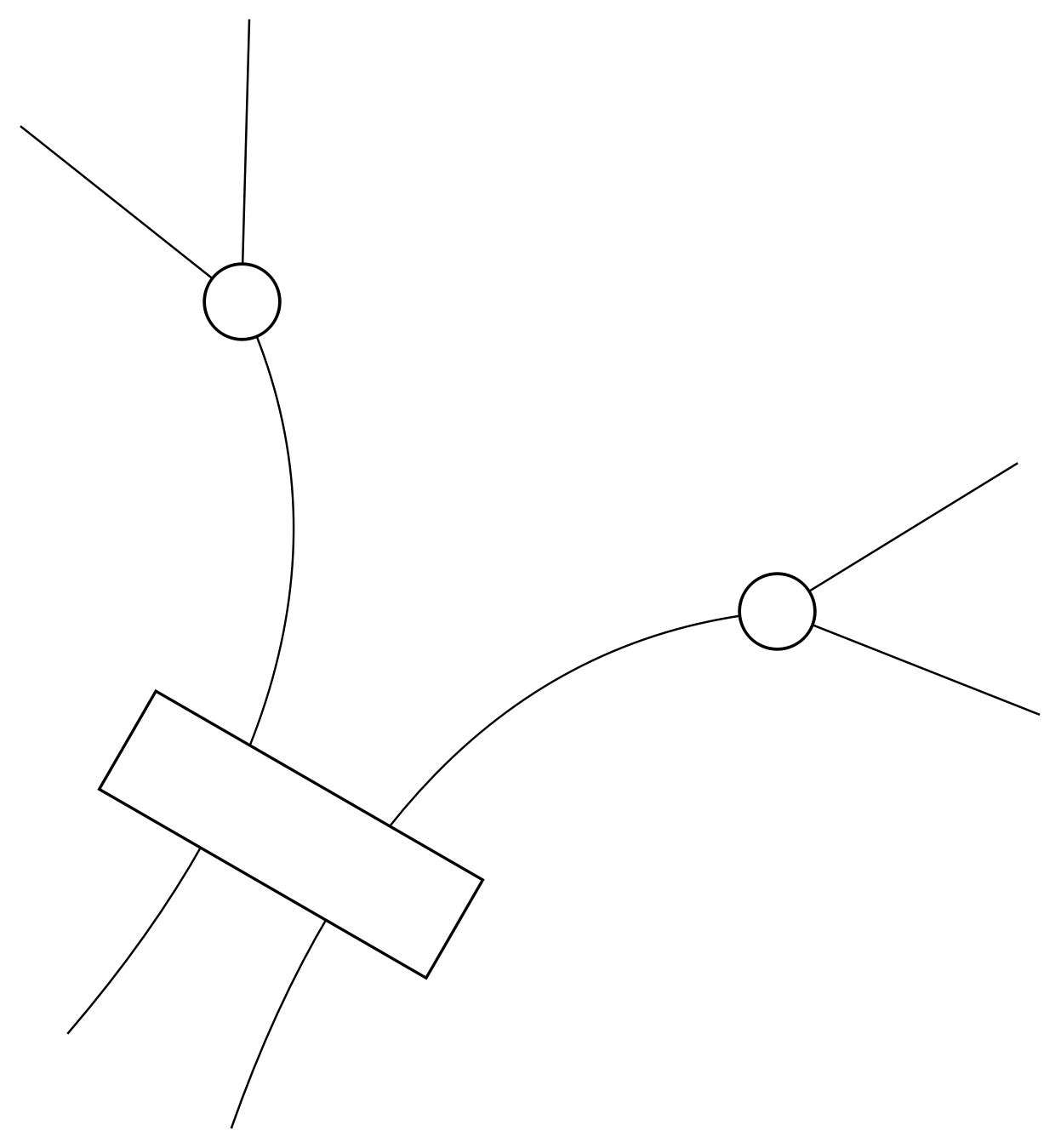}} 
\cr
\hline
\rule{0em}{2.5em}
T_{6} &
  \raisebox{-1.9em}{\includegraphics[scale=.12]{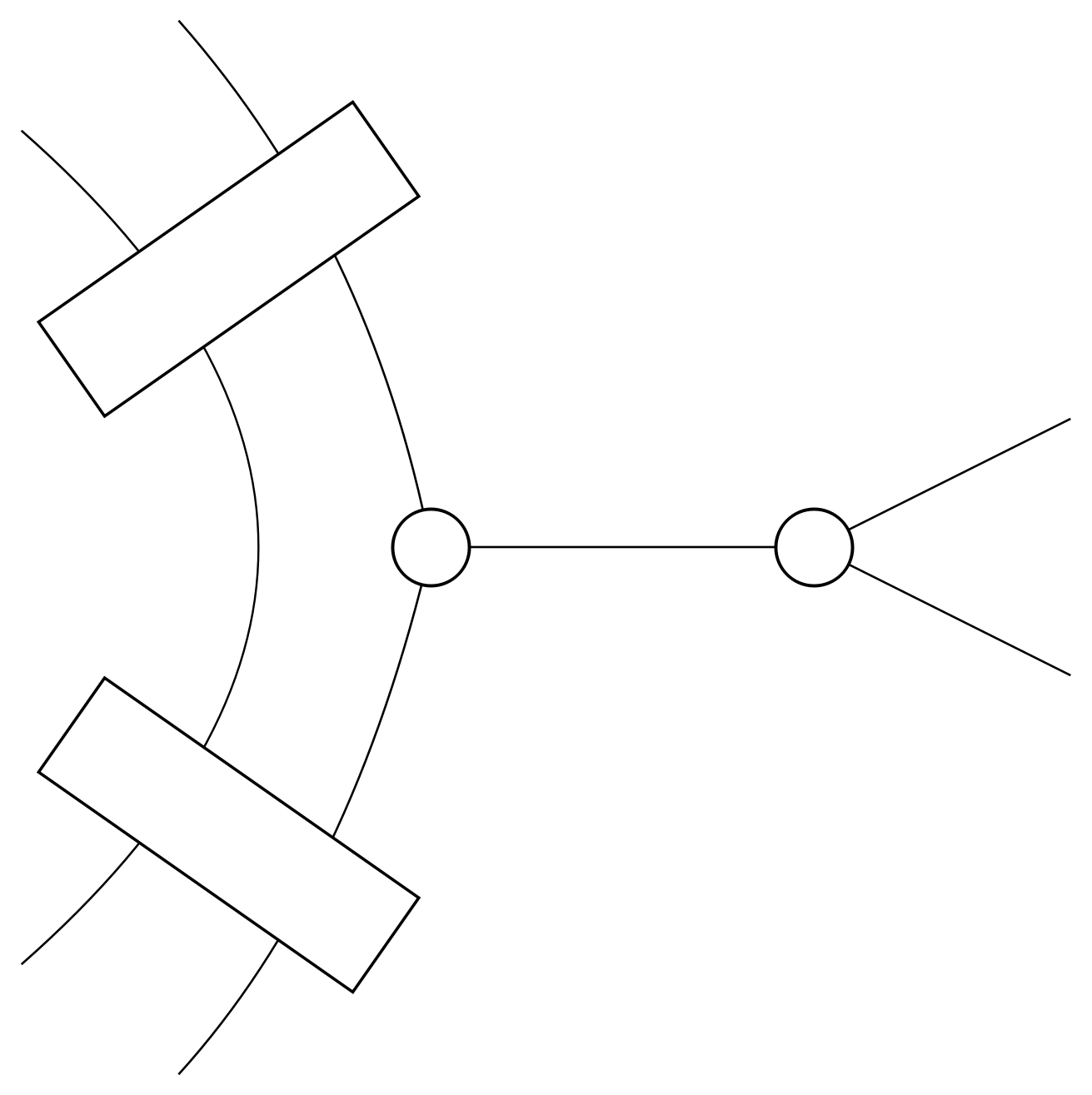}} +
  \raisebox{-1.9em}{\includegraphics[scale=.12]{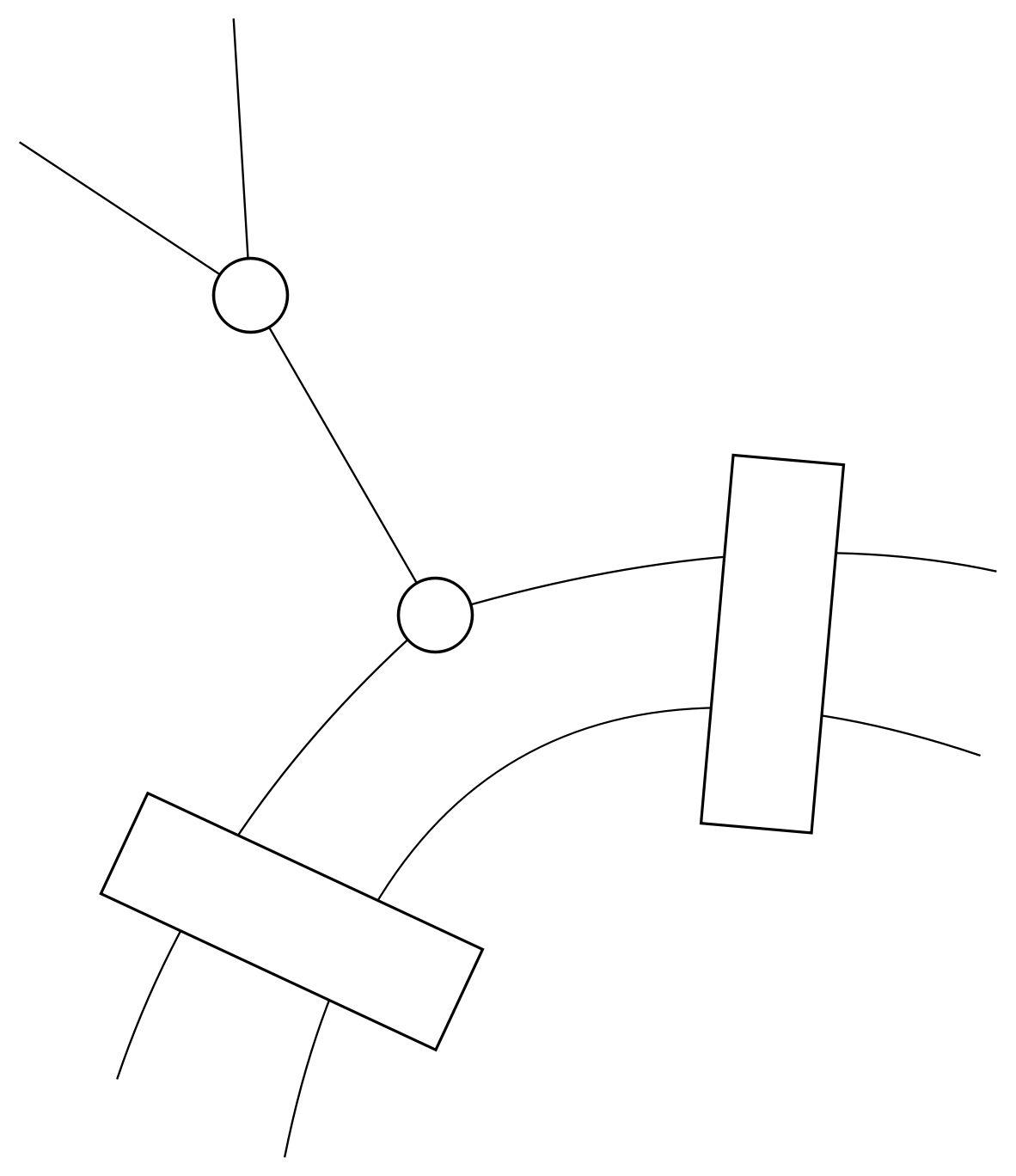}} +
  \raisebox{-1.9em}{\includegraphics[scale=.12]{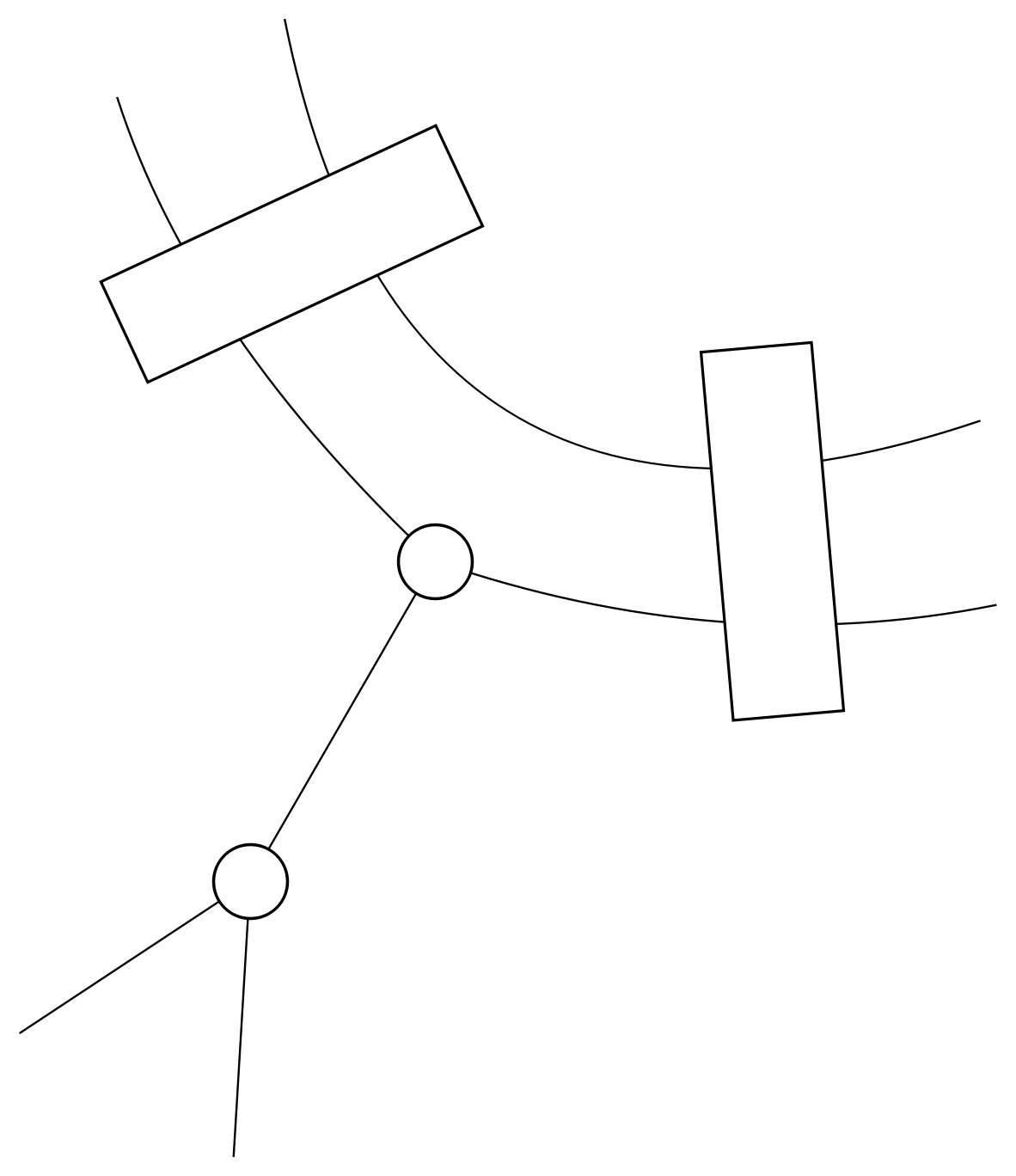}} 
\cr
\hline
\rule{0em}{2.5em}
T_{S}; T_{dd} &
  \raisebox{-1.7em}{\includegraphics[scale=.12]{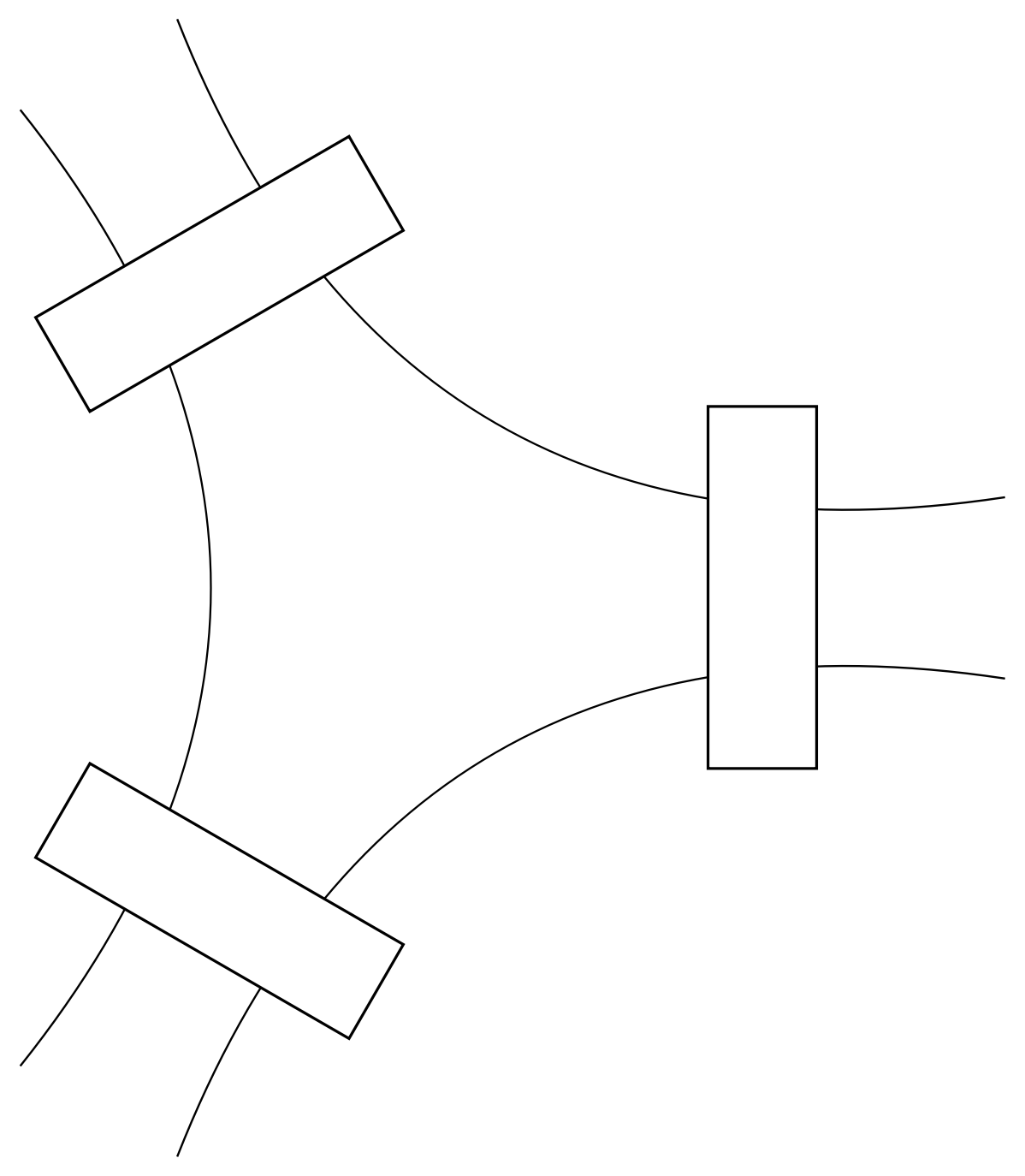}} \quad;\quad 
  \raisebox{-1.7em}{\includegraphics[scale=.12]{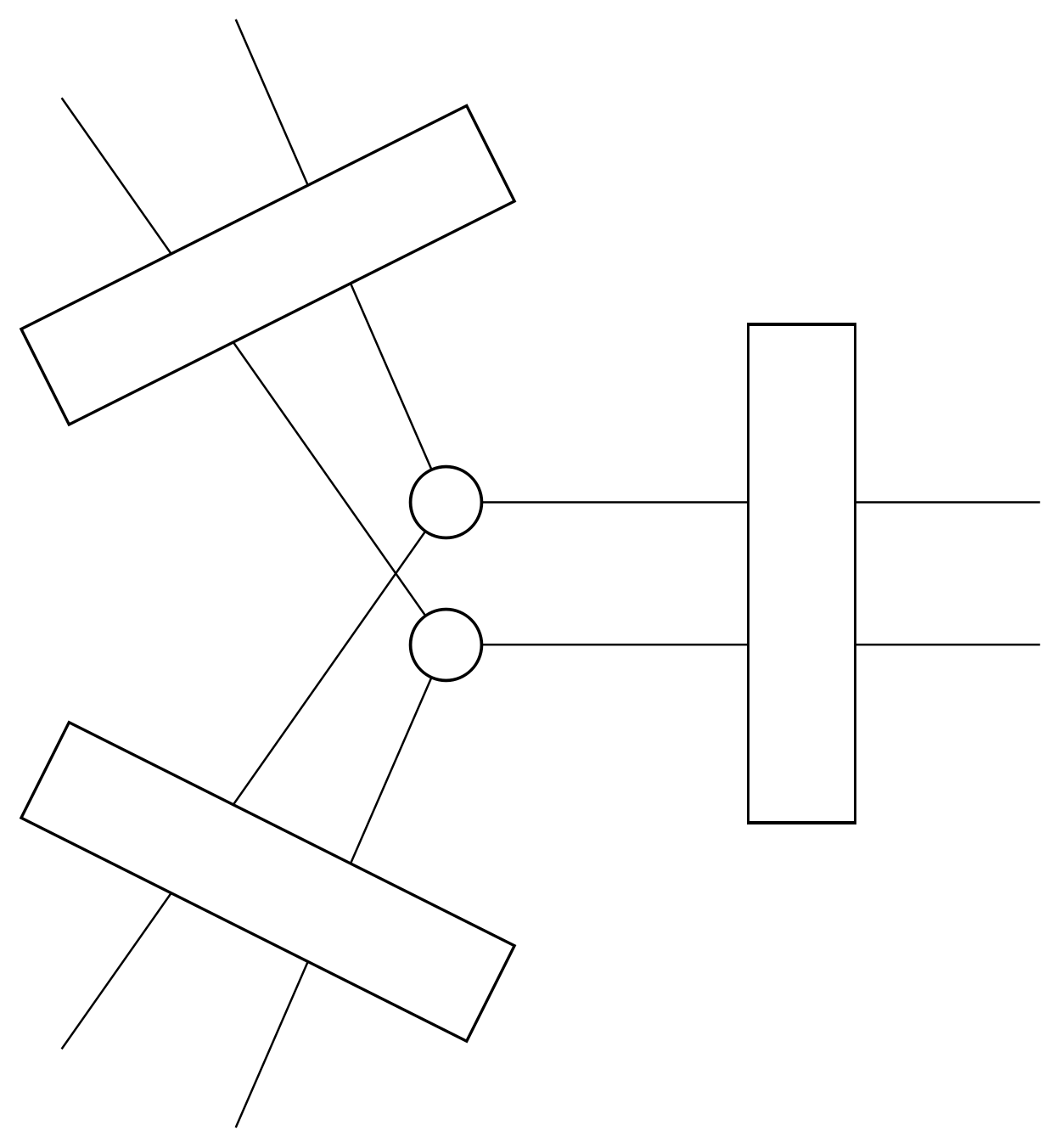}} 
\cr
\hline
\end{array}
\]
\caption{\label{tab:8.tensors}
Tensor basis of 8 tensors to expand $T_{f\!f}$ in \eqref{Tff}.
}
\end{table}

The identity we find is quite simple:
\eq{
\label{Tff.identity}
T_{f\!f}=T_{dd}-4T_{S}-\ums{3}T_{1}-\ums{2}T_2+\ums[2]{3}T_3+\ums[3]{2}T_{4a}-T_5+2T_6\,.
}
One can check, after some algebra, the syzygy \eqref{syzygy:ff} by contracting this identity with the adjoint vectors \eqref{8:udB}.
More generally, this identity allows us to write the square of any CP odd invariant of the form\,\cite{lebedev}
\eq{
\aver{[A,B]C}\,,
}
with $A,B,C$ being $3\times 3$ hermitean matrices, as a linear combination of (products of) CP even invariants defined by the tensor basis in Table~\ref{tab:8.tensors}.
An important remark about that basis is that, except for $T_{4a}$, all of its tensors are disconnected and these immediately leads to products of lower degree invariants when contracted to the adjoint vectors \eqref{8:udB}.
The syzygy \eqref{syzygy:ff}, however, contains no invariant which is not a product. 
Here, reduction of the contraction of $(\bs{8}_B)_k(\bs{8}_B)_{k'}$ to $T_{4a}$ is possible because the Yukawa $Y$, and consequently $YY^\dag$, is a rank one matrix which obeys \eqref{YY2=YY}.
Such a property will no longer hold for two or more VLQs and the identity \eqref{Tff.identity} will not lead to a syzygy but to a reduction of the higher degree CP even invariant associated to $T_{4a}$. 

We find the identity \eqref{Tff.identity} following the method of Ref.\,\cite{dittner}.
In this method, we expand
\eq{
T_{f\!f}=\sum_A x^A T_A\,,
}
in the basis tensors of Table\,\ref{tab:8.tensors} and then contract with each of the same tensors obtaining the linear equation
\eq{
\label{y=Qx}
y^B=Q_{BA}x^A\,,
}
with the $Q$-matrix defined by
\eq{
\label{def:Q}
Q_{BA}\equiv T_B\cdot T_A\,,
}
and $y^B=T_B\cdot T_{f\!f}$. The dot here indicates the full contraction of the six indices.
The expansion coefficients $x^A$ are obtained by solving \eqref{y=Qx}.
We can also check linear dependence by calculating the rank of $Q$ for any set of tensors.
If the rank is not full, the set is not linearly independent.
We use this method to check that the set in Table\,\ref{tab:8.tensors} is a basis for $T_{f\!f}$.
Further details for obtaining the identity \eqref{Tff.identity} can be seen in appendix \ref{ap:su3.ff}.

For comparison, we also write the same identity for $SU(2)$ where the cubic tensor $d'_{ijk}$ does not exist:
\eq{
\label{Tff.identity:su2}
T_{f\!f}=-4T_{S}-2T_{1}+2T_3\,.
}
We see that the tensors involving $d'_{ijk}$ are lacking and the remaining coefficients are different because one needs the inverse of the matrix $Q$. 
See details in appendix \ref{ap:su3.ff}.
This identity is the source of the unique syzygy for the CP odd invariant in the two family lepton sector with neutrino mass\,\cite{yu.zhou:inv.nu}.
For $SU(2)$, since the structure constant $C_{ijk}$ is proportional to the Levi-Civita tensor $\eps_{ijk}$, we can also obtain the identity above from reduction of $\eps_{ijk}\eps_{i'j'k'}$ in terms of Kronecker deltas. See, for example, eq.\,(6.28) in  Ref.\,\cite{cvitanovic}.

\section{Relations to physical parameters in VLQ mass basis}
\label{sec:phys.param}

Considering the transformation properties \eqref{spurions:Y} for the spurions $X_u,X_d,Y$, we can choose a weak basis where
\eq{
\label{phys.param}
X_u=\diag(y_u^2,y_c^2,y_t^2)\,,
\quad
X_d=V\diag(y_d^2,y_s^2,y_b^2)V^\dag\,,
\quad
Y=Y_0=\mtrx{Y_1\cr Y_2 \cr Y_3}\,,
}
where $Y_0$ depends on three moduli and two phases because a phase, e.g.\ of $Y_3$, can be removed by rephasing.
An analogous basis where $X_d$ is diagonal is
\eq{
\label{phys.param:Xu}
X_u=V^\dag\diag(y_u^2,y_c^2,y_t^2)V\,,
\quad
X_d=\diag(y_d^2,y_s^2,y_b^2)\,,
\quad
Y=\tY_0=\mtrx{\tY_1\cr \tY_2 \cr \tY_3}\,.
}
The two vectors $Y_0$ and $\tY_0$ are related by
\eq{
\label{Y0-tilde}
\tY_0=V^\dag Y_0\,.
}
The matrix $V$ is close to the CKM matrix of the SM and we can use a rephasing convention where the third row and third column of $V$ are real positive.
The number of physical parameters in $V$ is four, corresponding to three angles $\theta_{12},\theta_{23},\theta_{13}$ and one CP phase $\delta$ of the standard parametrization.
The total number of physical parameters contained in $X_u,X_d,Y$ amount to 15, which coincides with the number of factors in the denominator of the Hilbert series \eqref{HS:vlq.mass}.
This number is not yet the number of physical parameters in the Lagrangian \eqref{lag:vlq} because the VLQ mass $M$ is invariant under $G_F$ in \eqref{GF} and was disregarded as a spurion.

For the invariants involving only $X_u$ and $X_d$, the relation between invariants and physical parameters is the same as in the SM\,\cite{manohar:hilbert}: the 10 physical parameters are in one-to-one correspondence with the primary invariants $I_{20},I_{02},I_{40},I_{40},I_{60},I_{06},I_{22},I_{42},I_{24},I_{44}$.
The first six invariants give the six Yukawas in \eqref{phys.param} while the rest give information about the moduli $|V_{\alpha\beta}|^2$, among which we can choose, e.g., $|V_{ud}|^2,|V_{us}|^2,|V_{cd}|^2,|V_{cs}|^2$, as the independent ones.
The sign ambiguity of the CP violating $\delta$ can be eliminated by the CP odd Jarlskog invariant $I^-_{66}$.
With the information of the four independent moduli $|V_{\alpha\beta}|^2$ and one CP violation sign, one can completely recover the physical parameters of the matrix $V$\,\cite{branco.lavoura:|V|}.
Explicit formulas for the standard parametrization can be found in e.g. Ref.\,\cite{denton}.
A detailed discussion about quark flavor invariants within the SM, including RGE running, can be found in Ref.\,\cite{silva.trautner}.

Now let us discuss the parameters contained in $Y$ and its relation with the $K$-invariants.
From the denominator of the Hilbert series \eqref{HS:vlq.mass}, we still need primary invariants corresponding to the factors $(1-q^2)(1-q^3)^2(1-q^4)^2$.
We can choose
\eqali{
\label{phys.param:K}
K_{00}&=|Y_1|^2+|Y_2|^2+|Y_3|^2\,,
\cr
K_{20}&=y_u^2|Y_1|^2+y_c^2|Y_2|^2+y_t^2|Y_3|^2\,,
\cr
K_{02}&=y_d^2|\tY_1|^2+y_s^2|\tY_2|^2+y_b^2|\tY_3|^2\,,
\cr
K_{40}&=y_u^4|Y_1|^2+y_c^4|Y_2|^2+y_t^4|Y_3|^2\,,
\cr
K_{04}&=y_d^4|\tY_1|^2+y_s^4|\tY_2|^2+y_b^4|\tY_3|^2\,,
\cr
}
It is clear that the three moduli $|Y_1|^2,|Y_2|^2,|Y_3|^2$ and the three moduli $|\tY_1|^2,|\tY_2|^2,|\tY_3|^2$, can be determined from these five invariants, noting that
\eq{
\label{YYdag}
|Y_1|^2+|Y_2|^2+|Y_3|^2=|\tY_1|^2+|\tY_2|^2+|\tY_3|^2\,,
}
and once the physical values have nonzero $y_u^2y_c^2y_t^2(y_u^2-y_c^2)(y_c^2-y_t^2)(y_t^2-y_u^2)\neq 0$ and
$y_d^2y_s^2y_b^2(y_d^2-y_s^2)(y_s^2-y_b^2)(y_b^2-y_d^2)\neq 0$.

To be specific, let us focus on the basis \eqref{phys.param} where $(Y_0)_3=Y_3$ is real positive and $V$ has the third row and third column real positive.
We assume that $V$ is known from the $I$-invariants and the moduli $|Y_i|$ are known from $K_{00},K_{20},K_{40}$ in \eqref{phys.param:K}.
Then, the two physical phases $\alpha_1,\alpha_2$ in $Y_0$ remain to be determined:
\eq{
\label{Y0:param}
Y_0=\mtrx{|Y_1|e^{i\alpha_1}\cr |Y_2|e^{i\alpha_2}\cr |Y_3|}\,.
}
Now we consider $\tY_0$ as defined by \eqref{Y0-tilde} as an auxiliary variable whose absolute values of the components are determined by $K_{00},K_{02},K_{04}$ in \eqref{phys.param:K}.
Then, the phases $\alpha_i$ are determined, e.g., by
\eqali{
\Big|V_{11}^*|Y_1|e^{i\alpha_1}+V_{21}^*|Y_2|e^{i\alpha_2}+V_{31}^*|Y_3|\Big|^2&=|\tY_1|^2\,,
\cr
\Big|V_{13}^*|Y_1|e^{i\alpha_1}+V_{23}^*|Y_2|e^{i\alpha_2}+V_{33}^*|Y_3|\Big|^2&=|\tY_3|^2\,.
}
The analogous equation for $\tY_2$ is automatically satisfied from \eqref{YYdag} because of unitarity of $V$.
Equivalently, we could use the components of the normalized vector $Y_0/|Y_0|$.
Apart from the trivial sign ambiguity due to flipping of all the phases, including the $V\to V^*$ for the CKM, we check numerically that there are other discrete ambiguities even if we do not take the complex conjugate of the CKM.
These other ambiguities should be resolved from the knowledge of the other basic invariants of Table~\ref{tab:flavor_invariants} that are not considered as primary.
Based on the example of the analysis of the CP odd flavor invariants of the neutrino sector\,\cite{yu.zhou:inv.nu}, 
the necessity of some or all of the CP odd basic invariants may arise only on some special regions of parameter space such as when masses are degenerate\,\cite{yu.zhou:degenerate.nu}.
To cover all possibilities, all the basic invariants should be necessary.

We summarize the one-to-one relation between the chosen 15 primary invariants and the chosen 15 physical parameters as
\eqali{
\{I_{20},I_{40},I_{60}\}&\to \{y_u,y_c,y_t\}\,,
\cr
\{I_{02},I_{04},I_{06}\}&\to \{y_d,y_s,y_b\}\,,
\cr
\{I_{22},I_{42},I_{24},I_{44}\}&\to \{|V_{ud}|,|V_{us}|,|V_{cd}|,|V_{cs}|\}\,,
\cr
\{K_{00},K_{20},K_{40}\}&\to \{|Y_1|,|Y_2|,|Y_3|\}\,,
\cr
\{K_{02},K_{04}\}&\to \{\arg(Y_1),\arg(Y_2)\}\,.
}
The arrow means that the set of invariants determines the parameter up to discrete ambiguities that need to be resolved with the knowledge of other basic invariants.
Within the SM, analogous to the presence of $X_u,X_d$, it is known that the sign of the Jarlskog invariant $I^-_{66}$ resolves the ambiguity of the other CP even $I$-invariants.
With the addition of the VLQ, it is not clear what additional invariants are necessary to resolve the discrete ambiguities.
We should also remark that the values attained by the invariants corresponding to physical values of the parameters live in a space of nontrivial shape.
Such a space for the case of the SM is mapped out in Ref.\,\cite{silva.trautner}.

In analogy to the leptonic sector with Majorana neutrinos\,\cite{yu.zhou:inv.nu}, 
excluding some special points,
it is possible to establish simpler relations between physical parameters and flavor invariants by considering a mixture of primary and non-primary invariants.
But some ambiguities still need to be resolved by other invariants.\footnote{%
This point is not emphasized in Ref.\,\cite{yu.zhou:inv.nu}.}
Our simpler relations are schematically the following:
\subeqali[inv=param:simple]{
\label{inv=param:simple:1}
\{I_{02},I_{04},I_{06}\}&\to \{y_d,y_s,y_b\}\,,
\\
\label{inv=param:simple:2}
\{I_{20},I_{22},I_{24}\}&\to \{H_{11},H_{22},H_{33}\}\,,
\\
\label{inv=param:simple:3}
\{I_{40},I_{42},I_{44}\}&\to \{H_{12},H_{23},H_{31}\}\,,
\\
\label{inv=param:simple:4}
\{K_{00},K_{02},K_{04}\}&\to \{|\tY_1|,|\tY_2|,|\tY_3|\}\,,
\\
\label{inv=param:simple:5}
\{K^-_{22},K^-_{24},K^-_{26}\}&\to \{\sin(\phi_{12}),\sin(\phi_{23}),\sin(\phi_{31})\}\,.
}
We consider the basis \eqref{phys.param:Xu} of diagonal $X_d$ and apply rephasing on it so that $\tY_0$ has real positive components. The matrix $V$ now has three physical phases, still retaining rephasing freedom from the left.
To differentiate it from the usual CKM with four parameters, we denote this mixing matrix as $\tV$.
For definiteness, we can consider the same standard parametrization as of the PMNS matrix for Majorana neutrinos:
\eq{
\tV=
\left(
\begin{array}{ccc}
 1 & 0 & 0 \\
 0 & c_{23} & s_{23} \\
 0 & -s_{23} & c_{23} \\
\end{array}
\right)
\mtrx{ c_{13}&0&s_{13}e^{-i\delta}\cr
    0&1&0\cr
    -s_{13}e^{i\delta}&0&c_{13}}
\mtrx{ c_{12}&s_{12}&0\cr
    -s_{12}&c_{12}&0\cr
    0&0&1}
\mtrx{e^{i\alpha_1}&&\cr &e^{i\alpha_2}&\cr &&1}
\,,
}
where the usual shorthands $s_{ij}\equiv \sin\theta_{ij}, c_{ij}\equiv \cos\theta_{ij}$ are used and the phases $\alpha_1,\alpha_2$ are similar to \eqref{Y0:param} depending on the rephasing convention of $V$.
Concerning $X_u,X_d$, the situation is analogous to the case of the lepton sector with Majorana neutrinos\,\cite{yu.zhou:degenerate.nu}.
The rest of the parameters appearing in \eqref{inv=param:simple} is contained in
\eqali{
X_u=\tV^\dag\diag(y_u,y_c,y_t)\tV
=\mtrx{H_{11} & H_{12}e^{i\phi_{12}} & H_{31}e^{-i\phi_{31}} \cr
  \star  & H_{22} & H_{23}e^{i\phi_{23}} \cr
  \star  & \star & H_{33}
  }\,,
}
with six moduli $H_{11},H_{22},H_{33},H_{12},H_{23},H_{31}$ and three phases $\phi_{12},\phi_{23},\phi_{31}$ as independent parameters.
The knowledge of these 9 parameters is equivalent to the determination of the three Yukawas $y_u,y_b,y_t$,
and six parameters in $\tV$.

Let us now detail the relations in \eqref{inv=param:simple}.
The relation \eqref{inv=param:simple:1} is the same as in the SM.
The relation \eqref{inv=param:simple:2} is given by
\eqali{
\label{inv=param:Hii}
I_{20}&= H_{11}+H_{22}+H_{33}\,,
\cr
I_{22}&= y_d^2H_{11}+y_s^2H_{22}+y_b^2H_{33}\,,
\cr
I_{24}&= y_d^4H_{11}+y_s^4H_{22}+y_b^4H_{33}\,.
}
When the Vandermonde determinant $\Delta^d_{12}\Delta^d_{23}\Delta^d_{31}$, with $\Delta^d_{ij}=y^2_{d_i}-y^2_{d_j}$, is nonvanishing, these relations can be inverted to give $H_{ii}$.
The relation \eqref{inv=param:simple:3} comes from 
\eqali{
\label{inv=param:Hij}
I_{40}&= H^2_{11}+H^2_{22}+H^2_{33}+2(H_{12}^2+H_{23}^2+H_{31}^2)\,,
\cr
I_{42}&= y_d^2H_{11}^2+y_s^2H_{22}^2+y_b^2H_{33}^2
+(y_d^2+y_s^2)H_{12}^2+(y_s^2+y_b^2)H_{23}^2+(y_b^2+y_d^2)H_{31}^2
\,,
\cr
I_{44}&= y_d^4H_{11}^2+y_s^4H_{22}^2+y_b^4H_{33}^2
+(y_d^4+y_s^4)H_{12}^2+(y_s^4+y_b^4)H_{23}^2+(y_b^4+y_d^4)H_{31}^2
\,.
}
We note that the choice of invariants in \eqref{inv=param:Hii} and \eqref{inv=param:Hij} are simpler than in Ref.\,\cite{yu.zhou:inv.nu}. 
The relation \eqref{inv=param:simple:4} was given in \eqref{phys.param:K} and determines $\tY_i=|\tY_i|$.
Finally, the relation \eqref{inv=param:simple:5} is determined by
\eqali{
\label{simple:CPodd}
\frac{1}{2i}K^-_{22}&=
\Delta^d_{12}\tY_1\tY_2H_{12}\sin\phi_{12}
+\Delta^d_{23}\tY_2\tY_3H_{23}\sin\phi_{23}
+\Delta^d_{31}\tY_3\tY_1H_{31}\sin\phi_{31}
\,,
\cr
\frac{1}{2i}K^-_{24}&=
\Delta^d_{12}(y^2_{d_1}+y^2_{d_2})\tY_1\tY_2H_{12}\sin\phi_{12}
+\Delta^d_{23}(y^2_{d_2}+y^2_{d_3})\tY_2\tY_3H_{23}\sin\phi_{23}
+\Delta^d_{31}(y^2_{d_3}+y^2_{d_1})\tY_3\tY_1H_{31}\sin\phi_{31}
\,,
\cr
\frac{1}{2i}K^-_{26}&=
\Delta^d_{12}y^2_{d_1}y^2_{d_2}\tY_1\tY_2H_{12}\sin\phi_{12}
+\Delta^d_{23}y^2_{d_2}y^2_{d_3}\tY_2\tY_3H_{23}\sin\phi_{23}
+\Delta^d_{31}y^2_{d_3}y^2_{d_1}\tY_3\tY_1H_{31}\sin\phi_{31}
\,.
}
For nondegenerate $y_{d_i}$ and $\tY_1\tY_2\tY_3H_{12}H_{23}H_{31}\neq 0$,  
these invariants determine $\sin\phi_{ij}$ except for the sign ambiguity of $\cos\phi_{ij}$.
Hence, within this case, the vanishing of these three CP odd invariants establishes necessary and sufficient conditions for CP invariance.
For more special points, such as when some of $y_{d_i}$ are degenerate or when $\tY_1\tY_2\tY_3H_{12}H_{23}H_{31}=0$ the vanishing of these invariants cannot be used as sufficient conditions for CP conservation and other invariants needs to be sought. 
The alternative is to use all the 9 CP odd invariants in Table\,\ref{tab:flavor_invariants}; see discussion in the next section.
For the leading new sources of CP violation in the dimension six SMEFT operators, it is possible to establish necessary and sufficient conditions with invariants linear in the new coefficients\,\cite{grojean:cp}.

For the choice \eqref{inv=param:simple}, we know how to resolve the sign ambiguity of $\cos\phi_{ij}$ by using the invariants
\eqali{
K_{20}&=
\tY_1^2H_{11}+\tY_2^2H_{22}+\tY_3^2H_{33}
\cr&\quad
+2\tY_1\tY_2H_{12}\cos\phi_{12}
+2\tY_2\tY_3H_{23}\cos\phi_{23}
+2\tY_3\tY_1H_{31}\cos\phi_{31}
\,,
\cr
\frac{1}{2}K^+_{22}&=
y_d^2\tY_1^2H_{11}+y_s^2\tY_2^2H_{22}+y_b^2\tY_3^2H_{33}
\cr&\quad
+(y_d^2+y_s^2)\tY_1\tY_2H_{12}\cos\phi_{12}
+(y_s^2+y_b^2)\tY_2\tY_3H_{23}\cos\phi_{23}
+(y_b^2+y_d^2)\tY_3\tY_1H_{31}\cos\phi_{31}
\,,
\cr
\frac{1}{2}K^+_{24}&=
y_d^4\tY_1^2H_{11}+y_s^4\tY_2^2H_{22}+y_b^4\tY_3^2H_{33}
\cr&\quad
+(y_d^4+y_s^4)\tY_1\tY_2H_{12}\cos\phi_{12}
+(y_s^4+y_b^4)\tY_2\tY_3H_{23}\cos\phi_{23}
+(y_b^4+y_d^4)\tY_3\tY_1H_{31}\cos\phi_{31}
\,.
}
These relations can be solved for $\cos\phi_{ij}$.

\section{Relation to general weak basis and invariants}
\label{sec:SU4}

As $B_R$ has the same gauge quantum numbers as $d_{iR}$, the most general weak basis transformation is
\eq{
\label{quarks:U4}
\mtrx{d_{iR}\cr B_R}\to W \mtrx{d_{iR}\cr B_R}\,,
}
where $W\in U(4)$.
With this freedom, the general lagrangian can be written as
\eq{
\label{lag:vlq:su4}
-\lag = \overline{q}_{iL}\Tilde{H} Y^u_{ij}  u_{jR} + \overline{q}_{iL} H \YY_{i\alpha} d_{\alpha R}
      + \overline{B}_{L} \MM_{\alpha} d_{\alpha R}+ h.c.,
}
where now $\alpha=1,...,4$, $d_{4R}\equiv B_R$, $\YY\sim 3\times 4$ and $\MM\sim 1\times 4$.
We recover the lagrangian \eqref{lag:vlq} when
\eq{
\label{Y.M:vlq.mass}
\YY=\big(Y^d\mid Y^B\big)\,,
\quad
\MM=\big(0,0,0\mid M\big)\,.
}
Thus the lagrangian \eqref{lag:vlq:su4} still contains 16 physical parameters.

The flavor symmetry group in the absence of $Y^u,\YY,\MM$ is now
\eq{
U(3)_q\otimes U(3)_u\otimes U(4)_d\otimes U(1)_{B_L}\,,
}
instead of \eqref{GF:vlq:MB}, where the $U(1)$ factors correspond to rephasing of $B_L$.
Factoring $U(3)_u$ as usual by considering $X_u=Y^u{Y^u}^\dag$, 
the relevant flavor symmetry is
\eq{
\label{GF:su4}
SU(3)_q\otimes SU(4)_d\otimes U(1)_{B_L}\otimes U(1)_q\,.
}
The transformation of the spurions $X^u,\YY,\YY^\dag,\MM,\MM^\dag$ under this group is
\eqali{
\label{spurions:su4:transf}
X_u&\to U^qX_u{U^q}^\dag\,,
\cr
\YY&\to e^{i\theta_q}U^q\YY W^\dag\,,&\quad
\YY^\dag &\to W X_u{U^q}^\dag e^{-i\theta_q}\,,\quad
\cr
\MM&\to e^{i\theta_L}\MM W^\dag\,,&\quad
\MM^\dag &\to W \MM^\dag e^{-i\theta_L}\,,
}
with representations
\eq{
\label{spurions:su4}
X_u\sim (\bs{3}\otimes\bs{\bar{3}},\bs{1},0,0)\,,\quad
\YY\sim (\bs{3},\bs{\bar 4},0,+1)\,,\quad
\YY^\dag\sim (\bs{\bar 3},\bs{4},0,-1)\,,\quad
\MM\sim (\bs{1},\bs{\bar 4},+1,0)\,,\quad
\MM^\dag\sim (\bs{1},\bs{4},-1,0)\,.
}
Now these spurions depend on 16 parameters that match the 15 parameters contained in the spurions \eqref{spurions} in the VLQ mass basis, with addition of the mass $M$.

Considering the spurions \eqref{spurions:su4} as the same variable, $q\sim X_u,\YY,\YY^\dag,\MM,\MM^\dag$, we can write the Molien-Weyl formula for the unrefined Hilbert series as a residue integral, as shown in appendix \ref{ap:su4}.
The residue integral, however, involves seven integrals over seven complex variables which we were unable to solve.
Therefore, we can obtain the first few terms of the expanded Hilbert series by using the trick of Ref.\,\cite{lehman.1}: we first expand the integrand around $q=0$ and collect the residues of $z_i=0$.
The result up to $q^4$ is
\eq{
\HS(q)=1 +q +4q^2 +6q^3 +15 q^4+\cdots\,.
}
Up to the same power, we can also calculate the PL as
\eq{
\label{PL:su4}
\PL[H(q)]=q+3q^2+2q^3+3q^4+\cdots\,.
}
The basic invariants corresponding to these low degrees can be easily constructed and we list them in table\,\ref{tab:su4}.
\begin{table}[h]
\centering
\begin{tabular}{|c|c|c|c|}
\hline
Flavor Invariant & Phys. degree & degree in \eqref{PL:su4} & CP \\
\hline
$ I_{20} = \aver{X_u} $ & 2 & 1 &+ \\
$ \aver{\YY\YY^\dag} $ & 2 & 2 & + \\
$ \aver{\MM\MM^\dag} $ & 2 & 2 & + \\
$ I_{40} = \aver{X^2_u} $ & 4 & 2 & + \\
$ \aver{\YY\YY^\dag X_u} $ & 4 & 3 & + \\
$ I_{60} = \aver{X^3_u} $ & 6 & 3 & + \\
$ \aver{(\YY\YY^\dag)^2} $ & 4 & 4 & + \\
$ \aver{\YY\MM^\dag\MM\YY^\dag} $ & 4 & 4 & + \\
$ \aver{\YY\YY^\dag X^2_u} $ & 6 & 4 & + \\
\hline
\end{tabular}
\caption{\label{tab:su4}%
First few invariants in general weak basis along with their physical degree, the degree in the PL \eqref{PL:vlq.mass}, and CP parity. 
}
\end{table}

In fact, we can factor the $U(4)$ group of the space $(d_{iR},B_R)$ by defining the contracted quantities
\eq{
\label{def:XX}
\XX \equiv \YY\YY^\dag
\,,\quad
Z\equiv \YY\MM^\dag
\,,\quad
\MM\MM^\dag\,.
}
The last quantity is already an invariant by \eqref{GF:su4}.
So we can use as our nontrivial spurions,
\eq{
\label{spurions:su4:contracted}
X_u\sim (\bs{3}\otimes\bs{\bar{3}},0)\,,\quad
\XX\sim (\bs{3}\otimes\bs{\bar{3}},0)\,,\quad
Z\sim (\bs{3},-1)\,,\quad
Z^\dag \sim (\bs{\bar{3}},+1)\,,
}
of the group $G_F=SU(3)_q\otimes U(1)_{B_L}$, which is equivalent to the flavor group in the VLQ mass basis, cf.\,\eqref{GF}.
Surprisingly, the transformation properties of the spurions in \eqref{spurions:su4:contracted} 
coincide with the ones of the spurions in 
\eqref{spurions} in the VLQ mass basis.
Therefore the unrefined Hilbert series \eqref{HS:vlq.mass} for $X_u,\XX,Z,Z^\dag\to q$ is the same 
and our invariants constructed in Table \ref{tab:flavor_invariants} carry over directly to this 
case with the replacements
\eq{
\label{replacement}
X_d\to \XX\,,\quad
Y\to Z\,.
}
Note that we can consider $Z,Z^\dag$ as independent quantities from $\XX$.
Now as $Z$ has the same physical degree as $X_u$, the physical degree in Table \ref{tab:flavor_invariants} needs to be adapted by simply multiplying the PL degree by two.
The result is a total of 29 basic invariants of the full weak basis transformations including $U(4)$ in \eqref{quarks:U4}: 28 that can be constructed from Table \ref{tab:flavor_invariants} by applying the replacement \eqref{replacement} and one more which is $\MM\MM^\dag$.
The latter was also confirmed by the analysis leading to table \ref{tab:su4}.

We can also consider weak basis invariants after electroweak symmetry breaking.
In this case, the relevant quantity is the full $4\times 4$ down-type mass matrix
\eq{
\cM^d=
\mtrx{\frac{v}{\sqrt{2}}\YY\cr \MM}\,.
}
Its singular values in
\eq{
\cV^\dag \cM^d W=\diag(m_d,m_s,m_b,m_B)\,,
}
give the physical masses, where $m_B$ is the heavy VLQ mass.
In the basis where $Y^u$ is diagonal, the mixing matrix $\cV$ contains the physical $3\times 4$ CKM matrix.
In turn, this matrix enters in the flavor changing current coupling to the $Z$. 
See Ref.\,\cite{vlq:review} for a review.

Flavor invariants in this manner can be studied and CP odd invariants were found long ago\,\cite{vlq.inv}
while a choice of primary invariants, closely related to the physical parameters contained in the quark masses and in $\cV$, were found in Ref.\,\cite{albergaria} without the use of Hilbert series.
The latter are CP even and can be written in terms of the couplings in \eqref{spurions:su4}.
The physical parameters can be determined from these invariants up to discrete ambiguities that need to be resolved with other invariants much like our discussion in Sec.\,\ref{sec:phys.param}.
A similar problem is encountered for the SM with four quark families\,\cite{lavoura}.

Ref.\,\cite{vlq.inv} formulates necessary and sufficient basis invariant conditions for CP conservation in terms of the vanishing of 7 CP odd invariants which, adapted to our notation, are 
\eqali{
\label{inv:aguila-branco}
I_1 &=\aver{X_u\XX ZZ^\dag}-c.c.,\cr
I_2 &=\aver{X_u^2\XX ZZ^\dag}-c.c.,\cr
I_3 &=\aver{X_u^2[X_u,\XX]ZZ^\dag}-c.c.,\cr
I_4 &=\aver{X_u\XX^2 ZZ^\dag}-c.c.,\cr
I_5 &=\aver{X_u^2\XX^2 ZZ^\dag}-c.c.,\cr
I_6 &=\aver{X_u^2[X_u,\XX^2]ZZ^\dag}-c.c.,\cr
I_7 &=\aver{X_u^2\XX^2 X_u \XX}-c.c.,\cr
}
where we have used \eqref{def:XX}.

Here we show by an example that these 7 invariants do not cover all cases of CP violation.
We use the basis \eqref{Y.M:vlq.mass}, where our results and invariants are valid.
In such a basis,
\eq{
\XX=X_d+YY^\dag\,,\quad
ZZ^\dag=M^2 YY^\dag\,,
}
where $M$ is VLQ bare mass, cf.\,\eqref{lag:vlq}, and we only need to provide $X_u,X_d,YY^\dag$ to calculate the invariants.
For the example, we choose 
\eq{
\label{example}
X_u=\diag(y_u^2,y_u^2,y_t^2)\,,\quad
X_d=\mtrx{a&d&0\cr d&b&e \cr 0&e&c}\,,\quad
Y=\mtrx{|Y_1|\cr |Y_2|e^{i\alpha} \cr 0}
}
where $a,b,c,d,e$ are real.
Note the degeneracy $y_c=y_u$ and the vanishing $Y_3$; the zero in $X_d$ is not crucial but simplify the expressions.
It is clear that the Jarlskog invariant $I_{66}^-$ is zero due to a real $X_d$ and the degeneracy.
For this example, one can check that all invariants \eqref{inv:aguila-branco} vanish.
Many of the invariants are automatically zero by noting that $X_u$ and $YY^\dag$ commute, and only the latter contains a complex entry.
Analogously, all the 7 CP odd invariants (adapted to one down type VLQ) of Ref.\,\cite{albergaria} vanish as well.
To show that CP is still violated, we calculate the following two of our CP odd invariants:
\eqali{
\label{example:nonzero}
K^-_{26}&=2 i (y_t^2-y_u^2)d e^2 |Y_1| |Y_2| \sin\alpha\,,
\cr
K^-_{46}&=2 i (y_t^4-y_u^4) d e^2 |Y_1| |Y_2| \sin\alpha\,.
}
Hence, there is CP violation induced by the phase $\alpha$ which is not detected by the 7 invariants $I_k$.
This is in accordance with our basic invariants of Table\,\ref{tab:flavor_invariants} which contains 9 CP odd invariants.
All of them should be necessary to cover all possibilities of CP violation.
For the example \eqref{example}, the remaining 7 of our CP odd invariants other than \eqref{example:nonzero} vanish.
In general, we expect the number of the complete set of basic CP odd invariants to be equal or larger than the number of phases in some basis and parametrization. In turn, the number of phases is known to vary depending on the weak basis and/or parametrization\,\cite{less.phases,bastos.juca}.

To complete the discussion, we relate in the basis \eqref{Y.M:vlq.mass} the 7 invariants \eqref{inv:aguila-branco} with our 9 CP odd invariants of Table\,\ref{tab:flavor_invariants}:
\eqali{
\label{branco.ours}
I_1/M^2 &\sim K_{22}^-\cr
I_2/M^2 &\sim K_{42}^-,\cr
I_3/M^2 &\sim K_{62}^- +(\text{lower-degree}),\cr
I_4/M^2 &\sim K_{24}^- +(\text{lower-degree}),\cr
I_5/M^2 &\sim K_{44}^- +(\text{lower-degree}),\cr
I_6/M^2 &\sim K_{64}^- +(\text{lower-degree}),\cr
I_7/M^2 &\sim I_{66}^- + K_{64}^- +(\text{lower-degree}).\cr
}
The symbol $\sim$ means that we are suppressing numerical factors and (lower-degree) contains products of invariants of lower degree containing one CP odd invariant of the previous lines.
We note that all of our basic CP odd invariants appear with the exception of $K^-_{26},K^-_{46}$ in \eqref{example:nonzero}.

To conclude this section, necessary and sufficient conditions for CP invariance can be formulated in terms of 9 basic flavor invariants of the full basis transformations of the four fields $d_{iR},B_{R}$. These invariants can be quickly read from the 9 CP odd invariants of Table\,\ref{tab:flavor_invariants} by applying the replacement \eqref{replacement}.

\section{Summary}
\label{sec:summary}

For the SM adjoined with one singlet VLQ, we have found the complete list of basic flavor invariants with which all invariants can be written as a polynomial, i.e., the generating set.
These invariants are constructed with the aid of the Hilbert series calculated from the flavor symmetry properties of the noninvariant lagrangian parameters in the VLQ mass basis.
The generating set comprising 28 invariants, 19 CP even and 9 CP odd, is listed in Table\,\ref{tab:flavor_invariants}.
Among these, the 11 invariants depending solely on the SM Yukawas are equivalent to the ones of the SM.
Among the CP odd invariants, our invariant $K^-_{22}$ has physical degree 6 which is half of the degree 12 Jarlskog invariant $I^-_{66}$ of the SM.
Because of the lower degree, the addition of the VLQ may induce a much stronger new source of CP violation much needed for the explanation of the matter-antimatter asymmetry of the Universe.
We have also extended the analysis to the full weak basis transformations of the four righthanded quarks including the VLQ. The basic invariants are essentially the same in form with one additional one corresponding to the VLQ mass and hence CP even.
The total amounts to 29 basic invariants.

The Hilbert series through the plethystic logarithm also provided information about the nontrivial and nonlinear relations between the basic invariants, i.e., the syzygies.
The syzygy of lowest degree \eqref{syzygy:ff} corresponds to writing the square of our lowest degree CP odd invariant $K^-_{22}$ in terms of CP even invariants.
That this is always possible is expected from the CP properties of the invariants.
What is interesting in our case is that this syzygy is \emph{only valid} for one VLQ.
The generalized relation for more than one singlet VLQ is no longer a syzygy because a CP even invariant appears \emph{linearly}.
This last relation was uncovered because we were able to find the $SU(3)$ identity involving two structure constants $f_{ijk}f_{i'j'k'}$ underlying the syzygy, cf.\ eq.\,\eqref{Tff.identity}.
The $SU(3)$ relation is the expansion of $f_{ijk}f_{i'j'k'}$, a 6-legged tensor in the adjoint, in terms of a suitable basis that span such a tensor space.
Moreover, this identity allows us to write the square of any CP odd invariant of the form
\eq{
\nonumber
\aver{[A,B]C}\,,
}
with $A,B,C$ being $3\times 3$ hermitean matrices, as a linear combination of CP even invariants.
We have also studied six more syzygies of low degree and found out interestingly that two of these syzygies are generalizable to more than one VLQ while the rest are not.

To complete the study, we have established a one-to-one relation between the physical parameters and the primary invariants, except for discrete ambiguities.
Away from some special points, we also found simpler relations between invariants and physical parameters, in special, the CP violating phases. In this case, it is also shown how to resolve the sign ambiguities.
To cover all cases of CP violation, however, the complete list of 9 CP odd invariants should be used. We showed by an example that there is CP violation that is not detected by the CP odd invariants proposed in the literature so far.

\appendix
\section{Hilbert series in VLQ mass basis}
\label{app:HS}

Here we provide the details for obtaining the Hilbert series \eqref{HS:vlq.mass} through the Molien-Weyl formula which, in our case, reads\,\cite{gray.hanany,hanany.torri,lehman.1}
\begin{equation}
H(X_u,X_d,Y,Y^\dag)= \int_{T} d\mu\PE[X_{u},z_j]\PE[X_{d},z_j]\PE[Y,z_j]\PE[Y^{\dagger},z_j]\,,
\end{equation}
where the complex variables $z_j$, $j=1,\dots,r$, parametrizes the maximal torus $T$ of $G_F=SU(3)_q\otimes U(1)_{\rm VLQ}$, with $r$ being the dimension of $T$ and the rank of the group.
The spurions $X_u,X_d,Y,Y^\dag$ are considered as variables here and $\PE$ is the plethystic exponential, defined for a spurion $X$ transforming by the representation $R(g)$ of $g\in G$ as\,\cite{gray.hanany,hanany.torri,lehman.1}
\eq{
\label{PE}
\PE[X,z_j;R] \equiv 
\PE[X\chi_R(z_j)]
=\exp \left(\sum_{n=1}^\infty \frac{X^n \chi_R(z^n_j)}{n} \right)\,,
}
where $\chi_R$ is the character function of the representation $R$.
The PE \eqref{PE} is the generating function of symmetric combinations of the variable $X,z_j$ in $X\chi_R(z_j)$.
In general, for a function $f(t_1,\dots,t_r)$ vanishing at the origin, it is defined as
\eq{
\PE[f(t_1,\dots,t_r)]=\exp \left(\sum_{n=1}^\infty \frac{f(t_1^n,\dots,t_r^n)}{n} \right)\,.
}
For a group element $g$, the PE \eqref{PE} can be also written as
\eq{
\PE[X,g;R]= \frac{1}{\det(\id - XR(g))}
=\exp\big[-\tr\ln(\id - XR(g))\big]\,.
}

The Haar measure of our $G_F$ is
\begin{equation}
\label{Haar}
\int_{T} d\mu = 
\frac{1}{(2 \pi i)^3} \oint_{|z_1|=1} \frac{dz_1}{z_1} \oint_{|z_2|=1} \frac{dz_2}{z_2} (1-z_1 z_2) \left(1-\frac{z^2_1}{z_2} \right) \left(1-\frac{z^2_2}{z_1} \right)
\oint_{|z_3|=1} \frac{dz_3}{z_3}
\end{equation}
where $z_1,z_2$ are the variables for the maximal torus of $SU(3)$ while $z_3$ corresponds to $U(1)$.
These can be found e.g. in Refs.\,\cite{hanany.torri,lehman.1}.

Considering the spurion transformations \eqref{spurions}, the character functions are
\begin{equation}
\begin{aligned}
\chi[X_u]&=\chi[X_d]= z_1 z_2 +\frac{z_1^2}{z_2} + \frac{z_2^2}{z_1} + 3 + \frac{z_2}{z_1^2} + \frac{z_1}{z_2^2} + \frac{1}{z_1 z_2}\,,
\\
\chi[Y] &= \left( z_1 + \frac{1}{z_2} + \frac{z_2}{z_1}\right)\frac{1}{z_3} \,,
\\
\chi[Y^\dag] &= \left( z_2 + \frac{1}{z_1} + \frac{z_1}{z_2}\right)z_3 \,.
\end{aligned}
\end{equation}
Each character can be obtained from the more basic characters 
$\chi_{\bs{3}}(z_1,z_2)=z_1 + \frac{z_2}{z_1} + \frac{1}{z_2}$,
$\chi_{\bs{3}^*}(z_1,z_2)=\frac{1}{z_1} + \frac{z_1}{z_2} + z_2$,
for $\bs{3}$ and $\bs{3}^*$ of $SU(3)$,
and $\chi_{+1}(z_3)=z_3$ for charge $+1$ of $U(1)$.

Then, the PE for $X_u$ can be written as
\eqali{
\PE[X_{u},z_j] &=
  \exp\left\{\sum_{n=1}^{\infty} \frac{(X_u)^n}{n}\left[(z_1 z_2)^n + \left( \frac{z_2^2}{z_1} \right)^n+\left( \frac{z_1^2}{z_2} \right)^n + 3+ \left( \frac{z_2}{z_1^2} \right)^n+\left( \frac{z_1}{z_2^2} \right)^n+\left( \frac{1}{z_1 z_2}\right )^n \right]\right\}
  \\
  &=
\frac{z_1^4z_2^4}{(1 - X_u z_1 z_2)(z_1 - X_u z^2_2)(z_2 - X_u z^2_1)(1-X_u)^3 (z^2_1 - X_u z_2)(z^2_2 - X_u z_1)(z_1 z_2 - X_u)}\,,
}
while the PE for $X_d$ is the same function.
The PE for $Y$ is
\eqali{
\PE[Y,z_j] &= \exp{\left[ \sum_r \frac{Y^r}{r} \left(\frac{z_1^r}{z_3^r} + \frac{1}{z_2^r z_3^r}+ \frac{z_2^r}{z_1^r z_3^r}\right)\right]}, 
\\
    &= \Big[1 - \frac{Y z_1}{z_3}\Big]^{-1}\Big[1 - \frac{Y}{z_2z_3}\Big]^{-1}\Big[1 - \frac{Y z_2}{z_1z_3}\Big]^{-1}
}
while the PE for $Y^\dag$ can be obtaind by replacing $z_j\to z_j^{-1}$.

Instead of the unrefined Hilbert series \eqref{HS:vlq.mass} with variables $X_u\to q, X_d\to q, Y\to q, Y^\dag\to q$, we could instead perform the almost identical calculation with $X_u\to q^2, X_d\to q^2, Y\to q, Y^\dag\to q$, where the degree would match the physical degree directly.
The result is
\eq{
\label{HS:vlq.mass:q^2}
H(q)=
\frac{1+2 q^6+4 q^8+4 q^{10}+5 q^{12}+4 q^{14}+5 q^{16}+4 q^{18}+4 q^{20}+2 q^{22}+q^{28}}
{\left(1-q^2\right)^3 \left(1-q^4\right)^5 \left(1-q^6\right)^6 \left(1-q^8\right)}
\,,
}
with
\eq{
\label{PL:vlq.mass:q^2}
\PL[H(q)]=3 q^2+5 q^4+8 q^6+5 q^8+4 q^{10}+2 q^{12}-4 q^{14}-13 q^{16}+\cdots
\,.
}
Here we note the occurrence of the same cancellation reported in Ref.\,\cite{yu.zhou:inv.nu}: 
the coefficient of $q^{12}$ indicates two basic invariants but in Table~\ref{tab:flavor_invariants} we constructed three basic invariants.
The reason is clear: the missing $q^{12}$ term is cancelled by the presence of the single syzygy of degree 12 in 
Sec.\,\ref{sec:lowest} which contributes $-q^{12}$ in the PL.
Therefore, in this case, the unrefined Hilbert series in the form \eqref{HS:vlq.mass} gives a more precise information in the PL about the number and degree of the basic invariants than the series in \eqref{HS:vlq.mass:q^2}.
This example shows that calculating the unrefined Hilbert series with different powers of the single variable may provide additional information by reshuffling the polynomial degrees of the various invariants and syzygies.
Such a computation involves no more complexity than the usual unrefined case.

In contrast, the calculation of the graded (refined) Hilbert series requires more computational 
effort because the expressions in the intermediate steps and in the resulting series tend to be 
more complex.
For completeness, we provide it here with the simplified variables $X_u\to u, X_d\to d, YY^\dag\to x$:
\eq{
H(u,d,x)=\frac{N}{D}\,,
}
where 
\eqali{
D&=(1-d) \left(1-d^2\right) \left(1-d^3\right) (1-u) \left(1-u^2\right) \left(1-u^3\right) (1-d u)^2 \left(1-d^2 u\right) \left(1-d u^2\right) 
\cr&\quad
\times
(1-x)(1-d x) \left(1-d^2 x\right) (1-u x) \left(1-u^2 x\right) (1-d u x)\,,
\cr
N&=1-d u+d^2 u^2+d u x \left(1+2u+2d+u^2+d^2+d u-d^2 u-d u^2-d^2 u^2-d^3 u-d u^3\right)
\cr&\quad
+d^2 u^2 x^2 \left(d^2+d u+u^2+d^2 u+d u^2-d^3 u-d^2 u^2-d u^3-2 d^2 u^3-2 d^3 u^2-d^3u^3\right)
\cr&\quad
+d^4 u^4 x^3 \left(-1+d u-d^2 u^2\right)\,.
}
We gather $Y,Y^\dag$ as a single variable $x$ because they always appear in pairs.
The PL is
\eqali{
\PL[H(u,d,x)]
&=
u+d +u^2 +d^2 +ud +u^3+d^3 +u d^2 +u^2 d + u^2d^2 + u^3d^3 -u^6d^6
\cr&\quad
+x(1 + u+ d + u^2+ d^2  + 2 ud + 2 ud^2 + 2 u^2d+ u^3d+ ud^3 + 
 2 u^2d^2 + u^2d^3 + u^3d^2 - u^3d^4 - u^4d^3)
\cr&\quad
-x^2(u^2d^2 + 2 u^2d^3 + 2 u^3d^2 + 3 u^2d^4 + 3 u^4d^2 + 5 u^3d^3 )
+\cdots
}
We clearly identify the basic invariants in Table~\ref{tab:flavor_invariants} with positive 
coefficients while the syzygies in Sec.\,\ref{sec:lowest} and in Sec.\,\ref{sec:syz:14-16} appear 
with negative coefficients.
The first line corrresponds to the SM.
Here, differently from Ref.\,\cite{yu.zhou:inv.nu}, there is no cancellation in the graded PL 
because the basic invariants and the lowest degree syzygies have different degrees in $u,d,x$.
That is the reason why the trick of reshuffling the degree in the unrefined Hilbert 
series works in our case but would not work for \cite{yu.zhou:inv.nu}.

\section{Reduction of invariants in \eqref{inv.u3d3y2}}
\label{app:reduction}

Here we show how to reduce $K_{66}^{(4)}$ in \eqref{inv.u3d3y2}.
We first expand
\eq{
[u,d]^3=(ud)^3-(du)^3-[ud^2u^2d-du^2d^2u]
+\{ud,(du)^2\}-\{du,(ud)^2\}\,,
}
using the shorthand $u\equiv X_u$, $d\equiv X_d$.
If we insert this expansion in $\aver{[u,d]^3 x}$, with $x\equiv YY^\dag$, the first two terms are reducible by the CHT.
The third term leads to $K_{66}^{(4)}$.
Let us see that the last two terms in anticommutators also lead to $K_{66}^{(4)}$, modulo reducible terms, after applying the identity \eqref{manohar.id}.
To see that, note that the last term is the opposite of the hermitean conjugate of the second to last.
So we only need to consider the latter.
Its two terms can be reduced with \eqref{manohar.id} as
\eqali{
\aver{ud^2udux}
&=
\aver{\underbrace{u}_{A}\underbrace{d}_{B}\underbrace{u}_{A}\underbrace{xud^2}_{C}}
=-\aver{u^2dxud^2}-\aver{du^2xud^2}
+(\text{reducible})
\,,
\cr
\aver{dudu^2dx}
&=
\aver{\underbrace{d}_{A}\underbrace{u}_{B}\underbrace{d}_{A}\underbrace{u^2dx}_{C}}
=-\aver{d^2u^3dx}-\aver{ud^2u^2dx}
+(\text{reducible})
\,.
}
Both terms lead to the first piece of $K_{66}^{(4)}$.
We then get
\eq{
\aver{[u,d]^3x}=-3K_{66}^{(4)}+(\text{reducible})\,.
}
Considering the CHT in which $A^3=\ums{3}\aver{A^3}\id+\dots$, and $\aver{[u,d]^3}=3I_{66}^-$, we obtain
\eq{
K_{00}I_{66}^-=-3K_{66}^{(4)}+(\text{reducible})\,.
}
Using the reducibility of $\aver{A^2BAC}+\aver{ABA^2C}$, the equivalent expansion for $K_{66}^{(3)}$ has the same explicit coefficients while for $K_{66}^{(1)}$, $K_{66}^{(2)}$, we should replace $(-3)$ by $+3$.
Employing the method of Ref.\,\cite{yu.zhou:inv.nu}, we can find the explicit reduction of e.g. $K_{66}^{(3)}$ as
\eqali{
K_{66}^{(3)}
&=
+\frac{1}{4}K^-_{22}(I_{40}I_{02}^2 - I_{20}^2I_{02}^2 + I_{04} I_{20}^2+\ums{3}I_{40}I_{04}-\ums[4]{3}I_{44})
\cr&\quad
   +\frac{1}{3}K^-_{42}(I_{24} -I_{04} I_{20})
   -\frac{1}{3}K^-_{24}(I_{42}-I_{40}I_{02})
   +\frac{1}{3}K^-_{44}(I_{20}I_{02}+2I_{22})
\cr&\quad   
   +\frac{1}{2}K^-_{62}(I_{02}^2+\ums{3}I_{04})
   -\frac{1}{2}K^-_{26}(I_{20}^2 +\ums{3}I_{40})
  -\frac{2}{3}K^-_{64}I_{02}+\frac{2}{3}K^-_{46}I_{20} -\frac{1}{3}I^-_{66}K_{00}\,.
}
Note the antisymmetry in the exchange $u\leftrightarrow d$.

\section{Syzygy in eq.\,\eqref{syzygy:ff}}
\label{app:syzygy}

The full version for the syzygy \eqref{syzygy:ff} is
\eqali{
\label{syzygy:ff:full}
\big(K^{-}_{22}\big)^2
&=
  \big(K^+_{22}\big)^2 +4K^+_{22}(I_{20}I_{02}K_{00} -I_{20} K_{02} -I_{02} K_{20})
\cr&\quad
  +2K_{00}^2(I_{02}^2 I_{20}^2 - I_{02}^2 I_{40}-I_{20}^2I_{04}-2I_{02} I_{20}I_{22}
    +2 I_{02}I_{42}+2 I_{20} I_{24}+I_{04} I_{40}-2I_{44})
\cr&\quad
  +2 I_{20}^2(K_{02}^2-2I_{02}K_{00} K_{02}+K_{00}K_{04}) +2I_{40}(2I_{02}K_{00} K_{02}-K_{00}K_{04}-K_{02}^2)
\cr&\quad
  +2 I_{02}^2(K_{20}^2-2I_{20}K_{00}K_{20}+K_{00} K_{40}) +2I_{04}(2I_{20} K_{00}K_{20} -K_{00}K_{40} -K_{20}^2)
\cr&\quad
  +4I_{22}(K_{00}I_{20}K_{02} +K_{00}I_{02}K_{20} -K_{02}K_{20})
\cr&\quad
   +4 I_{20}I_{02}K_{20}K_{02}
   -4 K_{00}(I_{42} K_{02}+I_{24} K_{20})
   -4 K_{04} K_{40}
\cr&\quad
   +4K^+_{24}(K_{20} - I_{20}K_{00})
   +4K^+_{42}(K_{02} - I_{02}K_{00}) +4 K_{00} K^+_{44}\,.
}
This expression reduces to \eqref{syzygy:ff} after taking $I_{20}=I_{02}=0$ and replacing the
invariants by their tilde versions.
As pointed out in the text, both sides of the equality are the same wether we use the matrices 
$X_u,X_d$ or their traceless parts. 
This is obvious for the lefthand side but not evident for the righthand side.
The full syzygy \eqref{syzygy:ff:full}, however, is not valid if we use the traceless part 
of $X=YY^\dag$.
The lefthand side is unchanged but the righthand side is not. 
The reason is that $\aver{\tX}=0$ for traceless $\tX$  but $\aver{\tX^2}$ cannot vanish.
The rank one condition \eqref{YY2=YY}, however, implies $\aver{\tX^2}=\frac{2}{3}K_{00}^2$
and removing $K_{00}$ also removes $\aver{\tX^2}$.
The correct expression with $\aver{\tX^2}$ can be recovered from the $SU(3)$ 
identity \eqref{Tff.identity} but it is actually longer than \eqref{syzygy:ff}.

\section{The space $[(\bs{8}\otimes\bs{8})_s]^3$}
\label{app:21.tensors}

As shown in Ref.\,\cite{dittner}, the total number of $k$-legged independent tensors in $\bs{8}^{\otimes k}$ coincides with the number of invariants (singlets) in the branching of $\bs{8}^{\otimes k}$.
For $k=6$, this number is 145\,\cite{dittner} and listing a basis would be impractical.

For $[(\bs{8}\otimes\bs{8})_s]^3$, however, we can take advantage of the lower dimensional space coming from the symmetrization:
\eq{
\label{branch:8x8S}
(\bs{8}\otimes\bs{8})_s=\bs{1}+\bs{8}_s+\bs{27}\,.
}
The number of invariants in $[(\bs{8}\otimes\bs{8})_s]^3$ is just
\eq{
[(\bs{8}\otimes\bs{8})_s]^3=21(\bs{1})+\cdots
}
Furthermore, these invariants comes solely from the product of
\eq{
(\bs{1}+\bs{8}_s+\bs{27})\otimes(\bs{1}+\bs{8}_s+\bs{27})
=3(\bs{1})+8(\bs{8}_s)+10(\bs{27})+\cdots
}
with each of $\bs{1},\bs{8}_s,\bs{27}$ in the third $(\bs{8}\otimes\bs{8})_s$.
So it is not difficult to build most of the tensors in the basis using the projectors from $(\bs{8}\otimes\bs{8})_s\to (\bs{8}\otimes\bs{8})_s$:
\eqali{
\label{projectors:8^2}
P^1&=\frac{1}{d}~\raisebox{-.75em}{\includegraphics[scale=.13]{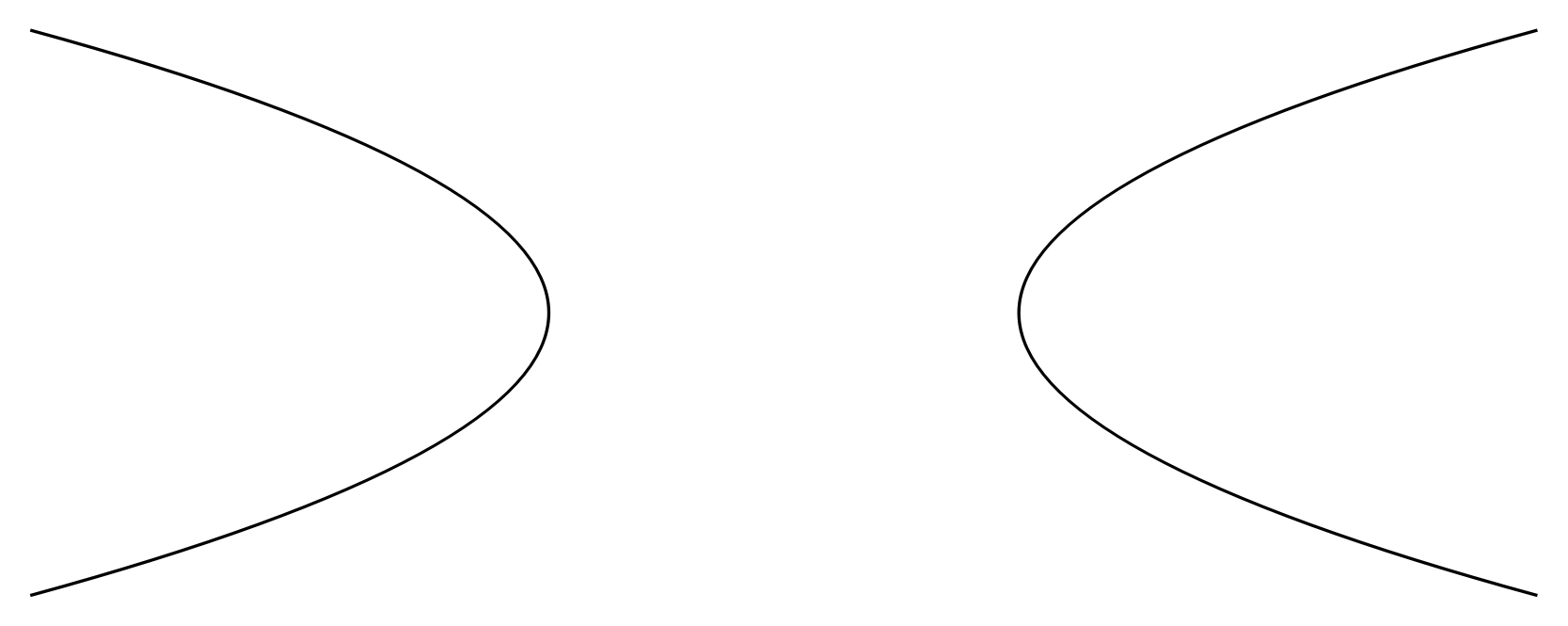}}
\quad,
\\
P^{8s}&=\frac{1}{C_D}\raisebox{-.75em}{\includegraphics[scale=.13]{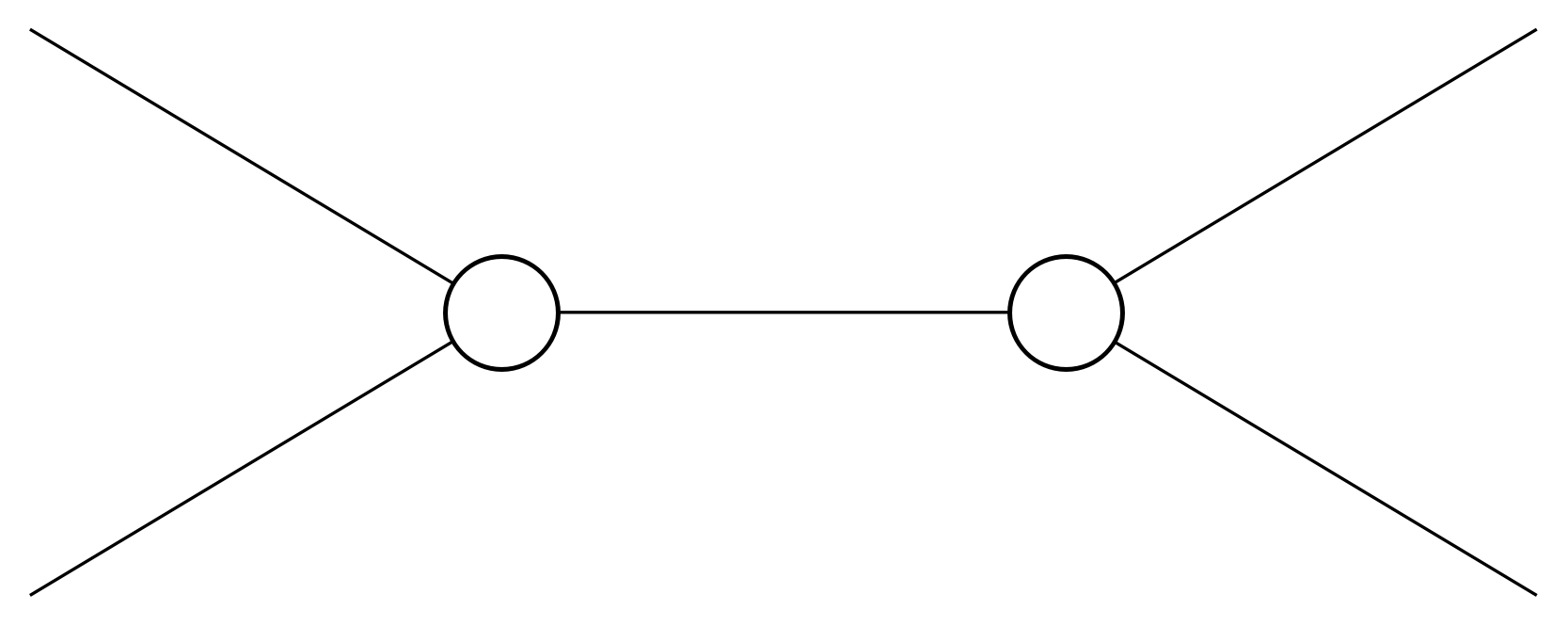}}
\quad,
\\
P^{27}&=\raisebox{-.75em}{\includegraphics[scale=.13]{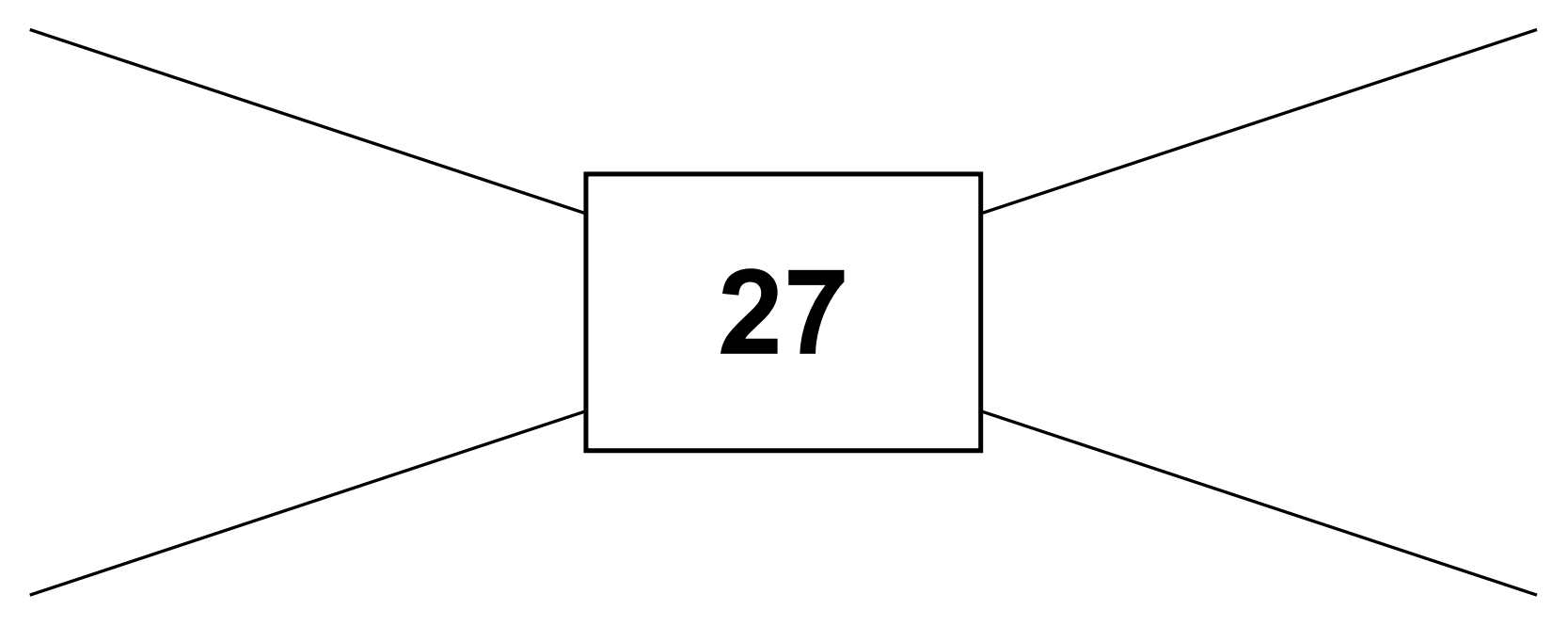}}
=\raisebox{-.9em}{\includegraphics[scale=.13]{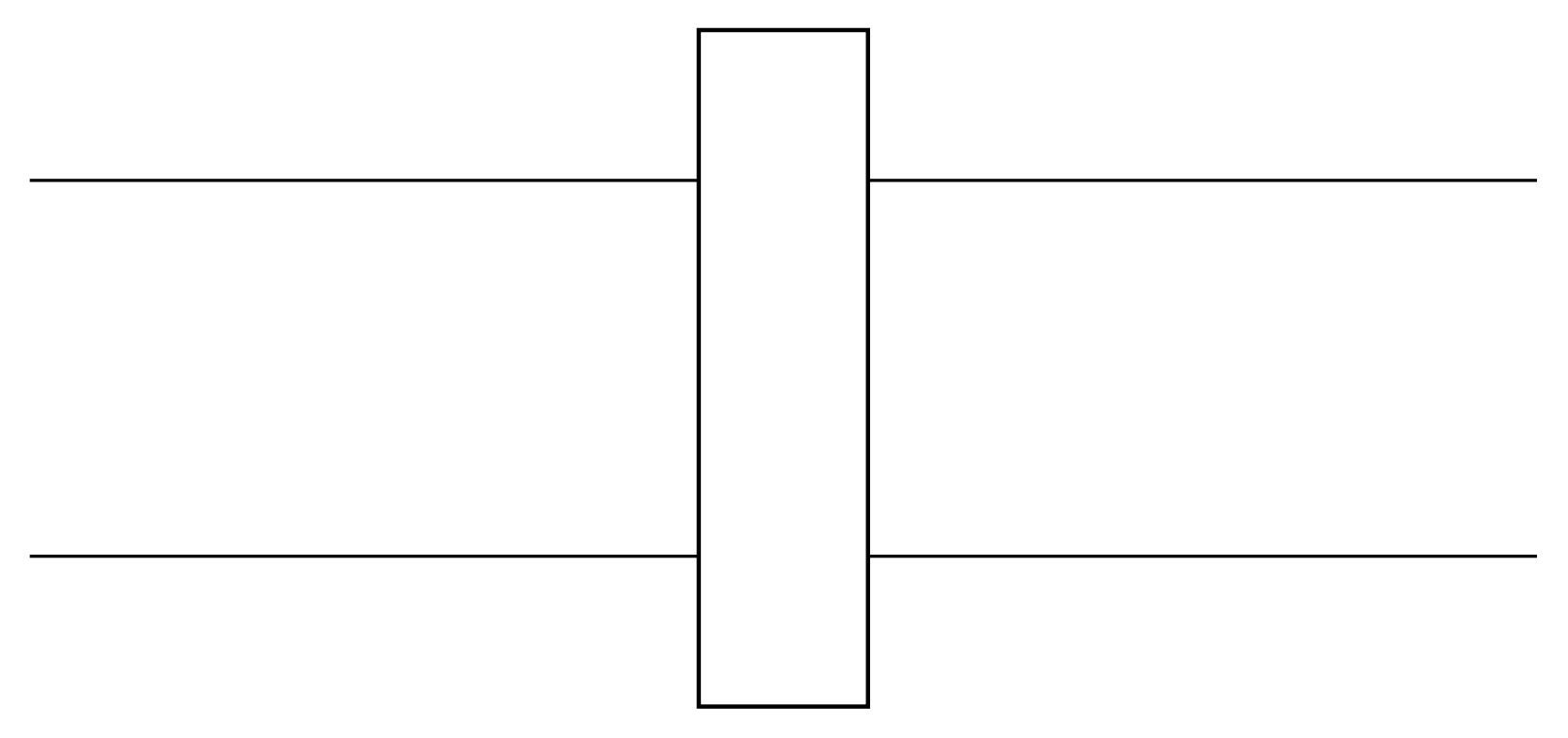}}
-P^1-P^{8s}
\,.
}
Here $d=n^2-1=8$ is the dimension of $SU(n)$ for $n=3$ and $C_D=2(n^2-4)/n$.
They are orthogonal and project out onto the irreps in \eqref{branch:8x8S}.
These projectors can be found in Refs.\,\cite{cvitanovic,keppeler}.
While $P^1$ and $P^{8s}$ can be generalized to $SU(n)$ with $n>3$, the projector $P^{27}$ here is specific to $SU(3)$.

We present in Table\,\ref{tab:21.tensors} a possible choice of basis tensors for $[(\bs{8}\otimes\bs{8})_s]^3$ of $SU(3)$.
They are all nonvanishing.
It is easy to see that due to the orthogonality of the projectors \eqref{projectors:8^2}, the 6-legged tensors in the different rows of the table are orthogonal. Most of them are also orthogonal within each row except for the last row.
We check they are independent using the method outlined in Sec.\,\ref{sec:su3.ff}.
\begin{table}[h]
\[
\begin{array}{|c|c|c|c|}
\hline
\text{Tensor(s)} & k=a & k=b & k=c
\\
\hline
\rule{0em}{2.5em}
T_{1k} & \raisebox{-1.9em}{\includegraphics[scale=.12]{figures/T1.png}}
&&
\cr
\hline
\rule{0em}{2.5em}
T_{2k} & \raisebox{-1.9em}{\includegraphics[scale=.12]{figures/T2a.png}} &
  \raisebox{-1.9em}{\includegraphics[scale=.12]{figures/T2b.png}} &
  \raisebox{-2em}{\includegraphics[scale=.12]{figures/T2c.png}} 
\cr
\hline
\rule{0em}{2.5em}
T_{3k} &
\raisebox{-2.2em}{\includegraphics[scale=.12]{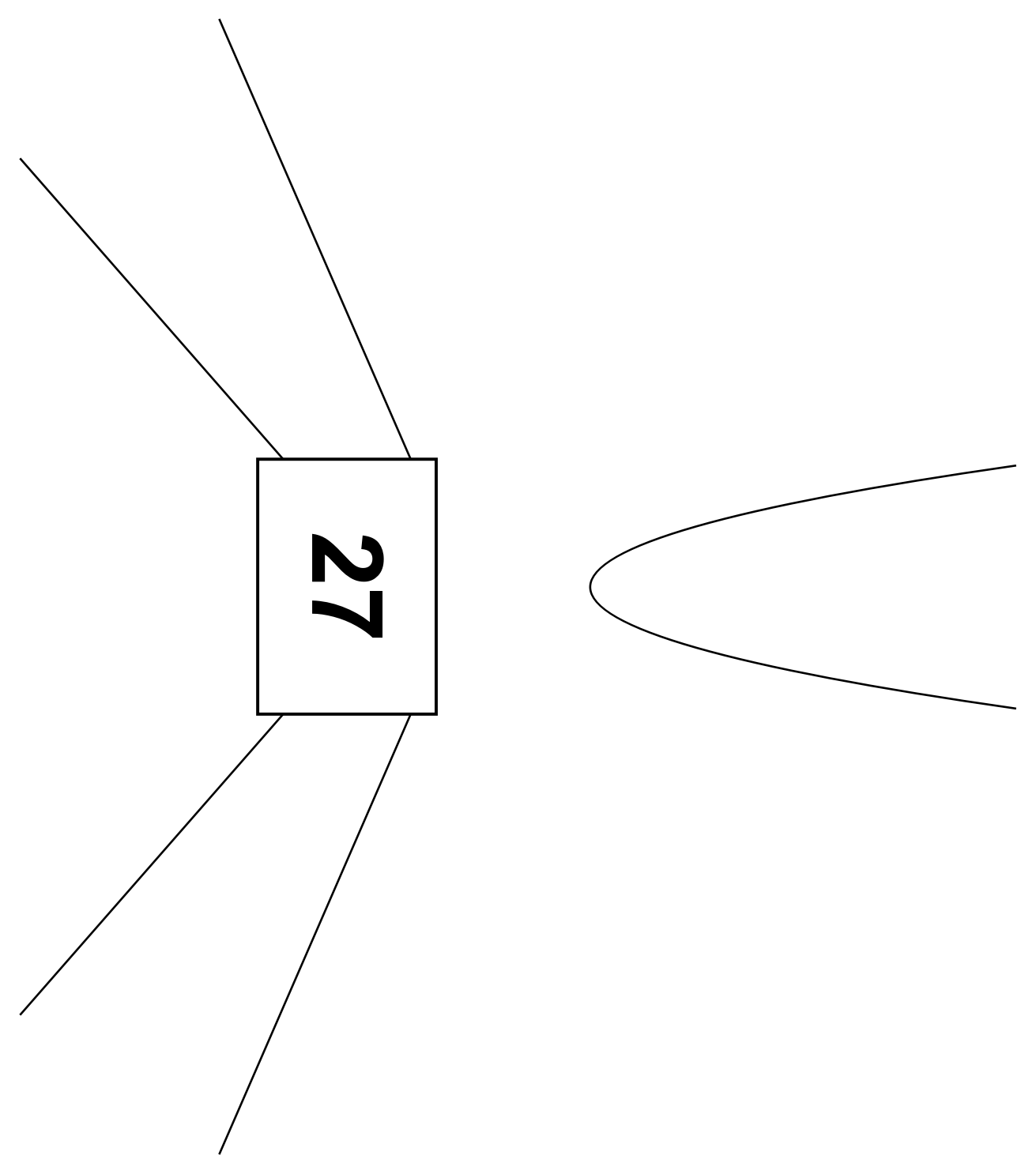}} &
  \raisebox{-1.9em}{\includegraphics[scale=.12]{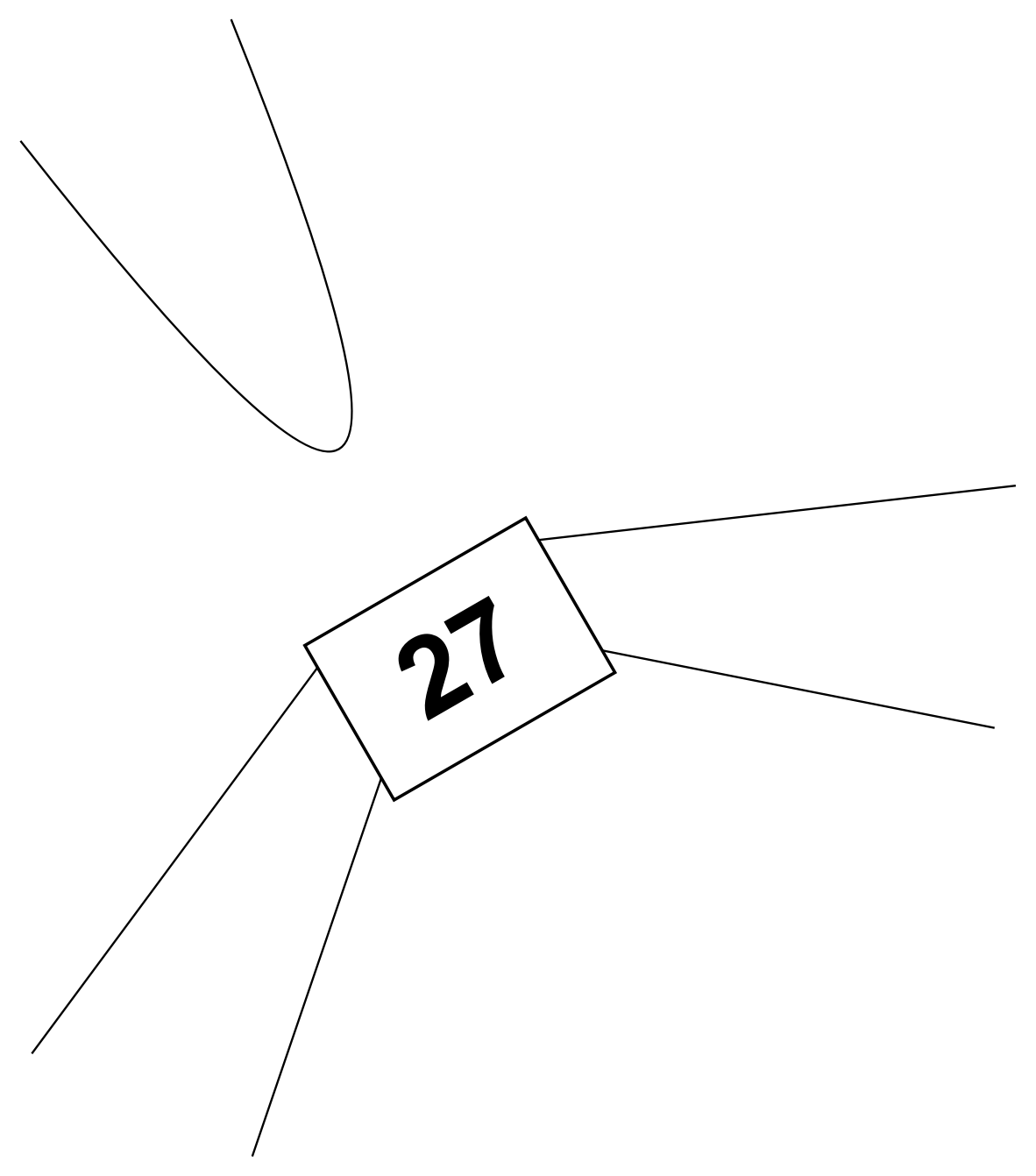}} &
  \raisebox{-1.9em}{\includegraphics[scale=.12]{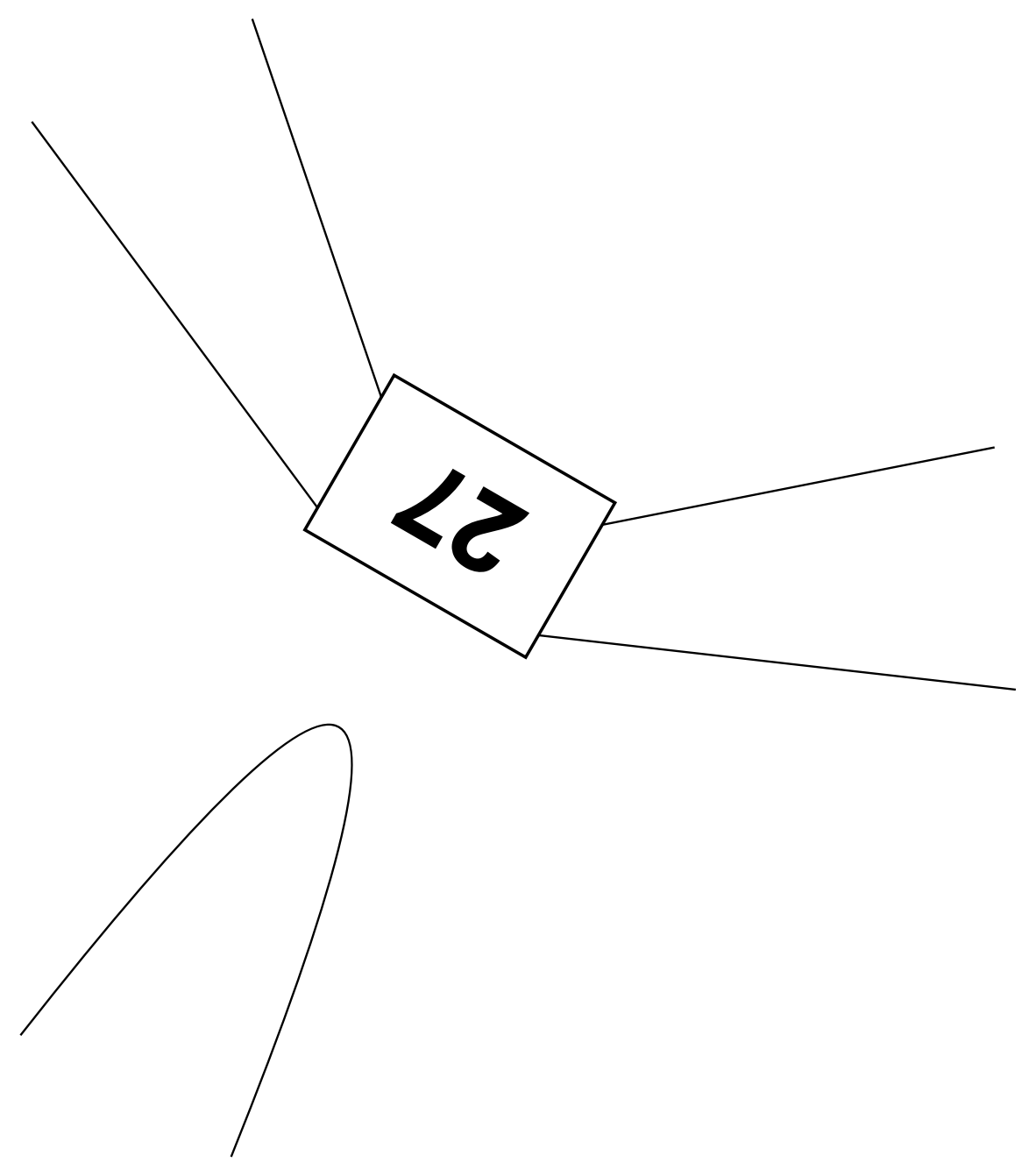}} 
\cr
\hline
\rule{0em}{2.5em}
T_{4k} &
\raisebox{-1.9em}{\includegraphics[scale=.12]{figures/T4a.png}} &
\raisebox{-1.9em}{\includegraphics[scale=.12]{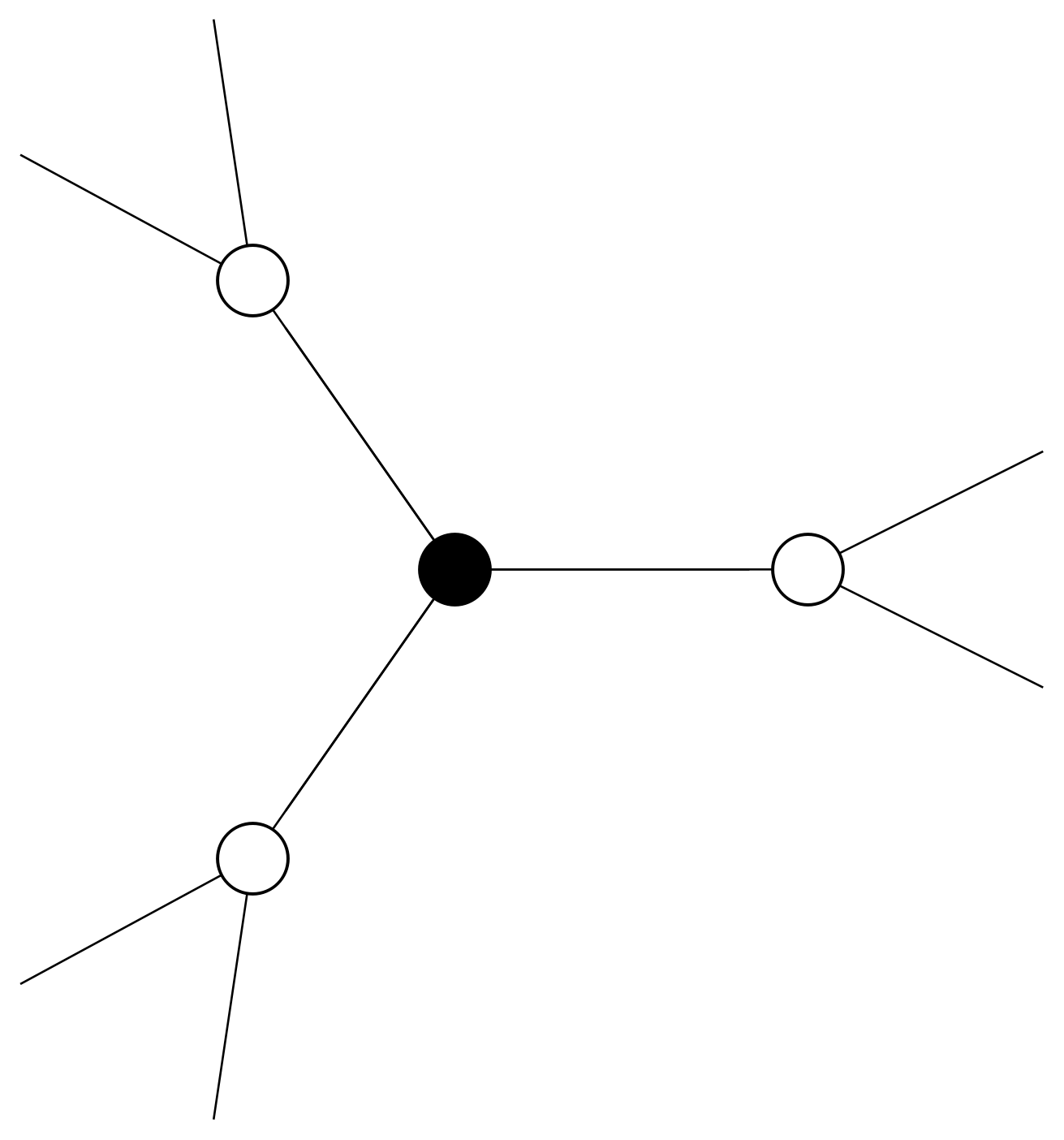}} &
\cr
\hline
\rule{0em}{2.5em}
T_{5k} &
  \raisebox{-1.9em}{\includegraphics[scale=.12]{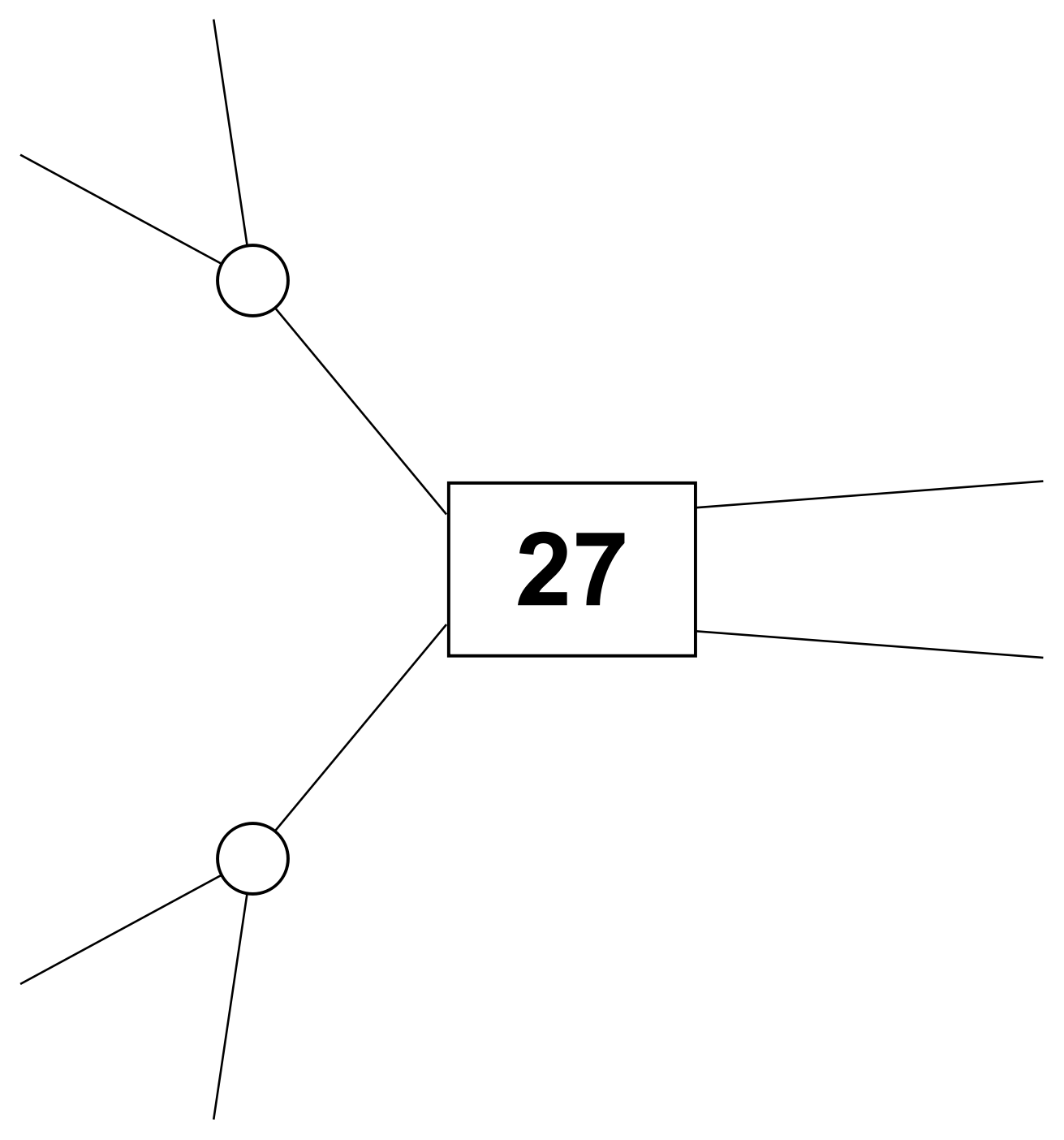}} &
  \raisebox{-1.9em}{\includegraphics[scale=.12]{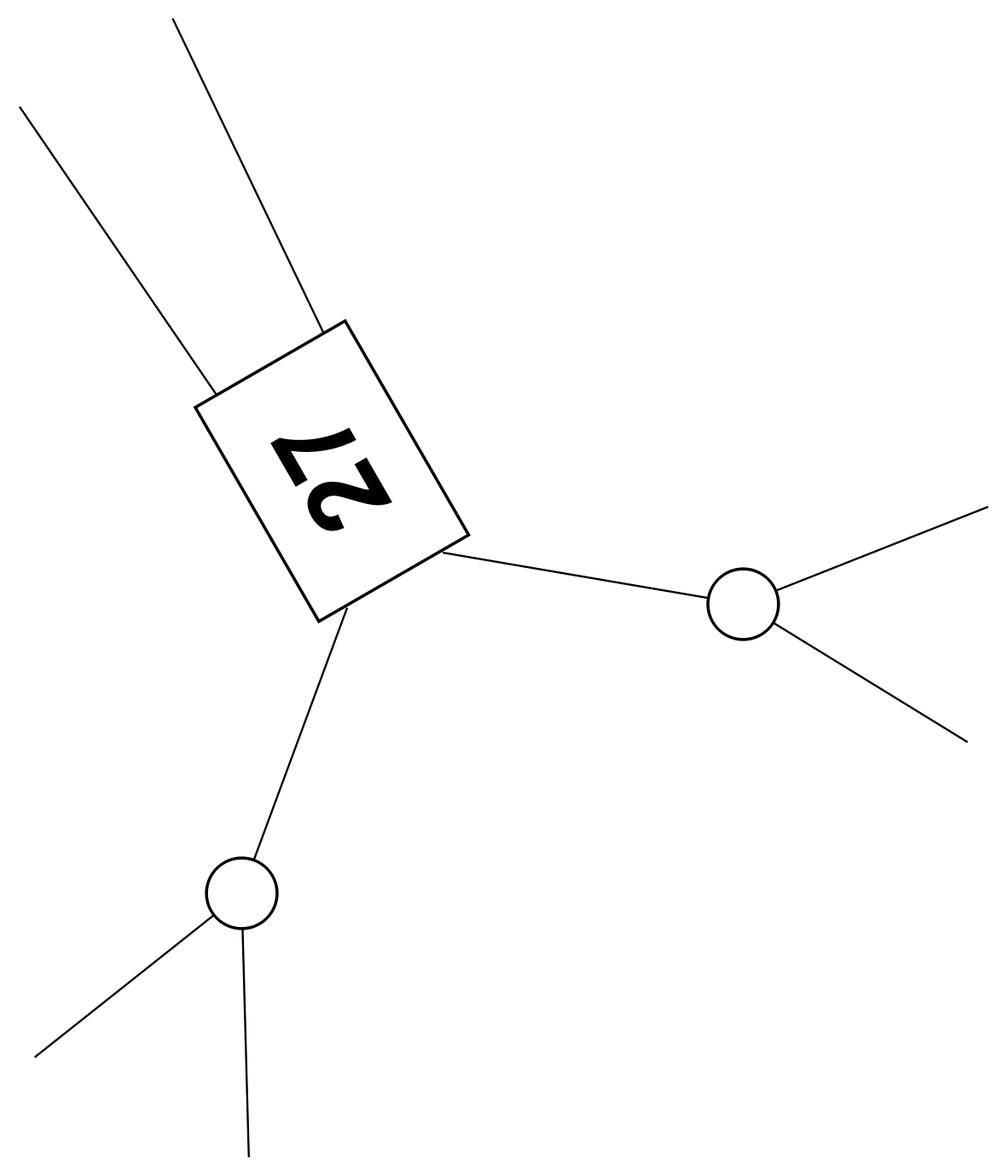}} &
  \raisebox{-1.9em}{\includegraphics[scale=.12]{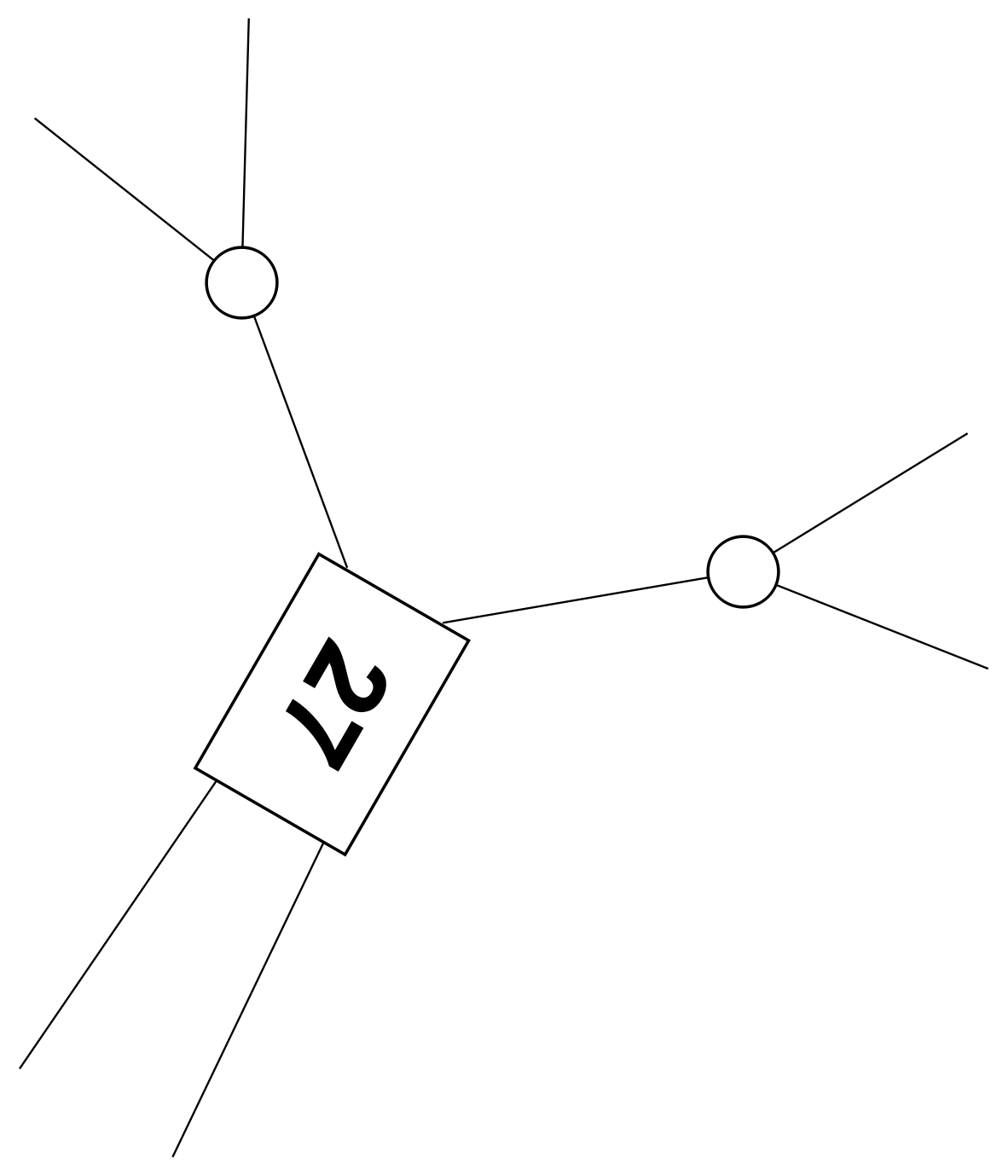}} 
\cr
\hline
\rule{0em}{2.5em}
T_{6k} &
  \raisebox{-1.9em}{\includegraphics[scale=.12]{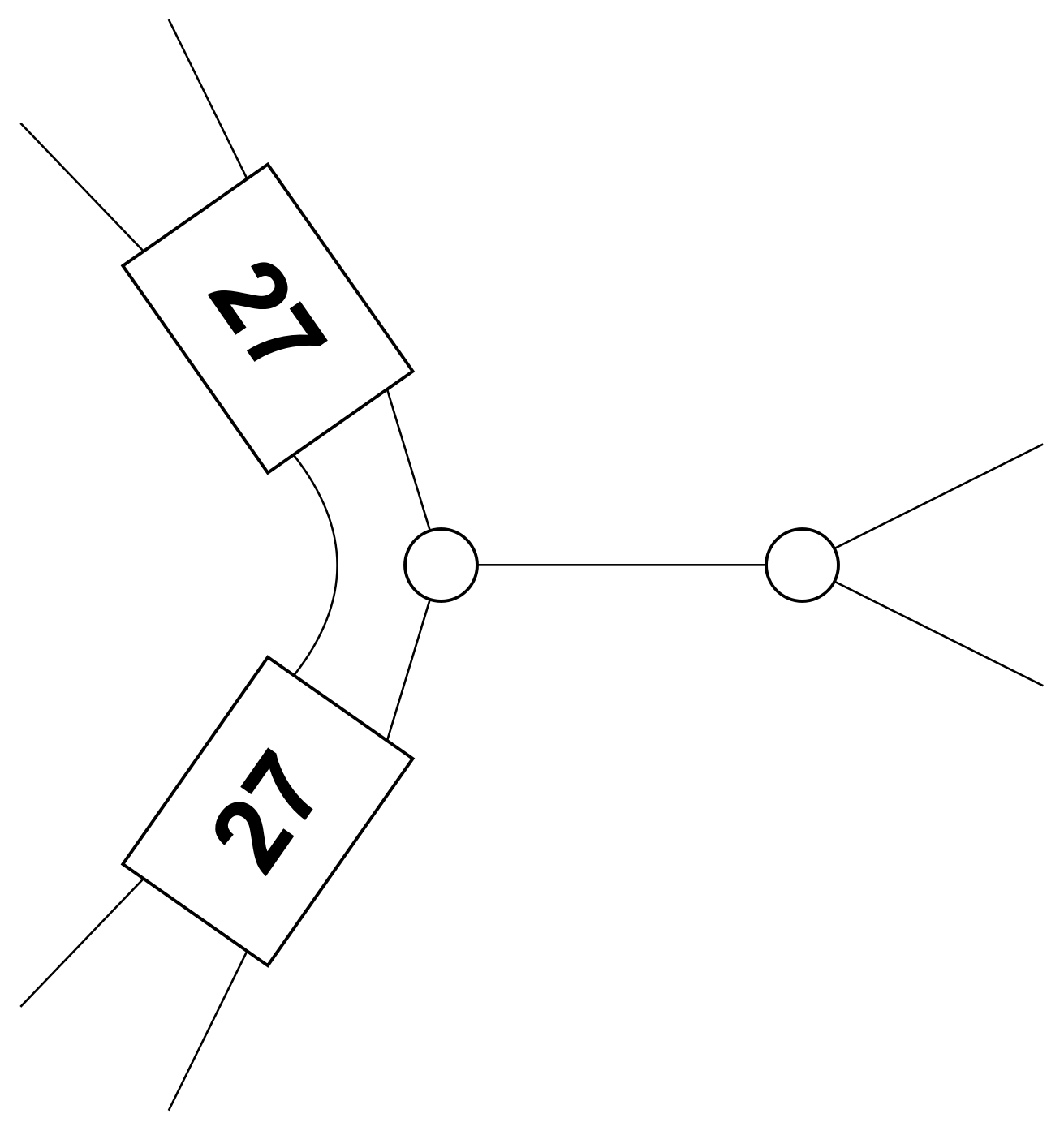}} &
  \raisebox{-1.9em}{\includegraphics[scale=.12]{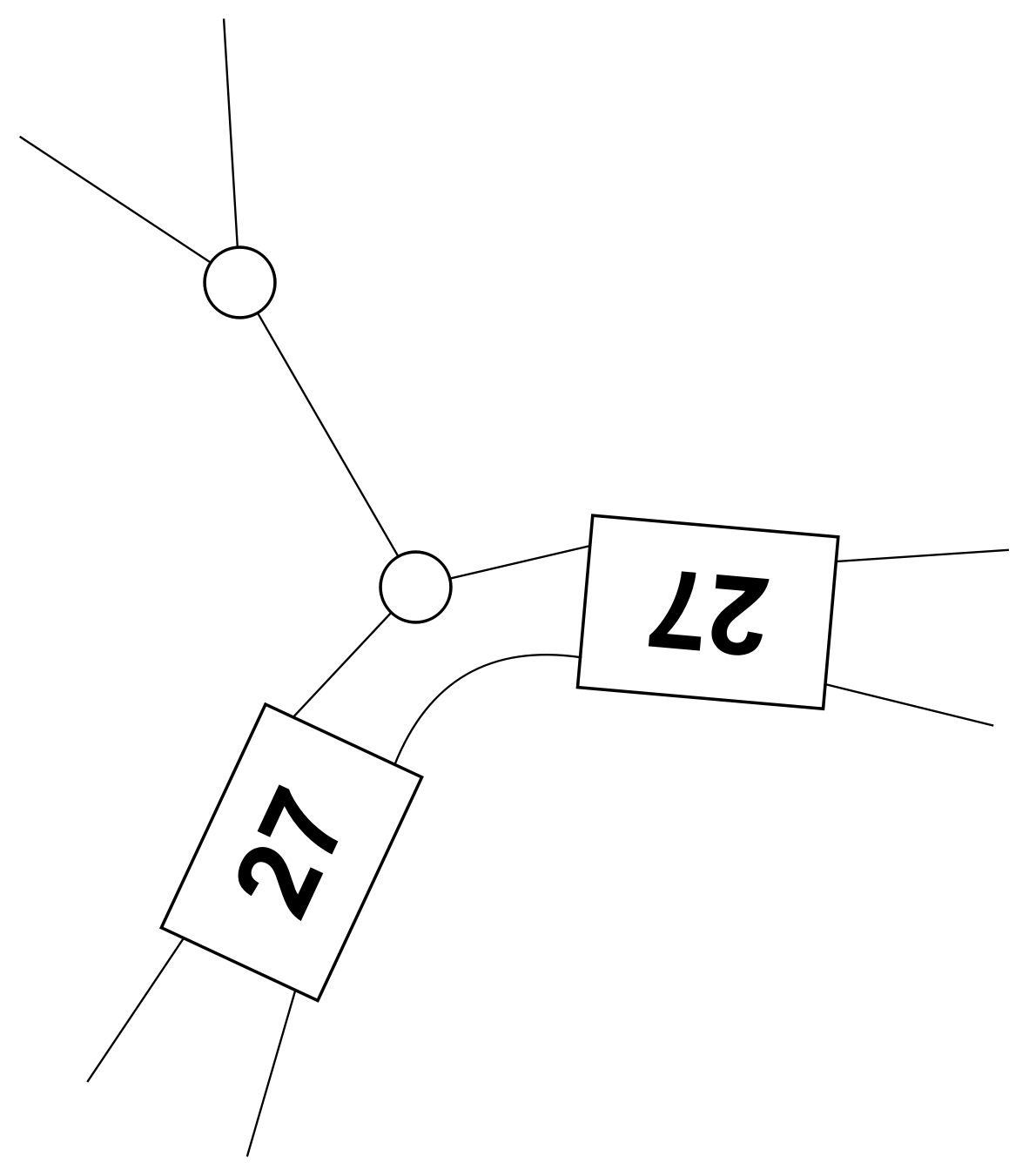}} &
  \raisebox{-1.9em}{\includegraphics[scale=.12]{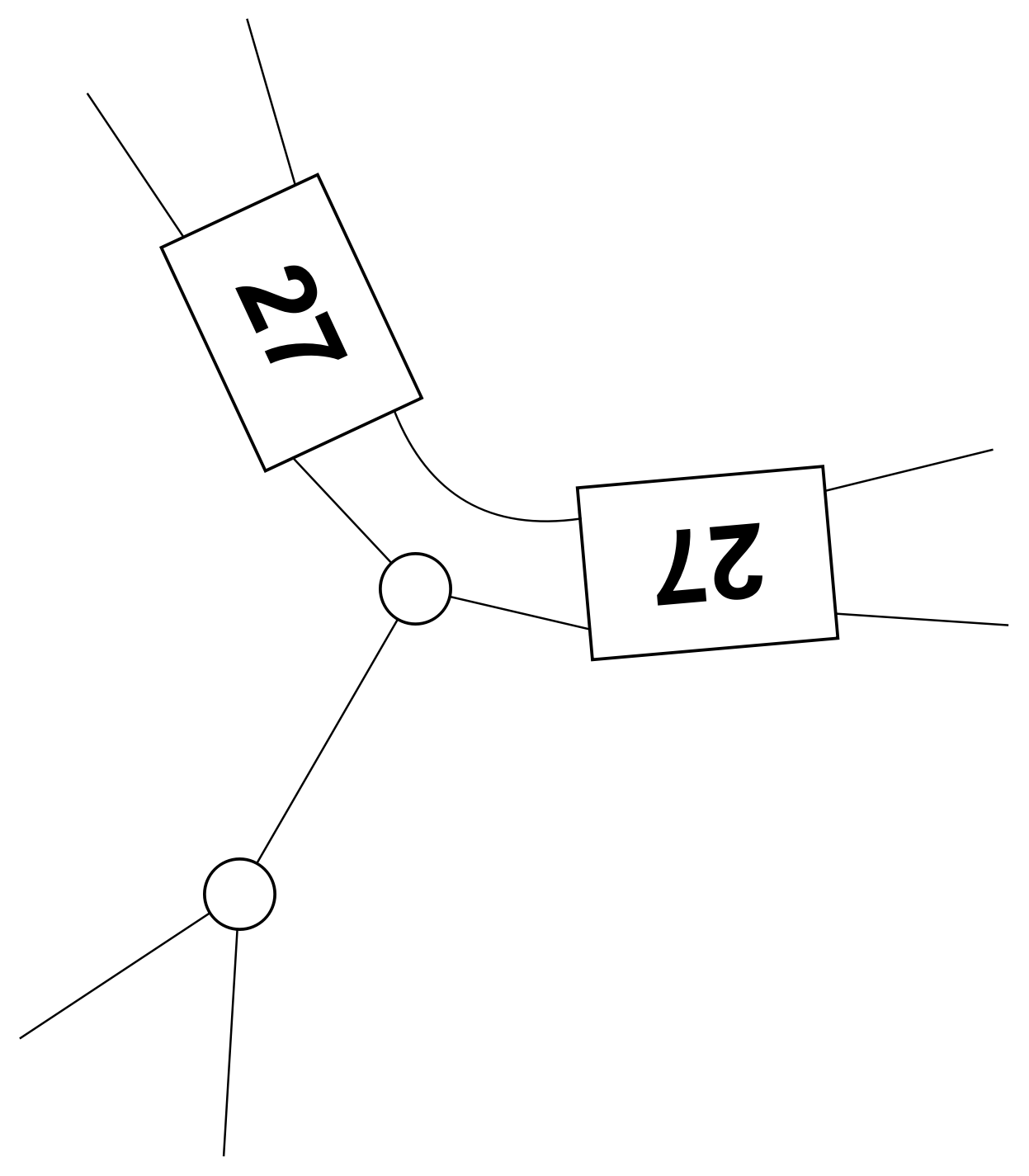}} 
\cr
\hline
\rule{0em}{2.5em}
T_{6'k} &
  \raisebox{-1.9em}{\includegraphics[scale=.12]{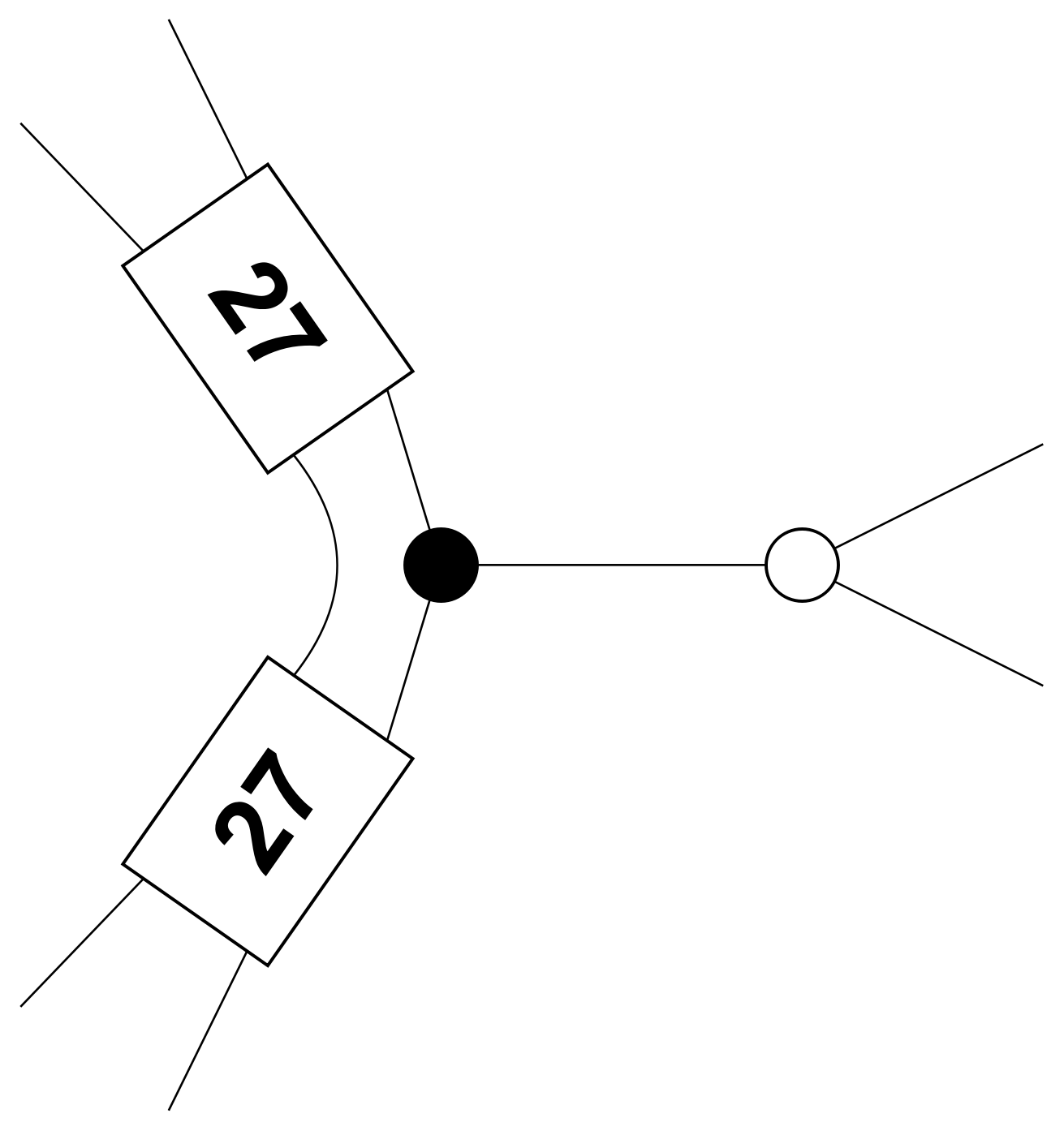}} &
  \raisebox{-1.9em}{\includegraphics[scale=.12]{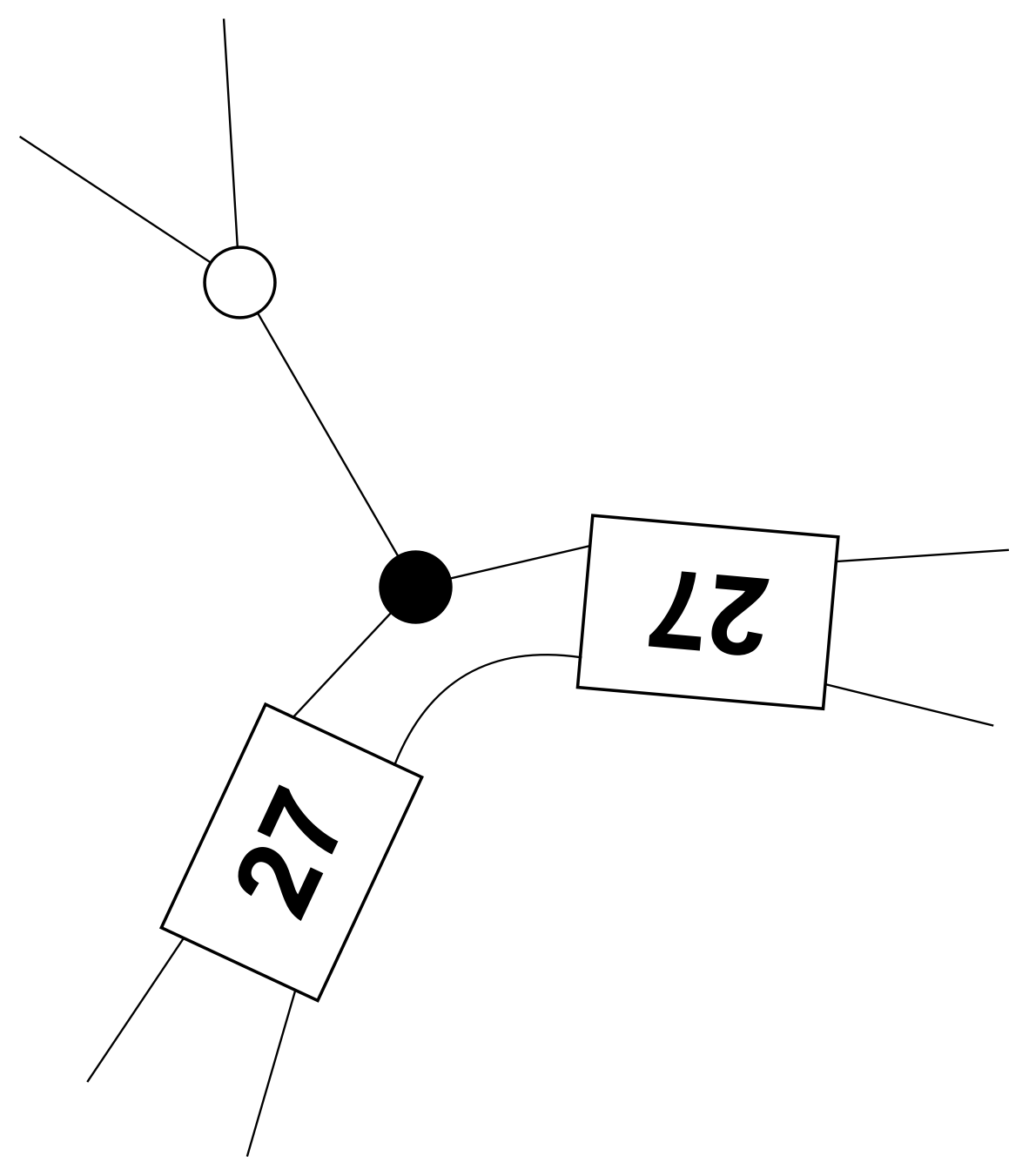}} &
  \raisebox{-1.9em}{\includegraphics[scale=.12]{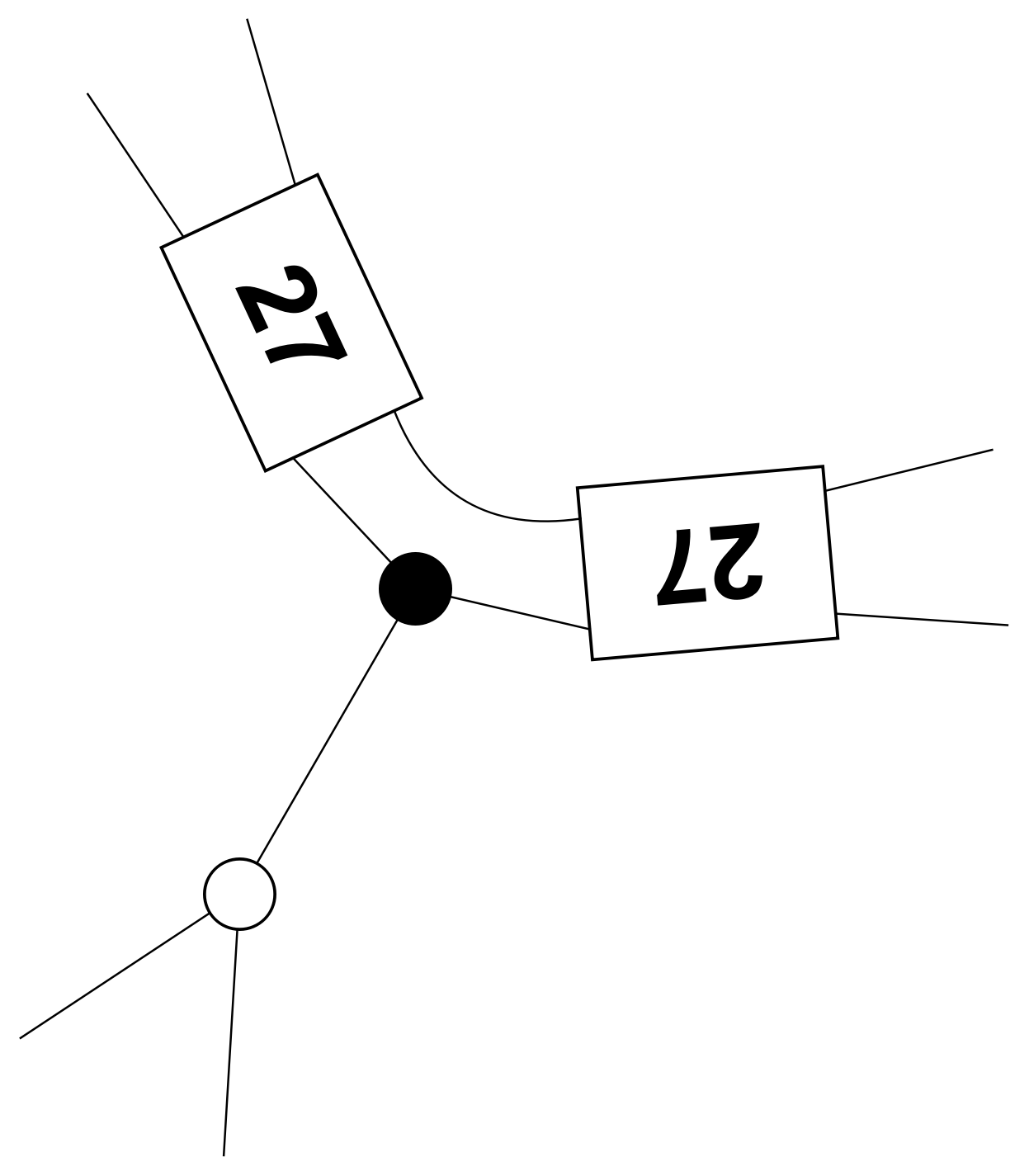}} 
\cr
\hline
\rule{0em}{2.5em}
T_{7k} &
  \raisebox{-1.7em}{\includegraphics[scale=.12]{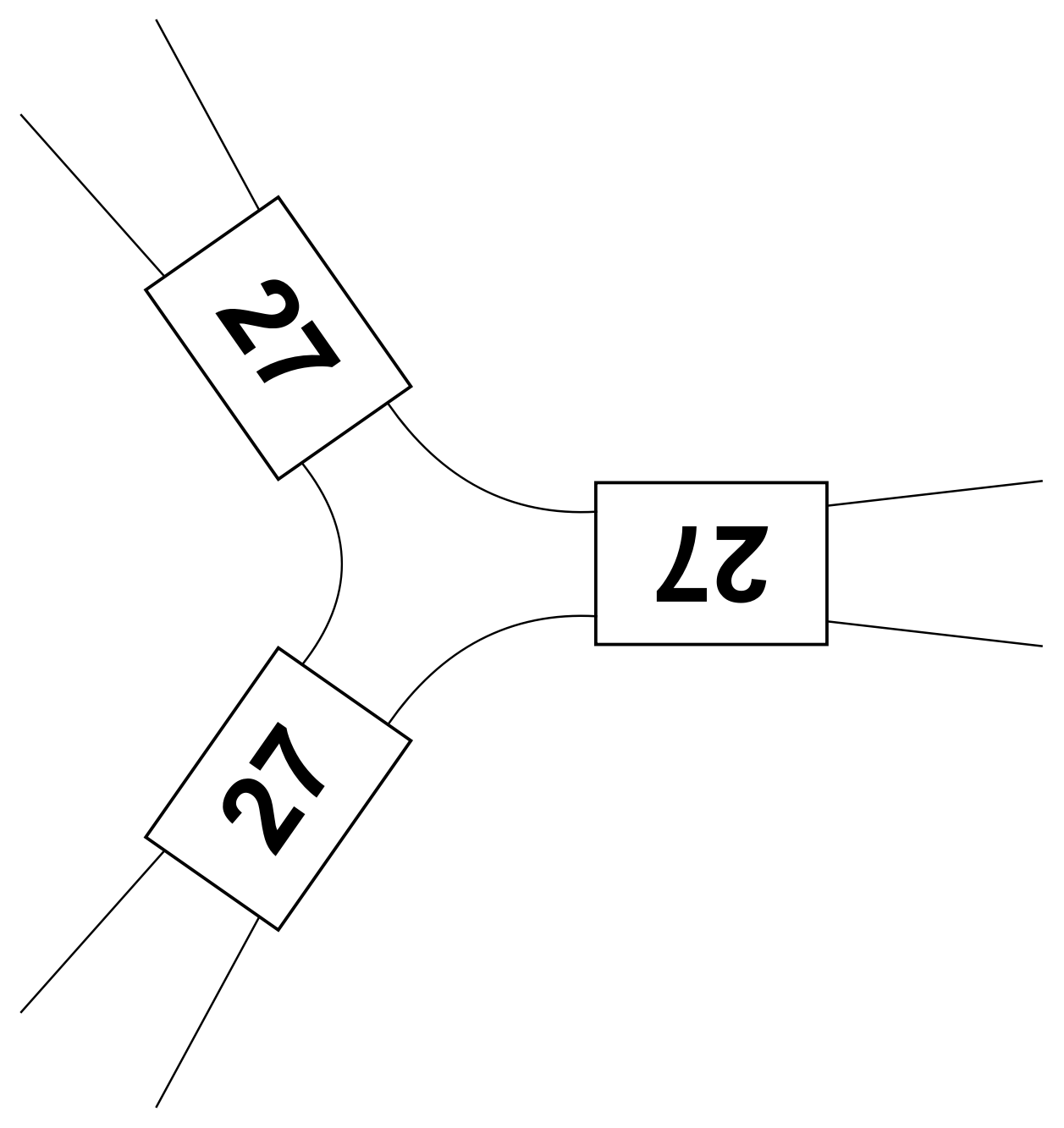}} &
  \raisebox{-1.7em}{\includegraphics[scale=.12]{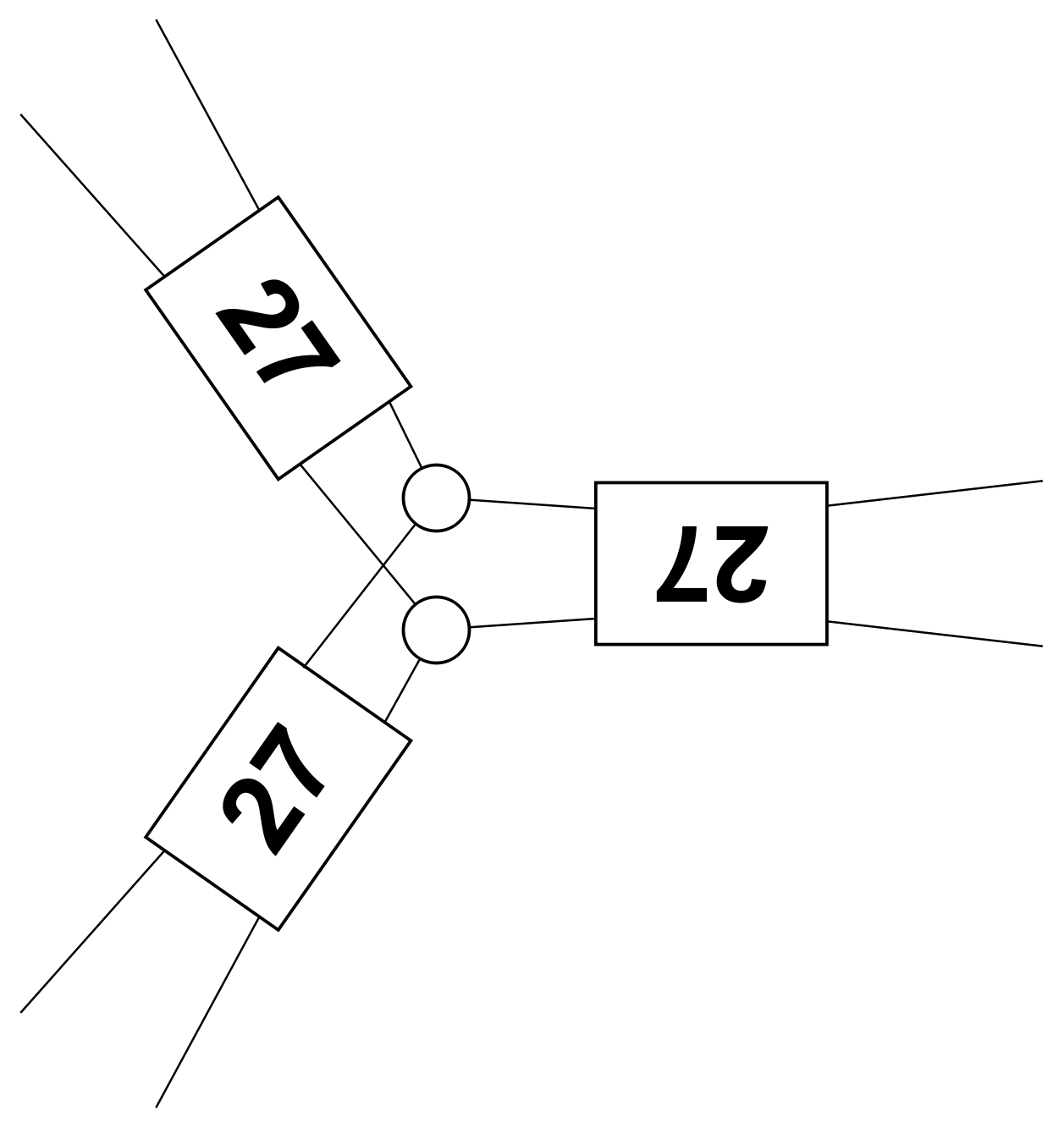}} &
  \raisebox{-1.7em}{\includegraphics[scale=.12]{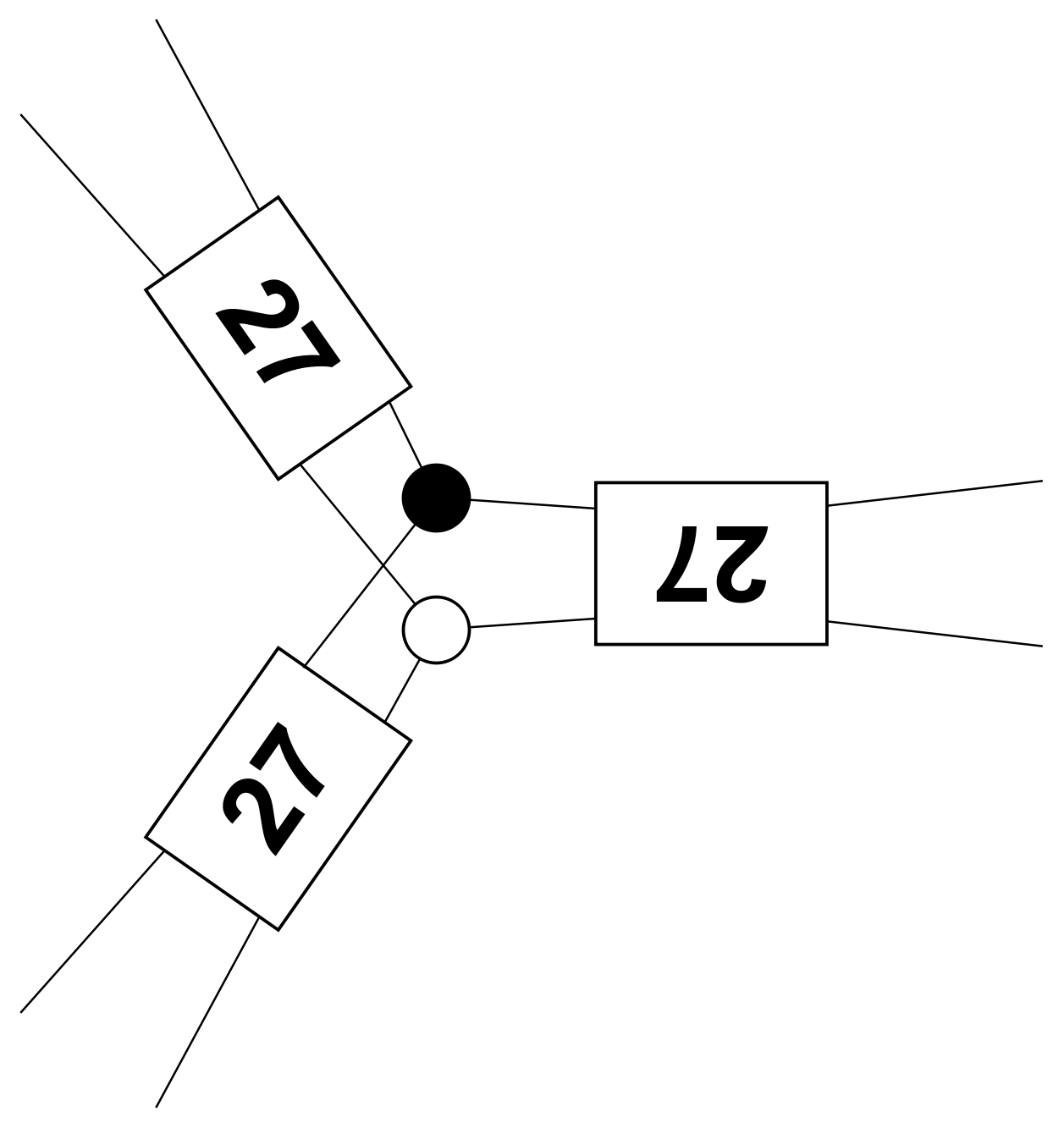}}
\cr
\hline
\end{array}
\]
\caption{\label{tab:21.tensors}
Tensor basis of 21 tensors for $[(\bs{8}\otimes\bs{8})_s]^3$.
}
\end{table}

\section{The $SU(3)$ identity \eqref{Tff.identity}}
\label{ap:su3.ff}

To get the $SU(3)$ identity \eqref{Tff.identity}, we could use the basis of 21 tensors in Table\,\ref{tab:21.tensors} and apply the method outlined in Sec.\,\ref{sec:su3.ff}.
To check, we have also applied the method to this basis of 21 tensors.
But we can easily recognize that the tensor $T_{f\!f}$ in \eqref{Tff} has a permutation symmetry where we can interchange any pair $(ii')\leftrightarrow (jj')$ or $(ii')\leftrightarrow (kk')$, or any combination thereof.
In contrast, some of the tensors in Table\,\ref{tab:21.tensors} are antisymmetric by such a kind of exchange and would be orthogonal to $T_{f\!f}$.
This reasoning eliminates the row $T_{6'k}$, $T_{4b}$ and $T_{7c}$.
Moreover, symmetry dictates that contraction of $T_{f\!f}$ with, e.g., $T_{2a}$ or $T_{2b}$ leads to the same result and only symmetric combinations would matter.
At last, given that the operator $P^{27}$ in \eqref{projectors:8^2} projects onto the subspace of $(\bs{8}\otimes\bs{8})_s$ orthogonal to the subspaces of irreps $\bs{8}_s$ and $\bs{1}$, we can replace $P^{27}$ by the symmetrizer \eqref{P:symm}, losing orthogonality but retaining independence.
With all these choices, we arrive at the 8 symmetric tensor combinations of Table\,\ref{tab:8.tensors}.

Now we can calculate the $Q$-matrix in \eqref{def:Q} for the tensors in Table\,\ref{tab:8.tensors}.
Using the same ordering of the table, we obtain
\eq{
Q=(T_A\cdot T_B)=
\left(
\begin{array}{cccccccc}
 512 & 0 & 192 & 0 & 0 & 0 & 8 & \frac{80}{3} \\
 0 & \frac{6400}{3} & 640 & 0 & \frac{800}{3} & \frac{1600}{3} & 80 & -80 \\
 192 & 640 & 912 & 0 & 80 & 160 & 108 & 40 \\
 0 & 0 & 0 & \frac{80000}{81} & \frac{8000}{9} & -\frac{800}{3} & -\frac{80}{3} & \frac{1280}{27} \\
 0 & \frac{800}{3} & 80 & \frac{8000}{9} & \frac{5200}{3} & \frac{560}{3} & 40 & \frac{800}{3} \\
 0 & \frac{1600}{3} & 160 & -\frac{800}{3} & \frac{560}{3} & \frac{2560}{3} & 200 & -40 \\
 8 & 80 & 108 & -\frac{80}{3} & 40 & 200 & 92 & \frac{20}{3} \\
 \frac{80}{3} & -80 & 40 & \frac{1280}{27} & \frac{800}{3} & -40 & \frac{20}{3} & \frac{2200}{9} \\
\end{array}
\right)\,.
}
The coefficients of the contraction to $T_{f\!f}$ are
\eq{
(y^B)=(T_B\cdot T_{f\!f})
=\Big(-48,-240,72,\frac{640}{3},0,120,-12,0\Big)^\tp
\,.
}
So the solution to \eqref{y=Qx} is
\eq{
(x^A)=(Q^{-1}y)
=\Big(-\frac{1}{3},-\frac{1}{2},\frac{2}{3},\frac{3}{2},-1,2,-4,1\Big)^\tp\,,
}
leading to \eqref{Tff.identity}.

The numbers above are specific to $SU(3)$. 
But some results not involving $P^{27}$ can be calculated for general $SU(n)$ by using birdtrack techniques\,\cite{cvitanovic}.
We list some of these general results below.
\eqali{
(T_1\cdot T_1)&=d^3\,,
&
(T_1\cdot T_3)&=3d^2\,,
&
(T_1\cdot T_S)&=d\,,
\cr
(T_3\cdot T_3)&=\frac{3d}{2}[d^2+d+4]\,,
&
(T_S\cdot T_S)&=\frac{d}{8}[d^2+3 d+4]\,,
&
(T_3\cdot T_S)&=\frac{3}{2}d(d+1)\,,
\cr
(T_{4a}\cdot T_{4a})&=dC_D^4\,,
&
(T_{2}\cdot T_2)&=3(C_D d)^2\,,
}
where $d=n^2-1$ and $C_D=2(n^2-4)/n$.
The results involving $d'_{ijk}$ are valid for $n\ge 3$ while the rest are valid for $n\ge 2$.
Together with
\eqali{
(T_{f\!f}\cdot T_1)&=-2nd\,,
&
(T_{f\!f}\cdot T_3)&=3nd\,,
&
(T_{f\!f}\cdot T_S)&=-\ums{2}nd\,,
}
we can set up equation \eqref{y=Qx} for $T_{f\!f}$ in $SU(2)$ and arrive at \eqref{Tff.identity:su2}.

\section{Hilbert series in general weak basis}
\label{ap:su4}

Considering all the spurions \eqref{spurions:su4} as $X_u,\YY,\YY^\dag,\MM,\MM^\dag \to q$, 
transforming under the group \eqref{GF:su4}, 
the Molien-Weyl formula for the unrefined Hilbert series is
\begin{align}
\label{Hilbert:su4} 
H(q) &= \frac{1}{(2\pi i)^7} \oint_{|z_1|=1} dz_1 \oint_{|z_2|=1} dz_2 \oint_{|z_3|=1} dz_3 \oint_{|z_4|=4} dz_1 \oint_{|z_5|=1} dz_5 \oint_{|z_6|=1} dz_6  \oint_{|z_7|=1} dz_7  \times \nonumber \\
    &\times (1-z_2 z_3)(z_3-z^2_2)(z_2-z^2_3)(1-z_4 z_6)(z_5-z^2_4)(z_6-z_4 z_5)(z_4 z_6-z^2_5) \times \nonumber \\
    &\times (z_4 - z_5 z_6)(z_5-z^2_6) z_1^3 z_2^{10} z_3^{10} z_4^5 z_5^5 z_6^5 z_7^{11}\PE'[q,z_j].
\end{align}
The PE, after factoring out some terms, is
\begin{align}
\PE'[q,z_j]&= \frac{1}{(1-q)^3 (1-q z_2 z_3)(z_3- qz_2^2)(z_2z_3- q)(z_3^2- qz_2)(z_2- qz_3^2)(z_2^2- qz_3)} \nonumber\\
       &\times \frac{1}{(z_4-q z_2 z_7)(z_5- q z_2z_4 z_7)(z_6- q z_2z_5z_7)(1- qz_2z_6 z_7)(z_3z_4- q z_7)(z_3z_5- qz_4 z_7)(z_3z_6-qz_5z_7)}  \nonumber\\
       &\times \frac{1}{(z_3-q z_6 z_7)(z_2z_4-q z_3 z_7)(z_2 z_5-q z_3 z_4 z_7)(z_2 z_6-q z_3z_5 z_7)(z_2- q z_3z_6 z_7)} \nonumber \\
       &\times \frac{1}{(z_7-qz_4z_3)(z_7 z_2-qz_4)(z_7z_3-qz_4z_2)(z_7z_4-qz_5z_3)(z_7z_4z_2-qz_5)(z_4z_3 z_7-qz_5z_2)(z_5z_7-qz_6z_3)} \nonumber\\
       &\times \frac{1}{(z_7z_5z_2-qz_6)(z_7z_5 z_3-qz_6z_2)(z_7z_6-qz_3)(z_2 z_6 z_7 -q)(z_7z_6z_3-qz_2)} \nonumber\\
       &\times \frac{1}{(z_4-qz_1)(z_5-qz_1z_4)(z_6-qz_1z_5)(1-qz_1z_6)} \nonumber\\
       &\times \frac{1}{(z_1-qz_4)(z_4z_1-qz_5)(z_5z_1-qz_6)(z_1z_6-q)}\,.
\end{align}


\end{document}